\documentclass[11pt]{article}

\usepackage{varioref}
\usepackage{amsmath,amsfonts,amsthm,mathrsfs}

\usepackage[normalem]{ulem}
\usepackage{multirow,color,graphics}
\usepackage{amsmath}
\usepackage{amsfonts}
\usepackage{amssymb}
\usepackage{jheppub}
\usepackage{enumerate}
\usepackage{fancyvrb}
\usepackage{verbatim}
\usepackage{wrapfig}
\usepackage{appendix}
\usepackage{amstext}
\usepackage{amssymb}
\usepackage{graphicx}
\usepackage{color}
\usepackage{varioref}
\usepackage{multirow,graphics}
\usepackage{epstopdf}
\newcommand{\nn}{\nonumber}

\numberwithin{equation}{section}

\def\[{\left[}
\def\]{\right]}
\def\({\left(}
\def\){\right)}
\def\d{\partial}

    \newcommand{\br}[1]{{\langle\!\!\langle} #1 {\rangle\!\!\rangle}}

    \newcommand{\beq}{\begin{equation}}
    \newcommand{\eeq}{\end{equation}}
    \newcommand\beqa{\begin{eqnarray}}
    \newcommand\eeqa{\end{eqnarray}}
\newcommand\bea{\begin{array}}
\newcommand\eea{\end{array}}

\newcommand{\bQ}{{\bf Q}}

\newcommand{\bP}{{\bf P}}

\newcommand{\vint}{\int_{{\bf |}}}
\newcommand{\la}[1]{\label{#1}}
\newcommand{\eq}[1]{(\ref{#1})}

    \def\bQ{{\bf Q}}
    \def\bP{{\bf P}}

\makeatletter
     \@ifundefined{usebibtex}{} {}
\makeatother

    \def\bQ{{\bf Q}}

        \def\bP{{\bf P}}

\newcommand{\cN}{\mathcal{N}}

\newcommand{\cO}{\mathcal{O}}

\title{Quantum Spectral Curve and Structure Constants in $\cN=4$ SYM: Cusps in the Ladder Limit}

\author[]{~~Andrea Cavagli\`a${}^{\,\displaystyle\measuredangle}$,}
\author[]{~~Nikolay Gromov${}^{\,\displaystyle\measuredangle,\,\angle}$,}
\author[1]{~~Fedor Levkovich-Maslyuk${}^{{\,\displaystyle\sphericalangle\,},\, }$\note{On leave from Institute for Information Transmission Problems, Moscow 127994, Russia}}

\affiliation[]{${}^{\displaystyle\,\measuredangle}$Mathematics Department, King's College London,
The Strand, London WC2R 2LS, UK}
\affiliation[]{${}^{\displaystyle\angle}$St.Petersburg INP, Gatchina, 188 300, St.Petersburg,
  Russia}
\affiliation[]{${}^{\displaystyle\,\sphericalangle}$Departement de Physique, Ecole Normale Superieure / PSL Research University, CNRS, 24 rue Lhomond, 75005 Paris, France\\ }

\emailAdd{andrea.cavaglia$\bullet$kcl.ac.uk}
\emailAdd{nikgromov$\circ$gmail.com}
\emailAdd{fedor.levkovich$\bullet$gmail.com}

\abstract{
We find a massive simplification in the non-perturbative expression for the structure constant of Wilson lines with $3$ cusps when expressed in terms of the key Quantum Spectral Curve quantities,
namely Q-functions. Our calculation is done for the configuration of $3$ cusps lying in the same plane with arbitrary angles in the ladders limit.
This provides strong evidence that the Quantum Spectral Curve 
is not only a  highly efficient tool for finding the anomalous dimensions but also
encodes correlation functions with all wrapping corrections taken into account to all orders in the `t Hooft coupling. 
We also show how to study the insertions of scalars coupled to the Wilson lines and
extend our results for the spectrum and the structure constants to this case. We discuss an OPE expansion of two cusps in terms of these states. Our results give additional support to the Separation of Variables strategy in solving the planar ${\cal N}=4$ SYM theory.
}

\begin{document}

\maketitle

\newpage

\section{Introduction}

Integrability is a unique tool allowing one to obtain exact non-perturbative results in fully interacting field theories even when the supersymmetry is of no use. 
The range of theories where  integrability is known to be applicable includes supersymmetric theories such as planar ${\cal N}=4$ SYM and ABJM theory, which are important from a holographic perspective.
Quite significantly, recently found examples of   integrable theories include a particular class of scalar models in 4D possessing no supersymmetry at all \cite{Gurdogan:2015csr,Caetano:2016ydc,Gromov:2017cja,Grabner:2017pgm,Kazakov:2018qbr}.  

Integrability methods of the type  used here started being developed in the seminal papers \cite{bfklint} in the QCD context and independently in \cite{Minahan:2002ve} for ${\cal N}=4$ SYM. After almost $20$ years of development it was shown that both approaches can be united by the Quantum Spectral Curve (QSC) formalism
\cite{Gromov:2013pga,Gromov:2014caa}\footnote{The QSC formalism was also developed for the ABJM model in 
\cite{Cavaglia:2014exa,Bombardelli:2017vhk}
} of which both are some particular limits~\cite{Gromov:2014caa,Alfimov:2014bwa}.

The QSC was initially developed with the primary goal of computing the spectrum of anomalous dimensions or, equivalently, two point correlators. 
The QSC is based on the Q-system, a system of functional equations on Q-functions (see \cite{Gromov:2017blm,Kazakov:2018ugh} for a recent review). At the same time, the Q-functions are known to play the role of the wave functions in the Separation of Variables (SoV) program initiated for  quantum integrable models in 
\cite{Sklyanin:1989cg,Sklyanin:1991ss,Sklyanin:1992eu,Sklyanin:1995bm} 
and recently generalized to $SU(N)$ spin chains in \cite{Gromov:2016itr}  leading to a new algebraic construction for the states (see also \cite{Smirnov2001,Chervov:2007bb}). In all these models the Q-functions (Baxter polynomials in this case) give the wave functions in separated variables \footnote{Some inspiring results were obtained in \cite{Lukyanov:2000jp,Negro:2013wga}.}.\footnote{Moreover, even without use of the QSC, the standard SoV approach has already given a number of results for correlators in $\cN=4$ SYM \cite{Sobko:2013ema,Jiang:2015lda,Kazama:2016cfl,Kazama:2015iua,Kazama:2014sxa,Kazama:2013qsa,Kazama:2013rya,Kazama:2012is,Kazama:2011cp} though without  finite size wrapping effects or at the classical level.
} From this perspective it is natural to expect that the Q-functions of the QSC
construction in ${\cal N}=4$ SYM contain much more information than the spectrum and should also play an important role for more general observables.

There are a few important lessons one can learn from the simple spin chains. In particular one should introduce ``twists" (quasi-periodic boundary conditions/external magnetic field) in order for the SoV construction to work nicely. One of the main reasons why the twists are important is that they break  global symmetry and remove degeneracy in the spectrum. This makes the map between the Q-functions and the states bijective.
Fortunately, one can rather easily introduce twists into the QSC construction \cite{Gromov:2013qga,Gromov:2015dfa,Kazakov:2015efa} (see also \cite{Klabbers:2017vtw}), however the interpretation of these new parameters is not always clear from the QFT point of view. The $\gamma$-deformation of $\cN=4$ SYM \cite{Frolov:2005dj,Alday:2005ww,Frolov:2005iq,Beisert:2005if} is one of the cases which is rather well understood, but only breaks the R-symmetry part (dual to the isometries of $S^5$ part of AdS/CFT) of the whole ${\rm PSU}(2,2|4)$ group.\footnote{Recently in \cite{Guica:2017mtd} it was understood how to study the spectrum for a more general deformation.}

\begin{figure}
	    \centering
	\includegraphics[scale=1.1]{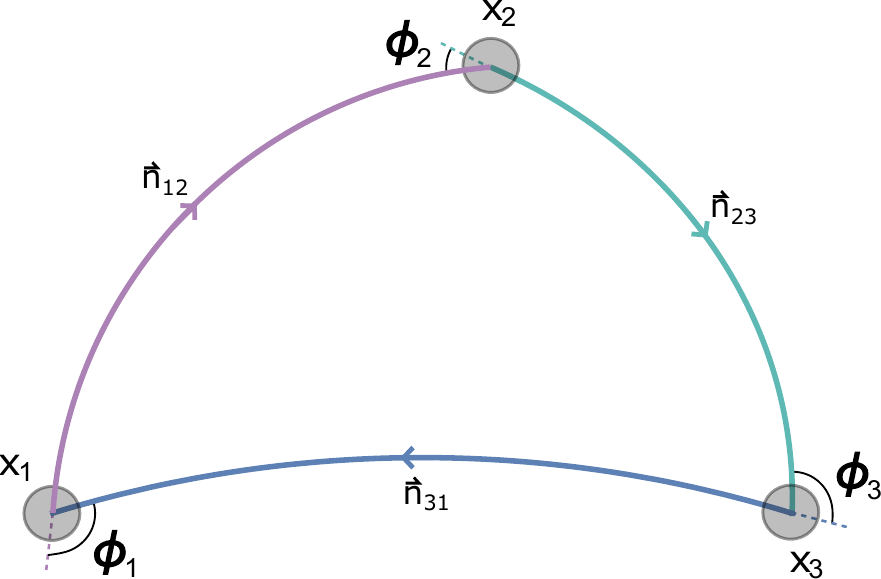}
	\caption{The Maldacena-Wilson loop with three cusps. The cusps are connected by circular arcs with $3$ different scalars $\vec \Phi\cdot\vec n_{ij}$ coupled to the three different arcs. The expectation value of this object behaves exactly in the same way as a three point correlation function of 3 local operators but provides additional $6$ parameters ($2$ for each cusp) $\phi_1,\;\phi_2,\;\phi_3$ and $\cos\theta_{1}=\vec n_{12}\cdot\vec n_{23},\;\cos\theta_{2}=\vec n_{23}\cdot\vec n_{31},\;\cos\theta_{3}=\vec n_{31}\cdot\vec n_{12}$, which are associated with twists in the QSC description.}
	\label{fig:triangleonsteroids}
\end{figure}

The situation where the twist in both $AdS_5$ and $S^5$ appears naturally is the cusped Maldacena-Wilson loop. In this paper we consider the correlation function of $3$ cusps for $3$ general angles (see Fig.~\ref{fig:triangleonsteroids}). We consider a ladders limit \cite{Erickson:1999qv, Erickson:2000af} where the calculation can be done to all loop orders starting from Feynman graphs. We observe that the result obtained as a resummation of the perturbation theory takes a stunningly simple form when expressed in terms of the Q-functions, which we produced from the QSC.

\paragraph{Set-up and the Main Results.}

The Maldacena-Wilson lines we consider are defined as
\beq
W={\rm Pexp}  \int\,d\tau  \(i A_\mu \dot x^\mu +\Phi^a n^a  | \dot x |\) ,
\eeq
where $n^a$ is a constant unit 6-vector parameterizing the coupling to the scalars $\Phi^a$ of $\cN=4$ SYM. The observable we study is the Wilson loop defined on a planar triangle made of three circular arcs\footnote{Each arc is the image of a straight line segment under a conformal transformation and thus is locally 1/2-BPS.}, see Fig.~\ref{fig:triangleonsteroids}. It is parameterized by three cusp angles $\phi_i$ at its vertices and also three angles $\theta_i$ between the couplings to scalars on the lines adjacent to each vertex. At each cusp we have a divergence controlled by the celebrated cusp anomalous dimension $\Gamma_{\rm cusp}(\phi_i,\theta_i)$ which can be efficiently studied via integrability \cite{Correa:2012hh,Drukker:2012de,Gromov:2015dfa} and is analogous to the local operator scaling dimensions in its mathematical description by the QSC. Due to this we will use notation $\Delta$ for the cusp dimension. To regularize the divergence we cut an $\epsilon$-ball at each of the cusps. The whole Wilson loop has a conformally covariant dependence on the cusp positions and defines the structure constant $C_{123}$ for  a 3-point correlator of three cusps.

We focus on the ladders limit in which $\theta_i\to i \infty$ while the 't Hooft coupling $g=\sqrt{\lambda}/({4\pi})$ goes to zero with the finite combinations
\beq
    \hat g_i=\frac{g}{2} e^{-i\theta_i/2}\la{lad}
\eeq
playing the role of three effective couplings. The perturbative expansion for $\Delta$ can then be resummed to all orders leading to a stationary Schr\"odinger equation \cite{Erickson:1999qv, Erickson:2000af,Correa:2012nk}. However, the 3-cusp correlator is much more nontrivial and depends on three couplings $\hat\lambda_i$ which we can vary separately. We have studied the case when two of them are nonzero, corresponding to the structure constant we denote by $C^{\bullet \bullet \circ}_{123}$. The result may be written in terms of the Schr\"odinger wave-functions but it is a highly complicated integral which does not offer much structure.  Yet once we rewrite it in terms of the QSC Q-functions  $q(u)$, we observe miraculous cancellations leading to a surprisingly simple expression
\beq\la{correlator}
\boxed{
C^{\bullet \bullet \circ}_{123} = 
\, \frac{\, \br{  q_{1} \, q_{2}\, e^{-\phi_3 u} }
}{\sqrt{ \br{ q_1^2}\br{ q_2^2}}  
} \ \ ,}
\eeq
where the bracket $\br{f(u)}$ is defined for the functions 
which behave as $\sim e^{u\beta}u^\alpha$ at large $u$ and are
analytic  for all ${\rm Re}\;u>0$ as 
\beq
\la{eq:thebracket}
\br{f(u) }\equiv \(2\sin\frac{\beta}{2}\)^\alpha\int_{c -i \infty}^{c+i\infty}  f(u)\frac{du}{2\pi i u}\;\;,\;\;c>0\;.
\eeq
The functions $q_1(u),q_2(u)$ describe the first and the second cusp, while $e^{-\phi_3 u}$ is just the Q-function at zero coupling corresponding to the third cusp. Each of the Q-functions solves a simple finite difference equation \eq{Bax2p}. This is precisely the kind of result one expects for an integrable model treated in separated variables.
Note that all the dependence on the angles and the couplings is coming solely through the Q-functions, which depend nontrivially on these parameters,
in particular at large $u$ we have $q_i(u)\simeq u^{\Delta_i}e^{\phi_i u}$.

We also found a very simple expression for the derivative of $\Delta$ w.r.t. the coupling $\hat g$ and the angle $\phi$
in terms of the bracket $\br{ \cdot}$
\beq\la{Cinsert}
-\frac{1}{4}\frac{\partial \Delta}{\partial\hat g^2}=\frac{\br{q^2\frac{1}u}}{\br{q^2}}\;\;,\;\;-2\, \frac{\d(\sin\phi\Delta)}{\d\phi}=\frac{\br{q^2 u}}{\br{q^2}}\;,
\eeq
which has the form very similar to \eq{correlator} with $q_1=q_2=q$ and different insertions in the numerator! These quantites 
can be interpreted as structure constants of two cusps with a local BPS operator \cite{Costa:2010rz}.

In the limit when the triangle collapses to a straight line, this configuration has recently attracted much attention as it defines a 1d CFT on the line \cite{Giombi:2017cqn,Beccaria:2017rbe,Kim:2017sju,Cooke:2017qgm,Kim:2017phs}. In particular the structure constants we consider were computed in \cite{Kim:2017sju} by resumming the diagrams using the exact solvability of the Schr\"odinger problem at $\phi=0$. Our results in the zero angle limit can be simplified further by noticing that for $\phi_i\to 0$ the integral is saturated by the leading large $u$ asymptotics of the integrand. This leads to $\br{ q_i q_j }\to {1/\Gamma(1-\Delta_i-\Delta_j)}$, reproducing the results of \cite{Kim:2017sju}.

As a byproduct, we also resolved the question of how to use integrability to compute the anomalous dimension for the cusp with an insertion of the same scalar as that coupled to the Wilson lines.
We propose that it simply corresponds to one of the excited states in the Schr\"odinger equation (and to a well-defined  analytic continuation in the QSC outside the ladders limit). We verified this claim at weak coupling by comparing with the direct perturbation theory calculation of \cite{Alday:2007he}\footnote{The result in that paper is for $\theta=0$, whereas we consider $\theta=i\infty$, however we expect the 1-loop result should not depend on $\theta$.}. Very recently the importance of the cusps with such insertions were further motivated in \cite{Bruser:2018jnc} where the $3$ loop result was extracted.

\begin{figure}
    \centering
    \includegraphics[scale=0.6]{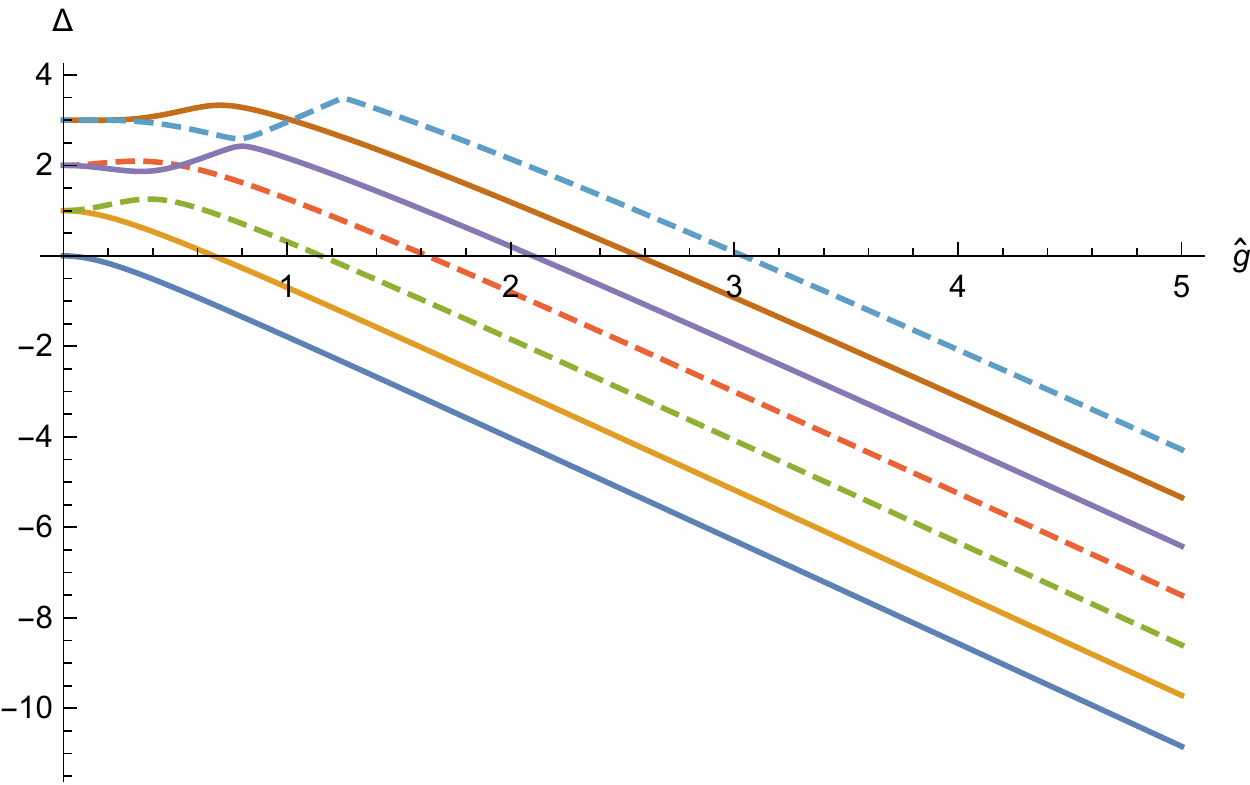}
    \includegraphics[scale=0.6]{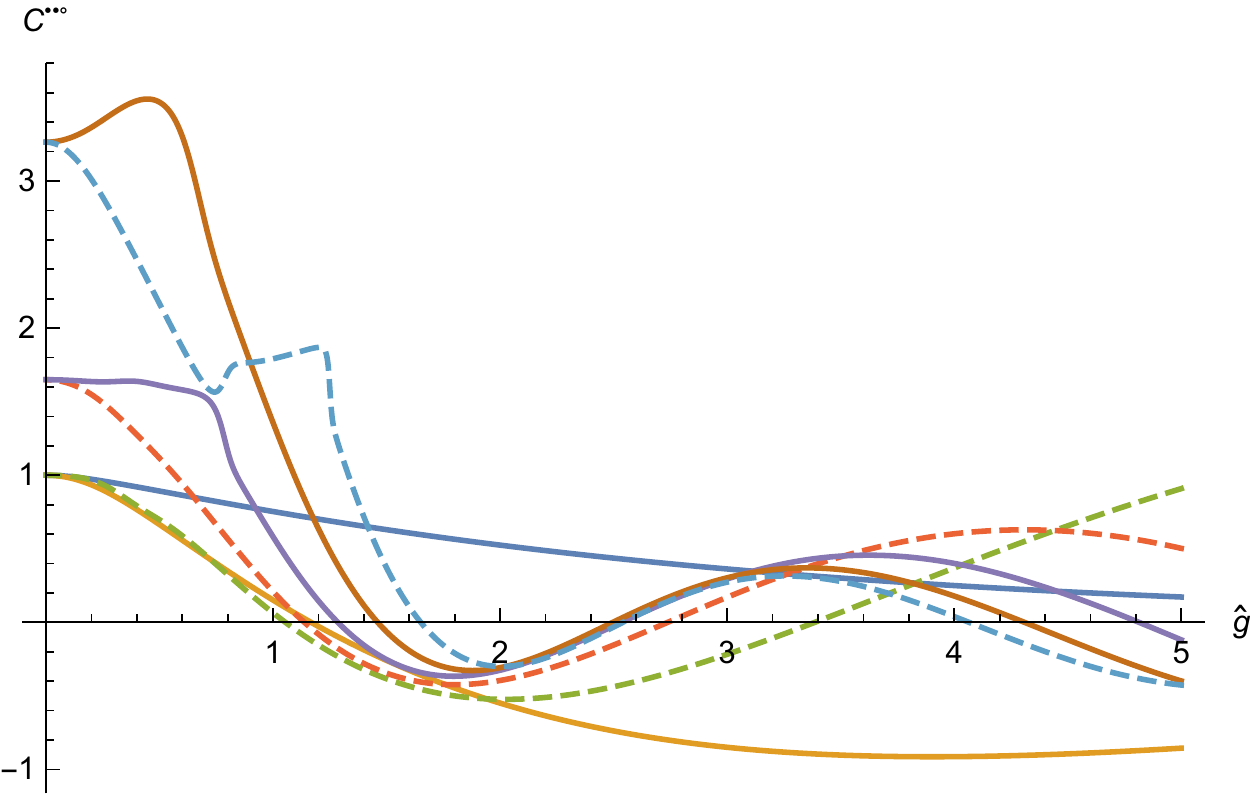}
    \caption{The spectrum (left) and the diagonal Heavy-Heavy-Light correlator given by \eq{correlator} (right) for the first several states ($n=0,1,\dots,7$), with all angles equal to $\phi=1 $. The solid blue line corresponds to the usual cusp, while others correspond to excited states with scalar insertions discussed in section~\ref{sec:excited}.}
    \label{fig:HHLnum2}
\end{figure}

We demonstrate 
some of our results in Fig.~\ref{fig:HHLnum2} where we show the plots of the spectrum and the structure constant for a range of the effective coupling $\hat g$.

\paragraph{Structure of the paper.}
The rest of the paper is organized as follows. In Sec. \ref{sec:qsclad} we briefly review the QSC and  present the Baxter equation to which it reduces in the ladders limit. We also derive compact formulas for the variation of $\Delta$ with respect to the coupling and the angle $\phi$. In Sec. \ref{sec:BS} we write the regularized 2-pt function in terms of the Schr\"odinger equation wave functions, in particular deriving the pre-exponent normalization which is important for 3-pt correlators. We also relate the wave functions to the QSC Q-functions via a Mellin transform. In Sec. \ref{sec:3pt1} we study the 3-cusp correlator and  derive our main result for the structure constant \eq{correlator}. In Sec. \ref{sec:excited} we describe the interpretation of excited states in the Schr\"odinger problem as insertions at the cusp. We generalize our results for 3-pt functions to the excited states and provide both perturbative and numerical data for their scaling dimensions. In Sec. \ref{sec:smallPhi} we describe the limit when the 3-cusp configuration degenerates, in particular reproducing the results of \cite{Kim:2017sju} when all angles become zero. In Sec. \ref{sec:num} and \ref{sec:weak} we present numerical and perturbative results for the structure constants. Finally in Sec. \ref{sec:ope} we interpret the regularized 2-pt function as a 4-cusp correlator for which we write an OPE-type expansion in terms of the structure constants, perfectly matching our previous results. In Sec. \ref{sec:concl} we present conclusions. The appendices contain various technical details, in particular the detailed strong coupling expansion for the spectrum.

\section{Quantum Spectral Curve in the ladders limit}

\label{sec:qsclad}

In this section we provide all necessary background for this paper about the Quantum Spectral Curve (QSC).
More technical details are given in Appendix~\ref{app:qsc}.

The QSC provides a finite set of equations describing non-perturbatively the cusp anomalous dimension $\Delta$ at all values of the parameters $\phi,\theta$ and any coupling $g$. Let us briefly review this construction and then discuss the form it takes in the ladders limit.
The QSC was originally developed in \cite{Gromov:2013pga,Gromov:2014caa} for the spectral problem of local operators in $\cN=4$ SYM. It was extended in \cite{Gromov:2015dfa} to describe the cusp anomalous dimension, reformulating and greatly simplifying the TBA approach of  \cite{Correa:2012hh, Drukker:2012de}. The QSC is a set of difference equations (QQ-relations) for the Q-functions which are central objects in the integrability framework. When supplemented with extra asymptotics and analyticity conditions, these relations fix the Q-functions and provide the exact anomalous dimension $\Delta$ (see \cite{Gromov:2017blm} for a pedagogical introduction and \cite{Kazakov:2018ugh} for a wider overview).

The QSC is based on 4+4 basic Q-functions denoted as $\bP_a(u)$, $a=1,\dots,4$ and $\bQ_i(u)$, $i=1,\dots,4$ which are related to the dynamics on $S^5$ and on $AdS_5$ correspondingly. The $\bP$-functions are analytic functions of $u$ except for a cut at $[-2g,2g]$. 
They can be nicely parameterized in terms of an infinite set of coefficients that contain full information about the state, including $\Delta$. Details of this parameterization are given in Appendix \ref{app:qsc}. The other 4 basic Q-functions $\bQ_i$ are indirectly determined by $\bP_a$ via the 4th order Baxter equation \cite{Alfimov:2014bwa}
\beqa\la{bax5}
\bQ^{[+4]}_iD_0
&-&
\bQ^{[+2]}
\[
D_1-\bP_a^{[+2]}\bP^{a[+4]}D_0
\]
+
\bQ
\[
D_2-\bP_a\bP^{a[+2]}D_1+
\bP_a\bP^{a[+4]}D_0
\]\\
&-&
\bQ^{[-2]}
\[
\bar D_1+\bP_a^{[-2]}\bP^{a[-4]}\bar D_0
\]
+\bQ^{[-4]}\bar D_0
=0\nn \ ,
\eeqa
where the coefficients $D_n, \bar D_n$ are simple determinants built from $\bP_a$ and are given explicitly in Appendix \ref{app:qsc}\footnote{The functions $\bP^a$ appearing here are defined by $\bP^a=\chi^{ab}\bP_b$ with the only non-zero entries of $\chi^{ab}$ being $\chi^{14}=-\chi^{23}=\chi^{32}=-\chi^{41}=-1\ $.}. Here we used the shorthand notation
\beq
    f^\pm=f(u\pm\tfrac{i}{2}), \ \ f^{[+a]}=f(u+\tfrac{ia}{2}) \ .
\eeq
Being of the 4th order, this Baxter equation has four independent solutions which precisely correspond to the four Q-functions $\bQ_i$. Different solutions can be identified by the four possible asymptotics
$	\bQ_i\sim u^{1/2\pm \Delta}e^{\pm u\phi}$
which uniquely fix the basis of four Q-functions up to a normalization if we also impose that
the solutions $\bQ_i(u)$ are analytic in the upper half-plane of $u$, which is always possible to do. 
Then they will have an infinite set of Zhukovsky cuts in the lower half-plane with branch points at $u=\pm 2g-in$ (with $n=0,1,\dots$).

Finally in order to close the system of equations
we need to impose what happens after the analytic continuation through the cut $[-2g,2g]$.
It was shown in  \cite{Gromov:2015dfa} that in order to close the equations one should impose the following ``gluing" conditions
\beqa\la{qtil}
&&\tilde { q}_1(u)={q}_1(-u)\\
&&\tilde { q}_2(u)={ q}_2(-u)\\
&&\tilde {q}_3(u)=a_1\sinh(2\pi u){ q}_2(-u)+{ q}_3(-u)\\
&&\tilde { q}_4(u)=a_2\sinh(2\pi u){ q}_1(-u)+{ q}_4(-u)\;,
\eeqa
where $q_i(u)=\bQ_i(u)/\sqrt{u}$ and $\tilde q_i$ is its analytic continuation under the cut. These relations fix both $\bP$- and $\bQ$-functions and allow one to extract the exact cusp anomalous dimension $\Delta$ from large $u$ asymptotics. The equations presented above are valid at any values of $g$ and the angles $\phi,\theta$. For the purposes of this paper we have to take the ladders limit of these equations. We will see that they simplify considerably.

\subsection{Baxter equation in the ladders limit}

\label{sec:baxter}

In the ladders limit \eq{lad} the coupling $g$ goes to zero and the QSC greatly simplifies as all the branch cuts of the Q-functions collapse and simply become poles. This limit was explored in detail in \cite{Gromov:2016rrp} for the special case $\phi=\pi$ corresponding to the flat space quark-antiquark potential. Here we briefly generalize these results to the generic $\phi$ case.

The key simplification is that the 4th order Baxter equation \eq{bax5} on $\bQ_i$ factorizes into two 2nd order equations, the first one being
\beq
\label{Bax2p}\boxed{
	\left(-2u^2 \cos \phi +2\Delta  u \sin \phi +4 \hat{g}^2\right){q}(u) +u^2
   { q}(u-i)+u^2 { q}(u+i)=0}
\eeq
and another equation obtained by $\Delta\to-\Delta$. 
This follows from the fact that coefficients $A_n,B_n$ entering $\bP$'s via \eq{cuspas}, \eq{fgAB} scale as $\sim 1$ in the ladders limit\footnote{We assumed this in analogy with the $\phi=\pi$ case and verified it by self-consistency}. Then as in \cite{Gromov:2016rrp} one can 
carefully expand the 4th order Baxter equation for $t\equiv e^{i\theta/2}\to 0$ and recover the 2nd order equation \eq{Bax2p}. As the large $u$ behaviour of $q(u)$ is fixed by the Baxter equation \eq{Bax2p}, we denote them as $q_+$ and $q_-$ according to the large $u$ asymptotics 
$q_\pm\sim e^{\pm\phi u} u^{\pm\Delta}$.
For example in the weak coupling limit $\hat g=0$ for $\Delta=0$ we see that $q_\pm$ are simply
\beq\la{weakq}
q^{(0)}_+=e^{+\phi u}\;\;,\;\;q^{(0)}_-=e^{-\phi u}\;.
\eeq
At finite $\hat g$ the Q-functions become rather nontrivial. 
While $q_\pm(u)$ are regular in the upper half-plane including the origin, 
they have poles in the lower half-plane at $u=-in, \ \ n=1,2,\dots$.

The equation \eq{Bax2p} is just an $sl(2)$ (non-compact) spin chain Baxter equation, similarly to \cite{Gromov:2017cja}. This is expected based on symmetry grounds. What is less trivial is the ``quantization condition" i.e. the condition which will restrict $\Delta$ to a discrete set. It was first derived in \cite{Gromov:2016rrp} for $\phi\to\pi$ and later generalized to the very similar calculation of two-point functions in the fishnet model \cite{Gromov:2017cja}. The derivation of the quantization condition for any  $\phi$ is
done in Appendix~\ref{app:qsc} and leads to the following result:
\beq
\label{qquant}
\boxed{
    	\Delta=-\frac{2\hat g^2}{\sin\phi}
	\frac{q_+(0)\bar q_+'(0)+\bar q_+(0)q_+'(0)}{q_+(0)\bar q_+(0)}  } \;.
\eeq
Together with the Baxter equation \eq{Bax2p}, this relation fixes $\Delta$ as well as $q_+$.

Note that the r.h.s.\! of \eq{qquant} contains $q_+$, which has to be found from the Baxter equation and thus also depends on $\Delta$ nontrivially. Due to this \eq{qquant} is a non-linear equation, which may have several solutions. Some intuition behind it becomes clearer after reformulating the problem in a more standard Schr\"odinger equation form as we will see in section \ref{sec:cuspdiv}.
At the same time we see that we only need $q_+$ to find the spectrum. For this reason we will simply denote it as $q(u)$ in the rest of the paper.

The meaning of the Q-functions from the QFT point of view is still a big mystery.
There is no known observable in the field theory which is known to correspond to them directly. However in the ``fishnet" theory, which is a particular limit of ${\cal N}=4$ SYM, such an object was recently identified~\cite{Gromov:2017cja}.
Here, in the ladders limit we will be able to relate $q(u)$ with a solution of the Bethe-Salpeter equation, which 
resums the ladder Feynman diagrams and thus has direct field theory interpretation.

\subsection{Scalar product and variations of $\Delta$}
In this section we demonstrate the significance of the 
bracket $\br{\cdot}$, which we defined in the introduction in \eq{eq:thebracket}.
In particular we will derive a closed expression for ${\d\Delta}/{\d\hat g}$ which 
can be considered as a correlation function of two cusps with the Lagrangian \cite{Costa:2010rz}.
Even though that seems to be the simplest application of the QSC 
for the computation of the $3$-point correlators,
it is not yet known how to write the result for $\partial\Delta/\d g$ for the general state in a closed form.
We demonstrate here that this is in fact possible to do at least in our simplified set-up.

First we rewrite the Baxter equation \eq{Bax2p} by defining the following finite difference operator
\beq
\hat O \equiv \frac{1}{u}\left[
(4\hat g^2-2u^2\cos\phi+2\Delta u\sin\phi)
+u(u-i)D^{-1}+u(u+i)D
\right]\frac{1}{u} 
\eeq
where $D$ is a shift by $i$ operator
so that the Baxter equation \eq{Bax2p} becomes
\beq
\hat O q(u)=0\;.
\eeq
Now we notice that this operator is ``self-adjoint" under the integration along the vertical contour to the right from the origin, meaning that
\beq
\vint q_1(u)\hat O q_2(u) du=
\vint q_2(u)\hat O q_1(u) du\;\;,\;\;\vint\equiv \int_{c-i\infty}^{c+i\infty} \ .
\eeq
where $c>0$\footnote{Due to the sign of the exponential factors in the asymptotics of $q(u)$ (where we assume $\phi > 0$ ), the integrals would vanish trivially if we chose an integration contour with $c < 0$. }. Indeed, consider the term with $D$:
\beq
\vint q_1(u)u(u+i) Dq_2(u) du=
\vint q_1(u)u(u+i) q_2(u+i) du=
\vint q_2(u)u(u-i)D^{-1} q_1(u) du
\eeq
which now became the term with $D^{-1}$ acting on $q_1(u)$.
In the last equality we changed the integration variable $u\to u-i$.
The fact that $\hat O$ has this property immediately leads to the great simplification for the expression for $\partial\Delta/\d g$. We can now apply the standard QM perturbation theory logic.

Changing the coupling and/or the angle $\phi$ 
will lead to a perturbation of both the operator $\hat O$ and the q-function
in such a way that the Baxter equation is still satisfied ,
\beq
(\hat O+\delta \hat O)(q+\delta q)=0\;\;,\;\;\delta \hat O = \frac{1}{u^2} (8\hat g\delta \hat g+2u\sin\phi \delta \Delta
+2u^2\sin\phi\delta \phi+2\Delta u\cos\phi\delta \phi
)\;.
\eeq
An explicit expression for $\delta q$ could be rather hard to find, but luckily we can get rid of it by contracting $(\hat O+\delta \hat O)(q+\delta q)$ with the original $q(u)$:
\beq
0=\vint q(\hat O+\delta \hat O)(q+\delta q)du=\vint(q+\delta q)(\hat O+\delta \hat O)q du=
\vint(q+\delta q)\delta \hat O q du 
\eeq
At the leading order in the perturbation we can now drop $\delta q$ to obtain
\beq
\vint q (8\hat g\delta \hat g+2u\sin\phi \delta \Delta
+2u^2\sin\phi\delta \phi+2\Delta u\cos\phi\delta \phi
) q \frac{du}{u^2} = 0 \ ,
\eeq
so that
\beq
\label{eq:dddg}\frac{\d \Delta}{\d \hat g}=-\frac{4\hat g}{\sin\phi}\frac{\vint\frac{q^2}{u^2} du}
{\vint\frac{q^2}{u} {du}}
\;\;,\;\;
\frac{\d \Delta}{\d \phi}=-\frac{\vint q^2 du}
{\vint\frac{q^2}{u} {du}}-\Delta\cot\phi
\;.
\eeq
In terms of the bracket $\br\cdot$ this becomes
\beq
\boxed{
-\frac{1}{4}\frac{\partial \Delta}{\partial\hat g^2}=\frac{\br{q^2\frac{1}u}}{\br{q^2}}\;\;,\;\;-2 \, \frac{\d(\sin\phi\Delta)}{\d\phi}=\frac{\br{q^2 u}}{\br{q^2}}\;.
}
\eeq
This very simple equation is quite powerful. For example by plugging the leading order $q=e^{u\phi}$ from \eq{weakq} and computing the integrals by poles at $u=0$ we  get
\beq
\frac{\d \Delta}{\d \hat g}=-\frac{4\hat g}{\sin\phi}2\phi+{\cal O}(\hat g^3)\;,
\eeq
which gives immediately the one loop dimension $\Delta=-\hat g^2\frac{4\phi}{\sin\phi}+{\cal O}(\hat g^4)$.

Furthermore, another interesting property of the bracket is that solutions with different $\Delta's$ are orthogonal to each other.
Indeed, consider two solutions $q_a$ of the Baxter equation with two different dimensions $\Delta_a$, such that $\hat O_1 q_1=\hat O_2 q_2=0$. Then
\beq
0=\vint q_1(u)(\hat O_1-\hat O_2)q_2(u) du =
(\Delta_1-\Delta_2)2\sin\phi\vint \frac{q_1(u)q_2(u)}{u} du \ ,
\eeq
from which we conclude that $\br{q_1(u)q_2(u)}=0$.

In the next section we relate the Q-function to the solution of the Bethe-Salpeter equation resumming the ladder diagrams for the two point correlator.

\section{Bethe-Salpeter equations and the Q-function}\label{sec:BS}
\begin{figure}
    \centering
    \includegraphics[scale=1.1]{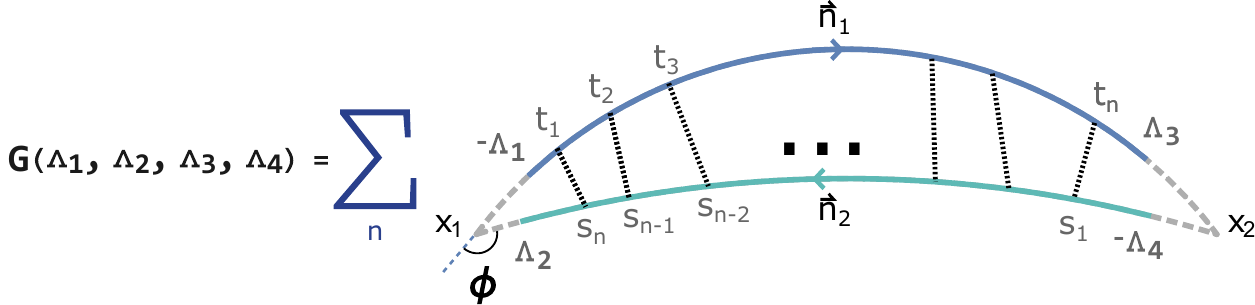}
    \caption{The two cusp correlator with four different cut-offs $\Lambda_a$, which can be considered as a particular case of $4$-cusp correlator. We take $n$ points along each of the circular arcs and connect them with scalar propagators. We have to integrate
    over the domain $-\Lambda_1<t_1<t_2<\dots<t_n<\Lambda_3$ and $-\Lambda_4<s_1<s_2<\dots s_n<\Lambda_2$. One should use a specific parameterization given in \eq{eq:param}.}
    \label{fig:G4}
\end{figure}

\label{sec:schr}
In this section we consider a two cusp correlator with amputated cusps shown on Fig.~\ref{fig:G4} which we denote by $G(\Lambda_1,\Lambda_2,\Lambda_3,\Lambda_4)$.
We derive an expression for it re-summing the ladder diagrams.
To do this we write a Bethe-Salpeter equation and then reduce it to a stationary Schr\"odinger equation, expressing 
$G$ in terms of the wave functions and energies of the Schr\"odinger problem.
After that we discuss the relation between the wave functions and the Q-functions introduced in the previous section.

\subsection{Bethe-Salpeter equation}\label{sec:cuspdiv}

Our goal in this section is reviewing the field-theoretical definition of the cusp anomalous dimension and its computation in the ladder limit, where it relates to the ground state energy of a simple Schr\"odinger problem.

First we define more rigorously the object from the  Fig.~\ref{fig:G4}. We are computing an expectation value
\beq
\label{G4WW}
G(\Lambda_1, \Lambda_2 , \Lambda_3, \Lambda_4 ) = \left\langle{\rm Tr\;} W_{\vec x_+(-\Lambda_1)}^{\vec x_+(+\Lambda_3)}(\vec n_1) \;\;W_{\vec x_-(-\Lambda_4)}^{\vec x_-(+\Lambda_2)}(\vec n_2)\right\rangle ,
\eeq
with 
\beq
W_{x}^{y}(\vec n)={\rm Pexp}\int_{x}^y \(iA_\mu d x^\mu+\Phi^a n^a  | dx |\) . \label{eq:Wxyop}
\eeq  
For simplicity we can assume that the contours belong to the $(*,*,0,0)$ two dimensional plane (which can be always achieved with a suitable rotation) and we use a particular ``conformal" parameterization of the circular arcs by
\beq\la{eq:param}
\vec{x}_\pm(s) = ( \text{Re}( \zeta_\pm(s) ) , \text{Im}(\zeta_\pm(s) ) , 0, 0 ) ,
\eeq
where
\beqa
\zeta_{_\pm}( s ) &=& z_1 + \frac{  (z_2 - z_1 ) }{1\mp i e^{\mp s + i (\chi\pm\phi)/2} } \label{eq:zazb}
\eeqa
such that $\vec x_1 \equiv ( \text{Re}( z_1 ) , \text{Im}(z_1) , 0, 0 ) =\vec x_\pm(\mp\infty)$ and $\vec x_2=( \text{Re}( z_2 ) , \text{Im}(z_2) , 0, 0 ) =\vec x_\pm(\pm\infty)$. Here $\vec x_+$ corresponds to the upper arc in Fig.~\ref{fig:G4}, and $\vec x_-$ to the lower one.
The configuration has one parameter $\chi$, which allows one to bend two arcs simultaneously keeping the angle between them fixed. This is the most general configuration of two intersecting circular arcs up to a rotation.

Next we notice that in the ladders limit we can neglect gauge fields so we get\footnote{Note also that in the ladders limit the orientation of the Wilson line is irrelevant, e.g. $\langle W^{\vec y}_{\vec x}(\vec n)\rangle=\langle W^{\vec x}_{\vec y}(\vec n)\rangle$.}
\beqa\la{eq:BS}&&
\d_{\Lambda_3}\d_{\Lambda_4} G(\Lambda_1, \Lambda_2 , \Lambda_3, \Lambda_4 ) =\\ \nn&& \left\langle{\rm Tr}\; W_{\vec x_+(-\Lambda_1)}^{\vec x_+(+\Lambda_3)}(\vec n_1) \;\;\Phi^a n_1^a |\dot{\vec x}_+(\Lambda_3)|\;\;\Phi^b n_2^b |\dot{\vec x}_-(-\Lambda_4)|\;\;W_{\vec x_-(-\Lambda_4)}^{\vec x_-(+\Lambda_2)}(\vec n_2)\right\rangle ,
\eeqa
which gives
\beq
\d_{\Lambda_3}\d_{\Lambda_4} G(\Lambda_1, \Lambda_2 , \Lambda_3, \Lambda_4 )=
 G(\Lambda_1, \Lambda_2 , \Lambda_3, \Lambda_4 )P( -\Lambda_4,  \Lambda_3) \ ,
\eeq
where the last term is the scalar propagator \beq
P( s,  t) =   4 \, \hat{g}^2  \, 
\frac{ | \dot{ \vec x}_- (s) | \, | \dot{ \vec x}_+ (t) |}{ | \vec x_+(t) - \vec x_-(s) |^2 } \label{eq:Pst}
\eeq
with  $\hat g^2=  g^2 \vec n_a \cdot \vec n_b/2$
(which is equivalent in the ladders limit to the definition of $\hat g$ in \eq{lad} as $\vec n_1 \cdot \vec n_2=\cos\theta$). The main advantage of the parameterization we used is that the propagator $P(s,t)$ is a function of the sum $s+t$:
\beq\la{eq:goodpropagator}
P( s,t) = \frac{2\hat{g}^2}{ \cosh(s+t) + \cos(\phi)}\; .
\eeq
Finally, we have to specify the boundary conditions. We notice that whenever one of the Wilson lines degenerates to a point the expectation value in the ladders limit becomes $1$, which implies
\beq
G(\Lambda_1,\Lambda_2,-\Lambda_1,\Lambda_4)=
G(\Lambda_1,\Lambda_2,\Lambda_3,-\Lambda_2)=1\;.
\eeq

\begin{figure}
    \centering
    \includegraphics[scale=1.6]{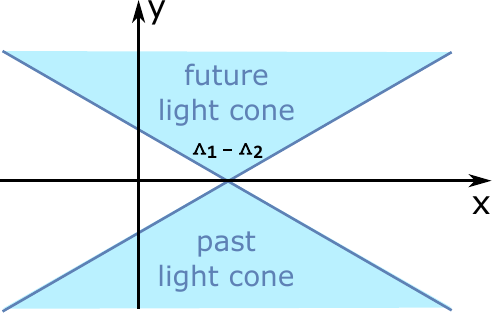}
    \caption{We have to impose the boundary condition $\tilde G_{\Lambda_1,\Lambda_2}(x,y)=1$ on the light-rays intersecting at $x=\Lambda_1-\Lambda_2$ and given by the equation $x=\Lambda_1-\Lambda_2\pm 2y$. The initial function $\tilde G_{\Lambda_1,\Lambda_2}(x,y)$ is only defined inside the future light cone. It can be extended to the whole plane by setting it to zero outside the light cone and
    imposing $\tilde G_{\Lambda_1,\Lambda_2}(x,y)=-\tilde G_{\Lambda_1,\Lambda_2}(x,-y)$ for negative $y$.}
    \label{fig:lc}
\end{figure}

\paragraph{Stationary Schr\"odinger equation.}
In order to separate the variables we introduce 
new ``light-cone" coordinates in the following way
\beq
x=\Lambda_4-\Lambda_3\;\;,\;\;y=\frac{\Lambda_1+\Lambda_2+\Lambda_3+\Lambda_4}{2}
\eeq
so that $\d_{\Lambda_3}\d_{\Lambda_4}=-\d_x^2+\frac{1}{4}\d_y^2$. We also denote
\beq
\tilde G_{\Lambda_1,\Lambda_2}(x,y)\equiv G(\Lambda_1, \Lambda_2 , \Lambda_3, \Lambda_4 )
\eeq
so that  \eq{eq:BS} becomes
\beq\la{eq:BS2}
\frac{1}{4}\d_y^2\tilde G_{\Lambda_1,\Lambda_2}(x,y)=
\[\d_x^2+
\frac{2\hat{g}^2}{ \cosh x + \cos\phi}\]
\tilde G_{\Lambda_1,\Lambda_2}(x,y)\;.
\eeq
In order to completely reduce this equation to the  stationary Schr\"odinger problem,
we have to extend the function $G_{\Lambda_1,\Lambda_2}(x,y)$ to the whole plane.
Currently it is only defined for $-\Lambda_1<\Lambda_3$ and $-\Lambda_2<\Lambda_4$ i.e. inside the future light-cone,
see Fig.~\ref{fig:lc}.
We extend $\tilde G_{\Lambda_1,\Lambda_2}(x,y)$ to the whole plane using the following definition:
\beqa
\tilde G_{\Lambda_1,\Lambda_2}(x,y)&=&-\tilde G_{\Lambda_1,\Lambda_2}(x,|y|)\;\;,\;\;y<0\\
\tilde G_{\Lambda_1,\Lambda_2}(x,y)&=&0\;\;,\;\;|y|>|x-\Lambda_1+\Lambda_2|/2\;.
\eeqa
With this definition it is easy to see that if \eq{eq:BS2} was satisfied in the future light cone, it will hold for the whole plane.

\begin{figure}[t]
    \centering
    \includegraphics[scale=0.8]{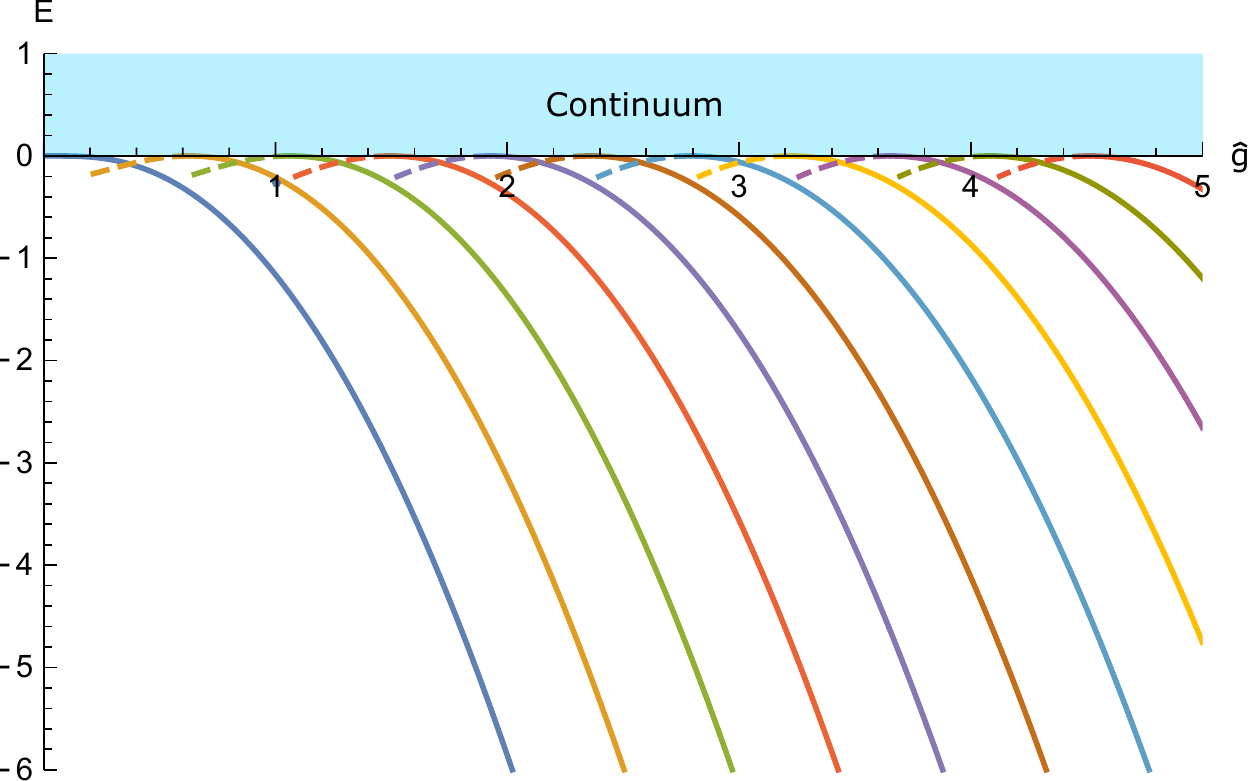}
    \caption{
    {\bf Spectrum of the Schr\"odinger problem} at  $\phi=1.5$ for a range of values of the coupling. Solid lines show numerical data for the first few bound states.
    For small $\hat g$ there is only one bound state in the spectrum, but their number grows linearly with the coupling. 
    Dashed lines show analytic continuation of the levels in the coupling  $\hat g$ beyond the point where they disappear from the bound state spectrum and become resonances (to be discussed in detail in section \ref{sec:qscexc}).
    }
    \label{fig:specschr}
\end{figure}

After that we can expand $\tilde G_{\Lambda_1,\Lambda_2}(x,y)$ in the complete basis of the eigenfunctions of the Schr\"odinger equation in the $x$ direction,
\beq
\tilde G_{\Lambda_1,\Lambda_2}(x,y)=\sum_n a_n(y) F_n(x)
\eeq
where 
\beq
4\[-\d_x^2-
\frac{2\hat{g}^2}{ \cosh x + \cos\phi}\]
F_n(x)={E_n} F_n(x)\;
\eeq
and $a_n(y)$ has to satisfy $a_n''(y)=-E_n a_n(y)$. Since $\tilde G(x,y)$ is odd in $y$ we get
\beq\la{Gxya}
\tilde G_{\Lambda_1,\Lambda_2}(x,y)=\sum\hspace{-5mm}\int_n C_n(\Lambda_1,\Lambda_2)\left(e^{\sqrt{-E_n} \, y}-e^{-\sqrt{-E_n} \, y}\right) F_n(x)\;.
\eeq
In the above expression we assume the sum over all bound states with $E_n < 0$ and  integral over the continuum $E_n>0$ (see Fig.~\ref{fig:specschr}).

Next we should determine the coefficients $C_n(\Lambda_1,\Lambda_2)$, for that we consider the small $y$ limit. 
For small $y$ we see that $G(x,y)$ is almost constant inside the light cone ($+1$ for $y>0$ and $-1$ for $y<0$) and is zero for $\Lambda_1-\Lambda_2-2y<x<\Lambda_1-\Lambda_2+2y$. In other words for small $y$ we have
\beq
\tilde G_{\Lambda_1,\Lambda_2}(x,y)\simeq 4 y\delta(x-\Lambda_1+\Lambda_2) \label{eq:tilGdelta}
\eeq
at the same time from the ansatz \eq{Gxya} we have, in the small $y$ limit
\beq
\tilde G_{\Lambda_1,\Lambda_2}(x,y)\simeq 2y\sum\hspace{-5mm}\int_n C_n(\Lambda_1,\Lambda_2)\sqrt{-E_n} F_n(x)\;.\label{eq:tilGexp}
\eeq
Contracting equations (\ref{eq:tilGdelta}) and (\ref{eq:tilGexp}) with an eigenvector $F_n(x)$ and comparing the results, we get
\beq
C_n(\Lambda_1,\Lambda_2)=\frac{2F_n(\Lambda_1-\Lambda_2)}{||F_n||^2\sqrt{-E_n}}\; .
\eeq
Which results in the following final expression for $G$
\beq\la{G1234}
G(\Lambda_1,\Lambda_2,\Lambda_3,\Lambda_4)
=\sum\hspace{-5mm}\int_n \frac{4F_n(\Lambda_1-\Lambda_2)F_n(\Lambda_4-\Lambda_3)}{||F_n||^2\sqrt{-E_n}}
\sinh\left(\sqrt{-E_n} \frac{\Lambda_1+\Lambda_2+\Lambda_3+\Lambda_4}{2}\right) \;.
\eeq
We will use this result in the next section to compute the two-point function in a certain regularisation 
including the finite part. This will be needed for normalisation of the 3-cusp correlator.

\subsection{Two-point function with finite part}\label{sec:2ptcusp}

\begin{figure}
    \centering
    \includegraphics[scale=1.4]{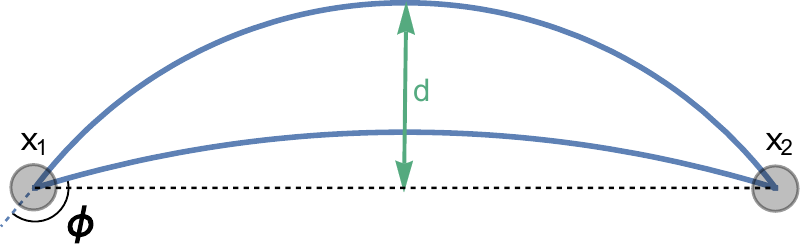}
    \caption{{\bf The 2-cusp correlator.} For regularisation we cut an $\epsilon$-ball around each of the cusps. The configuration is parameterised by the external angle $\phi$. The result does not depend on $d$ (or equivalently $\chi$ in \eq{eq:param}) and is only a function of $x_{12}=|x_1-x_2|,\;\phi,\;\Delta$ and the regulator $\epsilon$.}
    \label{fig:2pt}
\end{figure}

Now let us study the two-cusp configuration shown in Fig.~\ref{fig:2pt}, regularised by cutting $\epsilon$-balls around each of the cusps.
Here we show that the correlator has the expected space-time dependence of a two-point function with conformal dimension $\Delta = -\sqrt{-E_0}$.
In order to compute this quantity we need to work out which cut-offs in the  parameters $s$ and $t$ appearing in (\ref{eq:zazb}) correspond to the $\epsilon$-regularisation. By imposing
\beq
| \zeta_{+}(-\Lambda_1)-z_1|=\epsilon\;\;,\;\;
| \zeta_{+}(\Lambda_3)-z_2|=\epsilon\;\;,\;\;
| \zeta_{-}(+\Lambda_2)-z_1|=\epsilon\;\;,\;\;
| \zeta_{-}(-\Lambda_4)-z_2|=\epsilon
\eeq
we find (asymptotically for small $\epsilon$)
\beq
\Lambda_1=\Lambda_2=\Lambda_3=\Lambda_4=\log\(\frac{ x_{12}}{\epsilon} \), \;\;\;\; x_{12} = |z_1 - z_2 | ,\label{eq:lambdacutoff0}
\eeq
which allows us to write, using \eq{G1234}
\beqa
\langle W_{\epsilon, x_1, x_2} \rangle &=& G(\Lambda,\Lambda,\Lambda,\Lambda)\simeq 
 \frac{2F^2_0(0)e^{2\sqrt{-E_0} \Lambda}}{||F_0||^2\sqrt{-E_0}}
=
-\frac{2F^2_0(0)}{||F_0||^2\Delta_0}
\(\frac{\epsilon}{x_{12}} \)^{2\Delta_0}\; 
\eeqa
where we use that for large $\Lambda$ only the ground state contributes. We use the notation
\beq
\Delta_0\equiv-\sqrt{-E_0}\;
\eeq
so that $\Delta_0$ is the usual cusp anomalous dimension.
We see that the result for the $2$-cusp correlator takes the standard form $\frac{{\cal N}_{\hat g,\phi}^2}{x_{12}^{2\Delta_0}}$ with a rather non-trivial normalization coefficient 
\beq
{\cal N}_{\hat g,\phi}=\epsilon^{\Delta_0}\frac{F_0(0)}{||F_0||}\sqrt{\frac{2}{-\Delta_0}}\;,\label{NDelta}
\eeq
which we will use to extract the structure constant from the 3-cusp correlator.

\subsection{Relation to  Q-functions}
\label{sec:BaxtoSchrod}
Here we describe a direct relation between solutions of the Schr\"odinger equation and the Q-functions.
 From the previous section we can identify $\Delta=-\sqrt{-E}$ resulting in
\beq\label{eq:Schrodinger}
 F''(z) +\frac{2\hat g^2}{\cosh z+\cos\phi}F(z)=\frac{\Delta^2}{4}F(z)\;.
\eeq
In this section we will relate $F(z)$ with $q(u)$.
The relation is very similar to that found previously for the $\phi=\pi$ case in \cite{Gromov:2016rrp}.
For $\phi > 0$, the map is defined as follows
\beq
\frac{F(z)}{2\pi} = \, e^{-\Delta z/2 }  \vint \, q(u) \, e^{w_{\phi}(z) \, u} \, \frac{du}{2\pi i u}  \ \  ,\label{eq:qToF} 
\eeq
where 
\beq
\label{defw}
 e^{i w_{\phi}(z) } = \left(\frac{\cosh{\frac{z-i \phi}{2}} }{\cosh{\frac{z+ i \phi}{2}}}\right),
\eeq
 and $q(u)\equiv q_+(u)$ is one of the solutions of the Baxter equation (\ref{Bax2p}), specified by the  large $u$ asymptotics $q(u)\simeq u^\Delta e^{u\phi}$.
 We remind that we use the notation $\vint$ for the integration along a  vertical line shifted to the right from the origin. 
For negative $\Delta$ the integral in   \eq{eq:qToF} converges for any finite $z$, and we can shift the integration contour horizontally, as long as we do not cross the imaginary axis where the poles of $q(u)$ lie. 
Let us show that if $q$ satisfies the Baxter equation \eq{Bax2p}, then $F(z)$ computed from \eq{eq:qToF}
 satisfies the Schr\"odinger equation \eq{eq:Schrodinger}. Applying the derivative in $z$ twice to the relation (\ref{eq:qToF}) we find 
 \beqa \nn
&&F''(z) - \frac{\Delta^2}{4}F(z)\\ 
&&=  \frac{e^{-\Delta z/2 } }{2 (\cosh(z) + \cos\phi )}\, \vint  q(u) \,\left( ( D + D^{-1}  ) + \frac{2 \Delta \sin\phi}{u} - 2 \, \cos\phi  \right)[ u \, e^{u \, w_{\phi}(z)} ] \,du  \label{eq:rhs}
 \eeqa
 where $D$ represents the shift operator $D [ f(u) ] = f(u+i)$. Shifting the integration variable and using the Baxter equation (\ref{Bax2p}), the rhs of (\ref{eq:rhs}) simplifies leading to   (\ref{eq:Schrodinger}).  
 
Notice that this relation between the Baxter and Schr\"odinger equations holds  also off-shell, i.e. when $\Delta$ is a generic parameter and the quantization condition (\ref{qquant}) need not be satisfied. In Appendix \ref{app:quant} we show that the quantization condition (\ref{qquant}) 
is equivalent to the condition that $F(z)$ is a square-integrable function, so that it corresponds to a bound state of the Schr\"odinger problem.

\paragraph{Reality.}
Let us show that the transform (\ref{eq:qToF}) defines a real function $F(z)$. 
Here we assume the quantization condition to be satisfied.
Taking the complex conjugate of (\ref{eq:qToF}) we find
\beq
\frac{F^*(z)}{2\pi} =  \, e^{-\Delta z/2 }  \vint\,\bar{q}(u) \, e^{w_{\phi}(z) \, u} \, \frac{du}{2\pi i u}   .\label{eq:qToF*} 
\eeq
A precise relation between $q(u)$ and $\bar{q}(u)$ is discussed in appendix~\ref{app:qsc}. In particular, from (\ref{q1b}), (\ref{adif}) we see that, when the quantization conditions are satisfied, 
\beq
\bar{q}(u) = q(u) + \mathcal{O}( e^{- 2 \pi u} ) + \mathcal{O}( e^{ - \phi u } ), \label{qqbar}
\eeq
for large $\text{Re} \, u$. Shifting the contour of integration to the right we see that the contribution of the omitted terms in (\ref{qqbar}) is irrelevant, and therefore the integral transforms involving $\bar{q}(u)$ and $q(u)$ are equivalent. 
 This shows that $F^*(z) = F(z)$. 

\paragraph{Inverse map.}

The transform (\ref{eq:qToF}) can be inverted as follows:
\beq
\frac{q(u)}{u} = \frac{\sin{\phi}}{2 \pi} \, \int_{i \pi -i \phi}^{+\infty} \frac{dz \; e^{\Delta z/2 - w_{\phi}(z)  u }}{\cosh{z} + \cos\phi }  \, \, F(z) .\label{eq:Ftoq}
\eeq
The above integral representation converges for $\text{Im}(u) > 0$ and $\Delta < 0$. Assuming $F(z)$ is a solution to the Schr\"odinger equation with decaying behaviour $F(z) \sim e^{ \Delta z/2 }$ at positive  infinity $z \rightarrow + \infty$, this map generates the solution to the Baxter equation $q(u)$. When additionally $F(z)$ decays at $z \rightarrow - \infty$, $q(u)$ satisfies the quantization conditions.

\paragraph{Relation to the norm of the wave function.}
From the Schr\"odinger equation \eq{eq:Schrodinger} we can use the standard perturbation theory to immediately write
\beq\la{ddF}
\frac{ \d \Delta}{\d\hat g}=\frac{8\hat g}{\Delta}\frac{1}{||F||^2}\int \frac{F^2(z)}{\cosh z+\cos \phi}dz \ .
\eeq
We will rewrite the numerator in terms of the Q-function.
For that we use that $F_n(z)$ is either an even or an odd function depending on the level $n$, then we can write 
$F^2(z)=(-1)^n F(z)F(-z)$ and then use \eq{eq:qToF}. The advantage of writing the product in this way is that the factor $e^{+\Delta z/2 }$ in \eq{eq:qToF} cancels giving
\beq
\frac{1}{4\pi^2}\int \frac{F_n^2(z)}{\cosh z+\cos\phi} = (-1)^n \vint \frac{du}{2\pi i} \vint \frac{dv}{2\pi i}
\int_{-\infty}^{\infty}dz
\,\frac{ q_n(u) }{u}\frac{ q_n(v) }{v}\, \frac{e^{w_{\phi}(-z) \, u}e^{w_{\phi}(+z) \, v}}{{\cosh z+\cos\phi}} \;\;\;\; .\label{eq:qToF2}
\eeq
Next we notice that the integration in $z$ can be performed explicitly
\beq
K(u-v)\equiv \int_{-\infty}^\infty\frac{ e^{w_{\phi}(-z) \, u}e^{w_{\phi}(+z) \, v}}{\cosh z+\cos \phi}dz = 
\frac{ e^{\phi  (u-v)}-e^{\phi  (v-u)}}{(u-v)\sin\phi}\;.
\eeq
Note that the function $K(u-v)$ is not singular by itself as the pole at $u=v$ cancels.
We are going to get rid of the integral in $u$ in \eq{eq:qToF2},  for that we notice that we can move the contour of integration in $v$ slighly to the right from the integral in $u$, and after that we can split the two terms in $K(u-v)$.
The first term $\sim \frac{e^{\phi(u-v)}}{u-v}$ decays for ${\rm Re}\;v\to+\infty$ and we can shift the integration contour in $v$ to infinity, getting zero.
Similarly the second term $\sim \frac{e^{\phi(v-u)}}{u-v}$ decays for ${\rm Re}\;u\to+\infty$ and we can move the integration contour in $u$ to infinity, but this time on the way we pick a pole at $u=v$. That is, only this pole contributes to the result giving
\beq
\label{QFder2}
\frac{1}{4\pi^2}\int \frac{F_n^2(z)}{\cosh z+\cos\phi} =\frac{(-1)^n}{\sin\phi} \vint  \frac{q_n^2(v)}{v^2}\frac{dv}{2\pi i}\;.
\eeq

At the same time, above in \eq{eq:dddg} we have already derived an expression for $\d\Delta/\d\hat g$ in terms of the Q-function.
Comparing it with \eq{ddF} and using \eq{QFder2} we conclude that
\beq
\boxed{
\frac{1}{4\pi^2}||F_n||^2=-(-1)^n\frac{2}{\Delta_n}\vint \frac{q_n^2}{u}\frac{du}{2\pi i}\;. \label{eq:normQ}
}
\eeq
We will use the relations between $q$ and $F$ to rewrite the 3-cusp correlator in terms of Q-functions in the next section.

\section{Three-cusp structure constant}
\label{sec:3pt1}

In this section we derive our main result -- an expression for the structure constant.
First, we compute it for the case when only one of the $3$ couplings is nonzero.
We refer to this case as the Heavy-Light-Light (HLL) correlator \footnote{The name is justified since, in analogy with the case of local operators, the scaling dimensions of the cusps become  large at strong coupling.  }.
Then we generalize the result to two non-zero couplings, this case we call the Heavy-Heavy-Light (HHL) correlator. 
In both cases we managed to find an enormous simplification when the result is written in terms of the Q-functions.
We postpone the Heavy-Heavy-Heavy (HHH) case for future investigation.

\subsection{Set-up and parameterization}\label{sec:setup3pt}
\begin{figure}[t]
    \centering
    \includegraphics[scale=0.7]{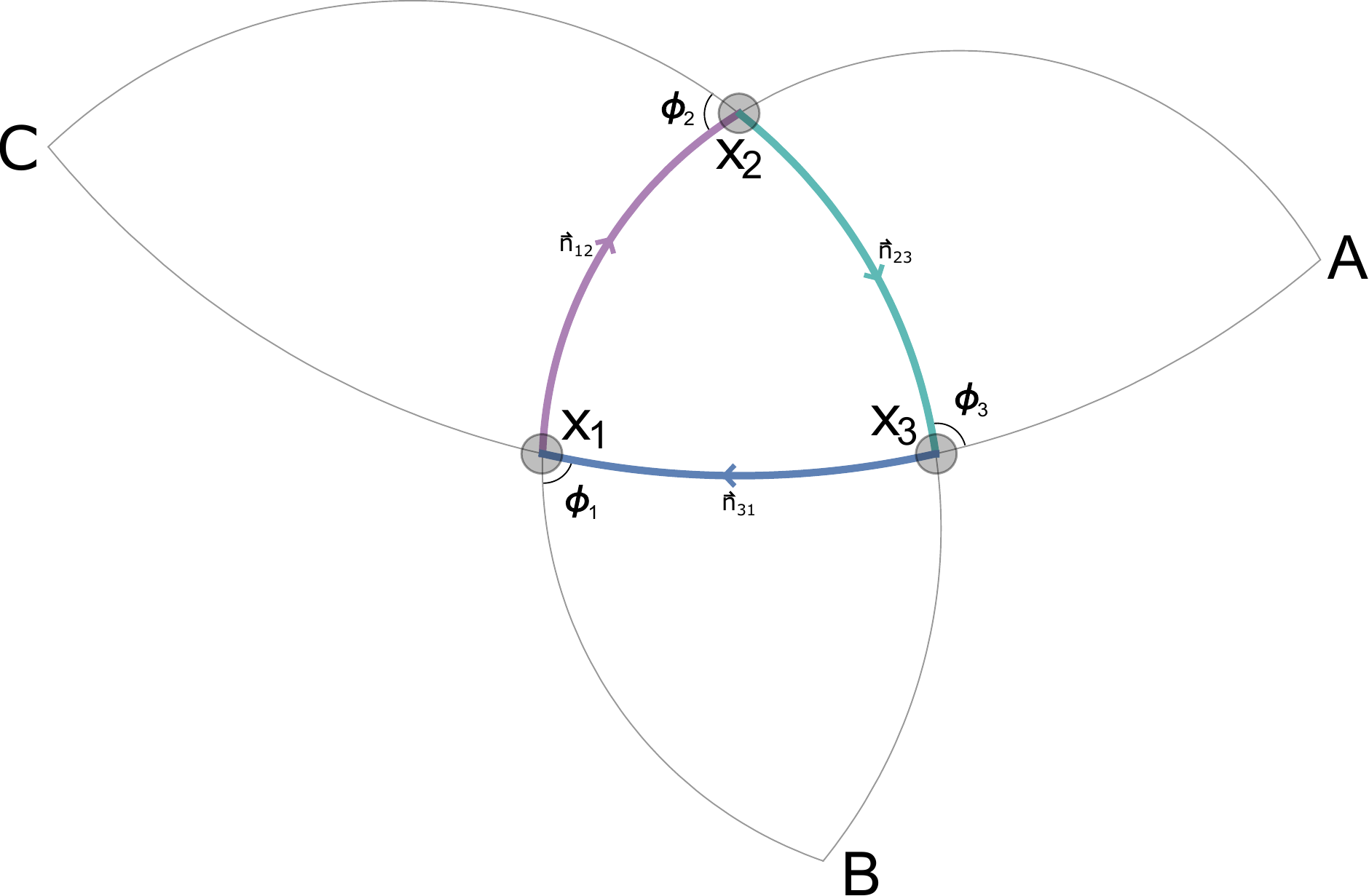}
    \caption{The general configuration of the Wilson loop we consider (the $x_1x_2x_3$ triangle) is built out of $3$ circular arcs belonging to the same plane. The configuration is parameterized by $3$ external angles $\phi_i$, coordinates of the vertices $x_i$ and $3$ scalar products of the unit vectors attached to the scalars inside the Maldacena-Wilson loop (or equivalently $3$ couplings $\hat g_a$). Pairs of arcs continued outside the triangle intersect again at $A$, $B$ and $C$. The renormalized  3-cusp correlator has the typical CFT dependence on the positions of the vertices, with a structure constant which depends only on the $3$ angles and $3$ couplings. In this paper we only consider the case with two non-zero couplings.}
    \label{fig:triangle}
\end{figure}

In this section we describe the $3$-cusp Wilson loop configuration, parameterization and regularisation, which we use in the rest of the paper. 
The Wilson loop is limited to a 2D plane and consists of $3$ circular arcs coming together at $3$ cusps (see Fig.~\ref{fig:triangle}). The $3$ angles $\phi_i$, $i=1,2,3$ can be changed independently. The  geometry is completely specified by the angles and the positions of the cusps $x_i$, $i=1,2,3$. 

In the rest of this paper, we consider the following ``triangular" inequalities on the angles:
\beqa
&&\phi_1 + \phi_2 > \phi_3  , \,\,\, \phi_3 + \phi_2 > \phi_1  , \,\,\, \phi_3 + \phi_1 > \phi_2  ,\,\,\, \label{eq:ineq}
 0 < \phi_i < \pi .
\eeqa
To understand the geometric meaning of these relations, consider the extension of the arcs forming the Wilson loop past the points $\vec{x}_i$: this defines three virtual intersections $A$, $B$, $C$ (see Fig.~\ref{fig:triangle}). The inequalities (\ref{eq:ineq}) mean that $A$, $B$, $C$ are all outside the Wilson loop. Our results will hold in this kinematics regime. In the limit where  we approach the boundary of the region (\ref{eq:ineq}) our result significantly simplifies and will be considered in Sec.~\ref{sec:smallPhi}, in particular we will reproduce the results of \cite{Kim:2017sju} for the case $\phi_1=\phi_2=\phi_3=0$.

Now we describe a nice way to parametrize the Wilson lines. 
Consider the two arcs departing from $\vec{x}_1$. Extending these arcs past the points $\vec{x}_2$, $\vec{x}_3$, they define a second intersection point $A$. 
By making a special conformal transformation, we map $A$  to infinity and both arcs connecting $x_1$ with $A$  to straight lines, which we can then map on a cylinder like in \eq{eq:param}.
The most convenient parametrization corresponds to the coordinate along the cylinder.
By mapping $A$ back to some finite position we get a rather complicated but explicit parametrization
like the one we used in Sec. \ref{sec:cuspdiv}.

It is again very convenient  to use complex coordinates, similarly to \eq{eq:param}, 
\beq
\vec{x} = ( \text{Re}(z) , \text{Im}(z) , 0, 0 ) ,
\eeq
so that the cusp points are  $\vec{x}_i = ( \text{Re}(z_i) , \text{Im}(z_i) , 0, 0 ) $, $i=1,2,3$. For the arcs departing from $z_1$ we obtain, as described above, the following representation  
\beqa\label{eq:zeta1213}
\zeta_{12}( s ) &=& z_1 - \frac{z_{12} \, z_{13} \, e^s}{ e^{s} \, z_{13}  +\frac{i}{2 \sin\phi_1 } \, z_{23} \, (1-e^{s}) \, ( -e^{i \phi_1} + e^{-i (\phi_3-\phi_2)} )} , \\
\zeta_{13}( t ) &=& z_1 -\frac{z_{12} \, z_{13} \, e^t}{ e^{t} \, z_{12} + \frac{i }{2 \sin\phi_1 } \, z_{23} \, (1-e^{t}) \, (-e^{-i \phi_1 } + e^{-i (\phi_3 - \phi_2 )} ) } ,\nn
\eeqa
where $z_{ab} = z_a - z_b$. 
Notice that we have slightly redefined the parameters such that $s=0$ and $t=0$ correspond to the other two cusp points:
$\zeta_{12}(0) = z_2$, $\zeta_{13}(0) = z_3$, while $\zeta_{12}(-\infty) = \zeta_{13}(-\infty) = z_1$, and $\zeta_{12}(\infty) = \zeta_{13}(\infty) = A$. 
By a cyclic permutation of all indices, we define similar parametrizations for the other arcs. Notice that, in this way, all arcs are parametrized in two distinct ways, e.g. the same arc connecting $\vec{x}_1$ and $\vec{x}_2$ is described by the functions $\zeta_{12}(s)$ and $\zeta_{21}(t)$, which are different.

The main advantage of the parametrization (\ref{eq:zeta1213}) is that the propagator between the two arcs is very simple:
\beq
\frac{|\dot {\vec{x}}_{12}(s_1)||\dot {\vec{x}}_{13}(t_1)|}
{|\vec{x}_{12}(s_1)-\vec x_{13}(t_1)|^2}=\frac{1/2}{ \cosh \left(s_1-t_1-\delta x_1\right)+\cos \phi_1} \ . \label{eq:propagator}
\eeq
However, since we decided to shift the parameters so that $s=0$ gives $\vec x_2$ and 
$t=0$ gives $\vec x_3$, the propagator appears to be shifted compared to \eq{eq:goodpropagator}
by the quantity
\beq
\delta x_1 = \log\frac{\sin \frac{1}{2} ({\phi_1}-{\phi_2}+{\phi_3})}{
   \sin \frac{1}{2} ({\phi_1}+{\phi_2}-{\phi_3})}\; ,\label{eq:deltax1}
\eeq
with $\delta x_2$ and $\delta x_3$  defined similarly by cyclic permutations of the indices $1,2,3$. We see now the importance of the inequalities \eq{eq:ineq} as they ensure $\delta x_i$ are real.

\paragraph{Notation.}

Below we consider correlators where the ladder limit is taken independently for the three cusps. Namely, by choosing appropriately polarization vectors $\vec{n}_i$ on the three lines, we define effective couplings  
\beq
\hat{g}_i^2 = g^2 \; \frac{(\vec{n}_{i-1,i} \cdot   \vec{n}_{i,i+1})}{2}, \;\;\;\;\;\; g \to 0 ,
\eeq
for the three cusps $i=1,2,3$.

Correspondingly, in this section we use the notation\footnote{This should not be confused with the notation for the scaling dimensions for excited states $\Delta_n$ used in other parts of the paper.} $\Delta_{i,0} $, $i=1,2,3$, to denote the scaling dimensions corresponding to the ground state for the three cusps (in the setup we consider we always have $\hat g_3=0$, $ \Delta_{3,0}=0$). 
The extension to excited states will be discussed in section~\ref{sec:excited}.

The Q-functions describing the ground state for the first and second cusps will be denoted as $q_i(u)$, $i=1,2$, respectively. 
 Explicitly, $q_i(u)$ is the solution of the Baxter equation $q_+(u)$, evaluated at parameters $\hat g= \hat g_i$, $\Delta = \Delta_{i,0}$ and $\phi = \phi_i$.

\subsection{Regularization}
The $3$ cusp correlator is UV divergent. To regularize the divergence we are going to cut $\epsilon$-circles around each of the cusps\footnote{See \cite{Dorn:2015bfa} for a general argument why the divergence depends on the geometry only  through the angles $\phi_i$.} -- the same way as we regularized the 2-cusp correlator in the previous section. This will set a range for the parameters $s_i$ and $t_i$  entering the parametrizations $\zeta_{ij}( s_i )$, $\zeta_{ij}( t_i )$ defined above. Namely from (\ref{eq:zeta1213}) it is easy to find that instead of running from $-\infty$ they now start from a cutoff:
\beq
s_i\in[-\Lambda_{s_i},0]\;\;,\;\;t_i\in [-\Lambda_{t_i},0]
\eeq
where
\beq
\Lambda_{s_1}=\log \left(\frac{x_{12} x_{13}\sin\phi _1}{x_{23} \epsilon  \sin \left(\frac{1}{2} \left(\phi _1-\phi
   _2+\phi _3\right)\right) }\right)\;\;,\;\;
\Lambda_{t_1}=\log \left(\frac{x_{12} x_{13}\sin\phi _1}{x_{23} \epsilon  \sin \left(\frac{1}{2} \left(\phi _1+\phi
   _2-\phi _3\right)\right) }\right)\;\;.\label{eq:LambdaHLL}
\eeq
All other $\Lambda_{s_i}$ and $\Lambda_{t_i}$ for $i=2,3$,
can be obtained by cyclic permutation of the indices $1,2,3$. We note that 
\beq
\la{lambdashift}
\Lambda_{s_i}+\delta x_i=\Lambda_{t_i}\;.
\eeq

\subsection{Heavy-Light-Light correlator}\label{sec:HLL}
\begin{figure}
    \centering
    \includegraphics{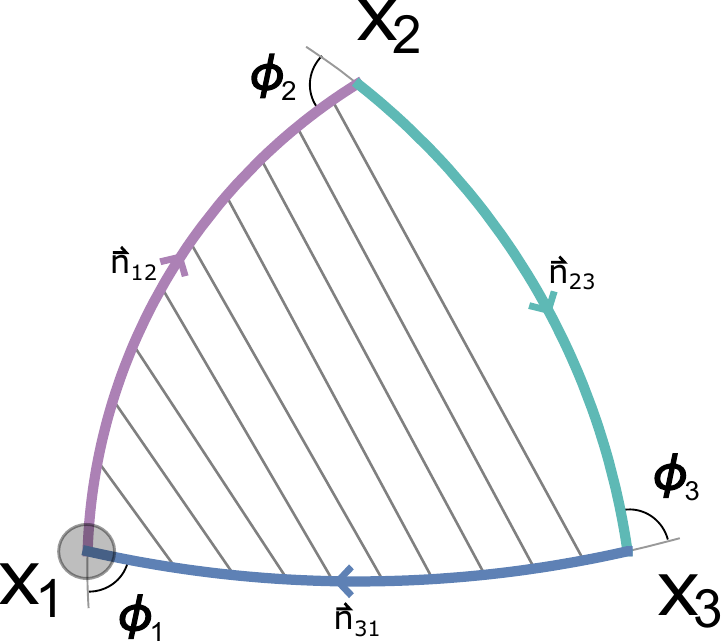}
    \caption{The HLL correlator corresponds to the situation when the couplings $\hat g_2$ and $\hat g_3$ are zero. In this case there is only one type of propagators to re-sum.}
    \label{fig:HLL}
\end{figure}
Now we consider the simplest example of three point function in the ladder limit, where we have only one non-vanishing effective coupling, $\hat{g}_1$ for the cusp at $\vec{x}_1$, with $\hat{g}_2 = \hat{g}_3=0$. Correspondingly, we will have $\Delta_{2,0} = \Delta_{3,0} = 0$, so that this can be considered as a correlator between one nontrivial operator and two protected operators (see Fig.~\ref{fig:HLL}). 
 For simplicity we will denote $\Delta_{1,0}$  as just $\Delta_0$ in this section.

 We start by defining a regularized correlator, which we denote as $Y_{\vec{x}_1 , \epsilon }( \vec{x}_2 , \vec{x}_3 )$, which is obtained by cutting the integration along the Wilson lines at a distance $\epsilon$ from $\vec{x}_1$. 
  To compute this observable we consider the sum of all ladder diagrams built around the first cusp and covering the Wilson lines $(12)$, $(13)$ up to the points $\vec{x}_2$, $\vec{x}_3$, respectively, see Fig.~\ref{fig:HLL}.  As discussed in section~\ref{sec:BS}, this is described by the Bethe-Salpeter equation, which takes a very convenient form using the parameterization introduced in the previous section for the Wilson lines departing from $\vec{x}_1$:  $\vec{\gamma}_{12}(s) = ( \text{Re}(\zeta_{12}(s) ), \text{Im}(\zeta_{12}(s) ), 0, 0 )$,  
 and 
   $\vec{\gamma}_{13}(t) = ( \text{Re}(\zeta_{13}(t) ), \text{Im}(\zeta_{13}(t) ), 0, 0 )$. The appropriate integration range for cutting an $\epsilon$-circle around $\vec{x}_1$ is $s \in [-\Lambda_{s_1}, 0]$, $t \in [- \Lambda_{t_1}, 0]$, with cutoffs defined in (\ref{eq:LambdaHLL}). However, in order to make a connection with 
   $G(\Lambda_1,\Lambda_2,\Lambda_3,\Lambda_4)$ defined in section~\ref{sec:BS}, we have to take into account the fact that the propagator in \eq{eq:propagator} is shifted by $\delta x_1$.  This  means that we have to redefine $s\to  s+\delta x_1$, which  will shift
   the range to $s \in [-\Lambda_{s_1}-\delta x_1,-\delta x_1]$, furthermore due to
   \eq{lambdashift} the range becomes
   $s \in [-\Lambda_{t_1},-\delta x_1]$
   . From that we read off the values of $\Lambda_k$ and find
\beq\la{Yground}
 Y_{\vec{x}_1, \epsilon}(\vec{x}_2 , \vec{x}_3 ) = G(\Lambda_{t_1},\Lambda_{t_1},-\delta x_1,0).
\eeq
Again, at large $\Lambda's$ only the ground state survives and we get
\beq
Y_{\vec{x}_1, \epsilon}(\vec{x}_2 , \vec{x}_3 ) \simeq
\frac{2F_0(0)F_0(\delta x_1)}{-||F_0||^2\Delta_0}
\exp\left(-\Delta_0 \frac{\Lambda_{t_1}+\Lambda_{s_1}}{2}\right) .
\eeq
Substituting the values for $\Lambda_{t_1}$
from \eq{eq:LambdaHLL} leads to
\beq
 Y_{\vec{x}_1, \epsilon}(\vec{x}_2 , \vec{x}_3 ) = \frac{2F_0(0)F_0(\delta x_1)}{-||F_0||^2\Delta_0}
 \, \epsilon ^{\Delta _0} \, \frac{
   ( L_{123}  )^{ {\Delta _0} } }{ x_{12}^{\Delta _0} x_{13}^{\Delta _0} x_{23}^{-\Delta _0}} ,\label{eq:unnormY}
\eeq
which naturally has the structure of the $3$-point correlator in a CFT, where we have defined
\beq
L_{123} = \frac{ \sqrt{ \sin\frac{1}{2}( \phi_1 + \phi_2 - \phi_3) \, \sin\frac{1}{2}( \phi_1 - \phi_2 + \phi_3)}
}{\sin\phi_1} .\label{eq:L123}
\eeq
Finally, to extract the structure constant we have to divide (\ref{eq:unnormY}) by the two point functions normalization \eq{NDelta}, 
$\mathcal{N}_{1} = \epsilon^{\Delta_0}\frac{F_0(0)}{||F_0||}\sqrt{\frac{2}{-\Delta_0}}$
, so we get:
\beq
C_{123}^{\bullet\circ\circ}  = \left(\frac{-2}{ \Delta_0 \, ||F_{0}||^2  }\right)^{\frac{1}{2} } \, ( L_{123} )^{ {\Delta _0}} \;  F_{0}(\delta x_1) .\label{eq:CHLL}
\eeq

Let us now write the result in terms of the Q-functions. 
Using (\ref{eq:qToF}) to evaluate the  shifted wave function in (\ref{eq:CHLL}), we already notice a nice simplification:
\beq
w_{\phi_1}( \pm \delta x_1 )  =  \pm  (\phi_2 - \phi_3 ), \label{eq:wphi1}
\eeq
therefore (using also parity of the ground-state wave function)
\beq
F_{0}(-\delta x_1 ) = F_{0}(+\delta x_1 )=  -i \, e^{- \frac{ \delta x_1}{2} \, \Delta_{0} } \, \vint \frac{q_{1}(u)}{u} \, e^{(\phi_2 - \phi_3) u} \, du 
\eeq
and taking into account also the norm formula (\ref{eq:normQ}), we find
\beq
C^{\bullet \circ \circ}_{123}=  \,(K_{123} )^{\Delta_{0}} \, \frac{ -i \, \vint \frac{q_{1}(u)}{u} \, e^{(\phi_2 - \phi_3) u} \, du }{\left( -2 \pi i \, \vint \frac{q_{1}^2(u) \,  }{u}  \, du  \right)^{\frac{1}{2}}} ,
\eeq
where the constant $K_{123}$ is defined as
\beq
K_{123} = L_{123} \, e^{ \frac{\delta x_1}{2}} = \frac{ \sin\frac{1}{2}( \phi_1 + \phi_2 - \phi_3) 
}{\sin\phi_1} .\label{eq:K123}
\eeq
 Using the parity of the ground state wave function $F_0$, it can be verified that the result is symmetric in the two angles $\phi_2 \leftrightarrow \phi_3$.

We see that the result takes a much  simpler form in terms of the Q-functions. The structure becomes even more clear when written in terms of the bracket $\br\cdot$ defined in \eq{eq:thebracket}:
\beq
\boxed{
C^{\bullet \circ \circ}_{123}=\frac{\br{q_1e^{u(\phi_2-\phi_3)}}}{\sqrt{\br{q_1^2}}}
}\; ,
\label{CHLL1}
\eeq
which is amazingly simple!
 
\subsection{Heavy-Heavy-Light correlator}
\begin{figure}
    \centering
    \includegraphics{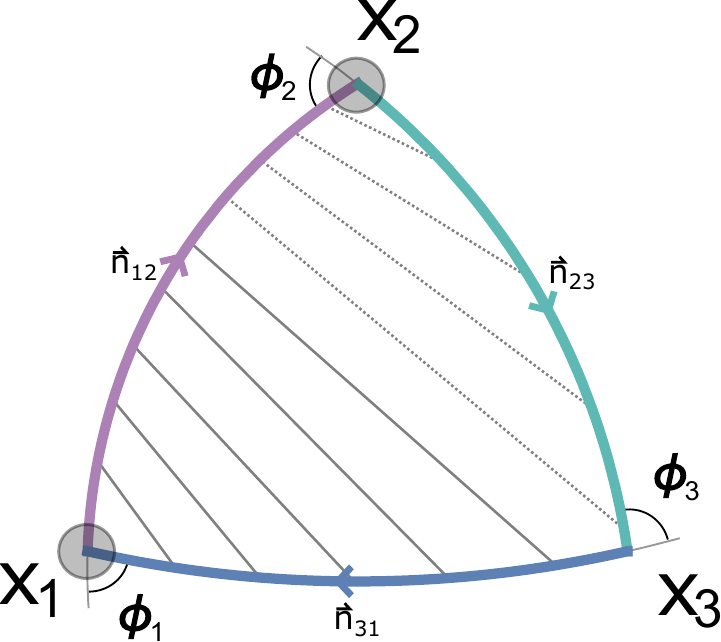}
    \caption{The HHL correlator. In this case there are two types of propagators since two couplings are
    non-zero.}
    \label{fig:my_label}
\end{figure}
Now, we switch on the effective couplings $\hat{g}_i$, $i=1,2$ for both the first and the second cusp. This means that  this observable is defined perturbatively by Feynman diagrams with two kinds of ladders built around the cusps $\vec{x}_1$ and $\vec{x}_2$, see Fig.~\ref{fig:my_label}. 

 As in the previous section let us denote by $
Y_{ \vec{x}_1 , \epsilon}( \vec{x}_2 , \vec{x}_3 ) $ 
 the sum of all ladders built  around the cusp point $\vec{x}_1$, with a cutoff at distance $\epsilon$ from the cusp.
 We introduce a similar notation for the ladders built around the second cusp. 

 The sum of all diagrams contributing to the $\epsilon$-regularized Heavy-Heavy-Light correlator can be organized as follows:

  \beq
 \label{Wsum}
 \begin{array}{rccccc}
 W^{{\bullet \bullet \circ} , \, \epsilon}_{123} &=& \underbrace{\sum\limits_{\text{propagators only around $2$}}}_{Y_{\vec{x}_2, \epsilon}(  \vec{x}_3 , \vec{x}_1 )} &+&\underbrace{\sum\limits_{\text{diagrams with at least one propagator around $1$}}}_{\left( W^{{\bullet \bullet \circ} , \epsilon}_{123} \right)_1}  
 \end{array}
 \eeq
 where the part $ \left( W^{{\bullet \bullet \circ} , \epsilon}_{123} \right)_1 $ represents the sum of all diagrams with at least one propagator around the cusp $x_1$. As we are about to show, the leading UV divergence comes only from the connected part, which behaves as $\sim \epsilon^{\Delta_{1,0} + \Delta_{2,0} }$. Since the disconnected contributions in (\ref{eq:sumHHL}) have a milder divergence $\sim \epsilon^{\Delta_{i,0}}$ $(i=1,2)$, we can drop them since they are irrelevant to the definition of the renormalized structure constant.
 
\begin{figure}
    \centering
    \includegraphics{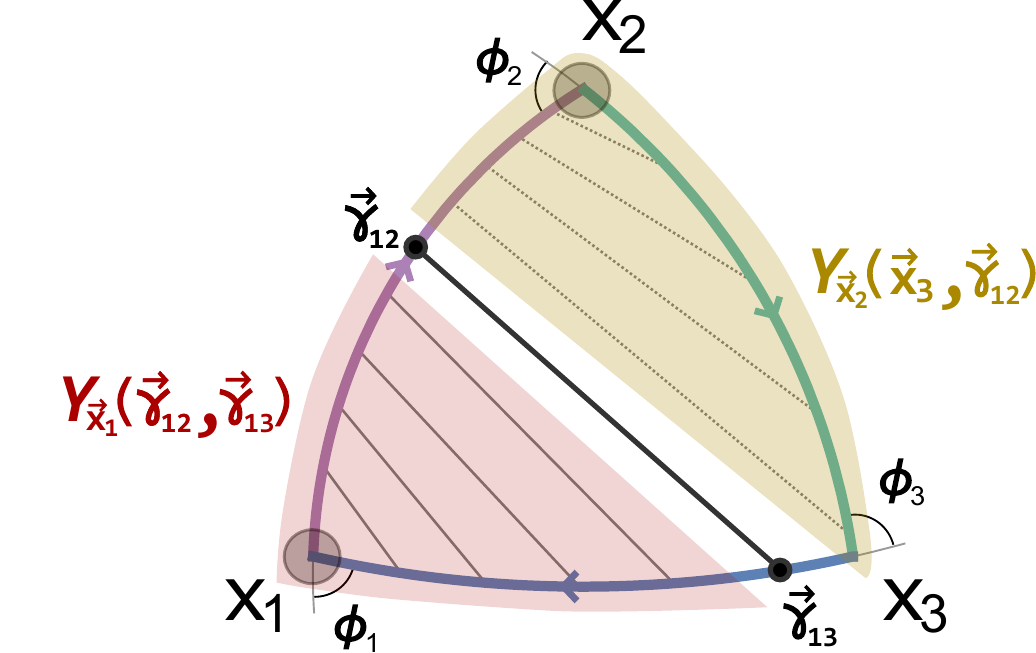}
    \caption{We split the propagators into two groups by explicitly writing the last propagator  between $\vec \gamma_{12}$ and $\vec \gamma_{13}$. 
    Then we re-sum the propagators surrounding cusp $x_2$ into $Y_{\vec x_2}(\vec x_3,\vec \gamma_{12})$
    and those around $x_1$ into $Y_{\vec x_1}(\vec \gamma_{12},\vec \gamma_{13})$. 
    }
    \label{fig:triangleHHLhl}
\end{figure} 
As illustrated in Fig.~\ref{fig:triangleHHLhl}, the main  contribution can be computed as follows:
\beqa
 \left( W^{{\bullet \bullet \circ} , \epsilon}_{123} \right)_1 =
\int_{ \vec{x}_1 + O(\epsilon) }^{ \vec{x}_2 + O(\epsilon) } d | \vec{\gamma}_{12} |  \int_{ \vec{x}_1 + O(\epsilon) }^{ \vec{x}_3 } d | \vec{\gamma}_{13}|\; Y_{ \vec{x}_1 , \epsilon }(  \vec{\gamma}_{12} , \vec{\gamma}_{13} ) \, \frac{1}{| \vec{\gamma}_{12} - \vec{\gamma}_{13} |^2 }\,   
Y_{ \vec{x}_2 , \epsilon}(\vec{x}_{3} , \vec{\gamma}_{12} ) , \nn \\
\label{eq:sumHHL}
\eeqa
where we are denoting with  $Y_{ \vec{x}_1 , \epsilon }(  \vec{\gamma}_{12} , \vec{\gamma}_{13} )$ the sum
of all ladder diagrams up to the points $\vec{\gamma}_{12}$, $\vec{\gamma}_{13}$ on the arcs $(12)$, $(13)$, respectively (and similarly for $Y_{ \vec{x}_2 , \epsilon }( \vec{x}_3 ,  \vec{\gamma}_{12})$).

To compute the connected integral explicitly we  choose the following parametrization for the arcs $(12)$, $(13)$:
\beqa
\vec{\gamma}_{12}(s) &=& \left( \text{Re}( \zeta_{12}(s) ), \text{Im}(\zeta_{12}(s) ) , 0, 0 \right), \label{eq:gammaexpli}\\
\vec{\gamma}_{13}(t) &=& \left( \text{Re}( \zeta_{13}(t) ), \text{Im}(\zeta_{13}(t) ) , 0, 0 \right),
\eeqa
where the functions $\zeta_{ij}$ are again the ones we defined above in section~ \ref{sec:setup3pt}. 
The function $Y_{\vec{x}_1, \epsilon}( \vec{\gamma}_{12}(s) ; \vec{\gamma}_{13}(t) )$ is given by the solution to the Bethe-Salpeter equation with shifted propagator (\ref{eq:propagator}), where the integration range is $s_1 \in [-\Lambda_{s_1} , s]$, $t_1 \in [-\Lambda_{t_1}, t]$.
 Exactly as described in section~\ref{sec:HLL}, redefining the parameters we find, in terms of the amputated four point function $G(\Lambda_1, \dots, \Lambda_4)$:
\beq
Y_{\vec{x}_1, \epsilon}( \vec{\gamma}_{12}(s) ; \vec{\gamma}_{13}(t) ) = G_{1}( \Lambda_{t_1} , \Lambda_{t_1} , s - \delta x_1 , 0 ), 
\eeq
where $\delta x_1$ is defined in (\ref{eq:deltax1}), 
and for $\epsilon \rightarrow 0$ we have
\beq
Y_{\vec{x}_1, \epsilon}( \vec{\gamma}_{12}(s) ; \vec{\gamma}_{13}(t) ) \sim \left(\frac{2F_{1,0}(0)}{-||F_{1,0}||^2\Delta_{1,0}} \right) \, \left( \frac{ \epsilon \, L_{123}  \; x_{23} }{x_{12} \, x_{13} } \right)^{\Delta_{1,0} } \, e^{-\frac{s + t}{2} \, \Delta_{1,0} } \, F_{1,0}(-\delta x_1 + s - t ),\label{eq:divY1}
\eeq
 where $L_{123}$ is defined in (\ref{eq:deltax1}). 

 The other ingredient appearing in (\ref{eq:sumHHL}) is $Y_{\vec{x}_2, \epsilon }( \vec{x}_3 , \vec{\gamma_{12}}(s) ) $. 
  Computing this quantity is slightly more complicated, since the ladders built around the second cusp point $\vec{x}_2$ are described most naturally in terms of a different parametrization, which uses the functions $\zeta_{21}(t_2)$, $\zeta_{23}(s_2)$ to parametrize the arcs $(12)$, $(23)$. In fact, it is only in the variables $s_2$ and $t_2$ that the propagator takes the simple form (\ref{eq:propagator}), with $\delta x_1 \rightarrow \delta x_2$. 
 Therefore we need to relate the two alternative parametrizations, $\zeta_{21}(t_2)$ vs $\zeta_{12}(s_1)$, for the line $(12)$. To this end we introduce the transition map $T_{12}(s)$:
\beq
\zeta_{12}(s) = \zeta_{21}( T_{12}(s) ),
\eeq
which is given explicitly by
\beq
e^{T_{12}(s) } = 
\frac{(1 - e^s)}{1 - e^s\frac{\cos\phi_3 - \cos( \phi_1 + \phi_2) }{ \cos\phi_3 - \cos( \phi_1 - \phi_2) } \, }
.\label{eq:T12}
\eeq
Using this map, we find that $Y_{\vec{x}_2, \epsilon }( \vec{x}_3 , {\vec\gamma_{12}}(s) ) $ is defined by the Bethe-Salpeter equation with propagator shifted by $\delta x_2 $ and integration ranges $s_2\in [-\Lambda_{s_2} , 0]$, $t_2 \in [-\Lambda_{t_2}, T_{12}(s) ]$. Taking into account the shift in the propagator, we have 
\beq
Y_{\vec{x}_2, \epsilon }( \vec{x}_3 , {\vec\gamma_{12}}(s) ) =
G_{2}( \Lambda_{t_2}, \Lambda_{t_2} , -\delta x_2 , T_{12}(s) ) , 
\eeq
which for small $\epsilon$ yields
\beq
Y_{\vec{x}_2, \epsilon }( \vec{x}_3 , \vec{\gamma_{12}}(s) ) \sim  \left(\frac{2F_{2,0}(0)}{-||F_{2,0}||^2\Delta_{2,0}} \right) \,\left( \frac{ \epsilon \; L_{231}\; x_{13} }{x_{23} \, x_{12} }\right)^{\Delta_{2,0}} \, e^{- \frac{T_{12}(s) }{2}\, \Delta_{2,0} } \, F_{2,0}( - \delta x_2 - T_{12}(s) ) , \label{eq:divY2} 
\eeq
where $L_{231}$ is defined applying a cyclic permutation to (\ref{eq:K123}). 
Combining (\ref{eq:divY1}), (\ref{eq:divY2}) in (\ref{eq:sumHHL}), we find, for the leading divergent part:
\beq
 W^{\bullet \bullet \circ , \, \epsilon}_{123}  = \frac{\epsilon^{\Delta_{1,0} + \Delta_{2,0}}( L_{123} )^{\Delta_{1,0}}\, ( L_{231} )^{\Delta_{2,0}} }{x_{12}^{\Delta_{1,0} + \Delta_{2,0} } \, x_{13}^{\Delta_{1,0}-\Delta_{2,0}} \, x_{23}^{\Delta_{2,0}-\Delta_{1,0}} } \,   \left(\frac{4 \, F_{2,0}(0) F_{1,0}(0) }{||F_{1,0}||^2 \, ||F_{2,0}||^2 \, \Delta_{1,0} \, \Delta_{2,0}} \right) \,\;\mathcal{N}^{\bullet \bullet \circ}_{123} ,\label{eq:3ptHHL}
\eeq
where $\mathcal{N}_{123}^{\bullet \bullet \circ}$ is a finite constant which can be written explicitly as\footnote{Notice that in this formula we have sent to infinite all the cutoffs defining the ranges of integration. Since the integrals in (\ref{eq:N123def}) are convergent, this does not change the leading  UV divergence of the correlator, which is enough to get to the final result for the OPE coefficient. A more detailed argument would show that, by sending the cutoffs to infinity in (\ref{eq:N123def}), we also restore the disconnected contributions with subleading divergences. }
\beqa
\mathcal{N}_{123}^{\bullet \bullet \circ} &=& 2 \hat{g}_1^2 \, \int_{-\infty }^{0} ds \, \int_{-\infty }^0 dt \,\frac{  
	F_{1,0}( -\delta x_1 +s - t) \, F_{2,0}( -\delta x_2 - T_{12}(s) ) \, e^{-\frac{s+t}{2} \, \Delta_{1,0}  - \frac{T_{12}(s)}{2} \, \Delta_{2,0} } }{ \cosh(s-t - \delta x_1 ) + \cos\phi_1 }. \nn\\
&&\label{eq:N123def}
\eeqa
Again, we see that (\ref{eq:3ptHHL}) has the correct space-time dependence for a CFT 3-point correlator.  Normalizing by the 2-pt functions factors $\mathcal{N}_{\Delta_i, \phi_i}$ defined in (\ref{NDelta}) for the two cusps, we get a finite expression for the structure constant:
\beq
C^{\bullet \bullet \circ}_{123} = 2 \, \frac{( L_{123}  )^{\Delta_{1,0}} \, (L_{231}  )^{\Delta_{2,0}}  }{ \sqrt{\Delta_{1,0} \, \Delta_{2,0} }\, ||F_{1,0}||\, ||F_{2,0}||} \, \mathcal{N}_{123}^{{\bullet \bullet \circ}} .
\eeq
Using the Schr\"odinger equation for $F_{1,0}$, we can simplify the expression for $\mathcal{N}_{123}^{\bullet \bullet \circ} $ further and remove one of the integrations:
\beqa
\mathcal{N}_{123}^{\bullet \bullet \circ}&=& \int_{-\infty }^{0} ds \, \int_{-\infty }^0 dt \,  \partial_s \partial_t \, \left(F_{1,0}( -\delta x_1 +s - t) \, e^{-\frac{s+t}{2}  \, \Delta_{1,0} } \right) \, F_{2,0}( -\delta x_2 - T_{12}(s) ) \, e^{- \frac{T_{12}(s)}{2} \, \Delta_{2,0} } \nn \\
&=&\int_{-\infty }^{0} ds \, \partial_s \left( F_{1,0}( -\delta x_1 +s ) \, e^{-\frac{s}{2} \, \Delta_{1,0} } \right) \, F_{2,0}( \,  -\delta x_2 - T_{12}(s) \, ) \, e^{ - \frac{T_{12}(s)}{2} \, \Delta_{2,0} } . \label{eq:NHHLexpl}
\eeqa

While (\ref{eq:NHHLexpl}) provides  an explicit result, it still appears rather intricate, especially since it contains the complicated transition function $T_{12}(s)$. 
We will now show that it can be reduced to an amazingly simple form in terms of the Q-functions.

First, applying the transform (\ref{eq:qToF}), and using parity of the ground state wave function, $F_{1,0}(z) = F_{1,0 }(-z)$, we can write
\beqa
F_{1,0}( s - \delta x_1 ) \, e^{-\frac{s-\delta x_1}{2} \Delta_{1,0} } &=& -i\vint \frac{du}{u} \, q_{1}(u) \times \text{exp}\left( u \, w_{\phi_1}(\delta x_1 - s )  \right) , \label{eq:inteF1} \\
F_{2,0}(  - \delta x_2 - T_{12}(s) ) \, e^{-\frac{T_{12}(s) + \delta x_2}{2} \Delta_{2,0} } &=& -i\vint \frac{du}{u} \, q_{2}(u) \times \text{exp}\left( u \,  w_{\phi_2}(- T_{12}(s)- \delta x_2 ) \right) .\nn\\\label{eq:inteF2}
\eeqa
We then plug these relations  into (\ref{eq:NHHLexpl}). We noticed a magic relation between the integrands of (\ref{eq:inteF1}) and (\ref{eq:inteF2}),
\beq
w_{\phi_1}( s-\delta x_1 ) = w_{\phi_2} ( -\delta x_2 - T_{12}(s) ) - \phi_3  \label{eq:magic} \ \ ,
\eeq
 which suggests that we switch to a new integration variable $\xi = w_{\phi_1}( s-\delta x_1 ) - \phi_3/2$. Notice that the integration measure is invariant, $ds \, \partial_s = d\xi \, \partial_{\xi}$.  Taking into account (\ref{eq:magic}) we get:
\beqa
\mathcal{N}_{123}^{\bullet \bullet \circ }&=& - e^{\frac{\delta_{12} }{2}} \, \vint \frac{du}{u} \vint \frac{dv}{v} \, q_{1}(u) \, q_{2}(v) \,\left[  \int^{-\phi_2 + \phi_3/2 }_{\phi_1 - \phi_3/2} d\xi \, \partial_{\xi} \left(  e^{- u \xi  - u \phi_3/2 } \right) \, e^{v \xi - v \, \phi_3/2 } \right] , \nn \\
\delta_{12} &=& -  {\delta x_1}{}\, \Delta_{1,0} + {\delta x_2 }{}\, \Delta_{2,0} \ \ ,
\eeqa
and remarkably we can do the integral explicitly and find
\beq
\mathcal{N}_{123}^{\bullet \bullet \circ }= e^{\frac{\delta_{12} }{2}} \, \vint du \, \vint \frac{dv}{v} \, q_{1}(u) \, q_{2}(v) \, \left(\frac{ e^{ (\phi_2 - \phi_3) u - \phi_2  v} -  e^{  -\phi_1 \, u + (\phi_1 - \phi_3 )  v} }{u-v}\right).\label{eq:dudv}
\eeq
We can simplify this expression further. In fact, notice that the integrand has no poles for $\text{Re}(u) > 0$, $\text{Re}(v) > 0$,  in particular there is no pole at $u\sim v$. Therefore we can shift the two integration contours independently.  Similarly to the trick used in section~\ref{sec:BaxtoSchrod}, we shift  the $v$ integration contour to the right so that $\text{Re}(v) > \text{Re}(u)$, and split the integral into two contributions. One of them vanishes since the $v$-integrand is suppressed and the integration contour can be closed at $\text{Re}(v) = \infty$:
\beq
\vint du \,  q_{1}(u) \, e^{ ( \phi_2 - \phi_3 ) u }  \,\left(\int_{ {{\bf |}}  \, + 0^+} \frac{dv}{v} \, q_{2}(v) \, \frac{ e^{- \phi_2 v } }{ u-v} \right)= 0 ,
\eeq
while for the second integral it is the $u$-integrand that is suppressed. Closing the contour we now pick a residue at $u \sim v$:
\beqa
\mathcal{N}^{\bullet \bullet \circ }_{123} &=& - e^{\frac{\delta_{12} }{2}} \, \int_{ {{\bf |}}  \, \, + \,0^+ } \frac{dv}{v} \,  q_{2}(v) \, e^{( \phi_1 - \phi_3 ) v } \,  \left( \vint du \, \frac{ q_{1}(u) \, e^{ - \phi_1 u }  }{u-v}\right) \\ 
&=& + e^{\frac{\delta_{12} }{2}} \,  ( 2 \pi i )\vint \frac{dv}{v} \,  q_{1}(v) \, q_{2}(v) \, e^{- \phi_3 v} .
\eeqa
Combining all ingredients, we get the final expression for the structure constant in terms of the Q functions:
\beq
 C^{\bullet \bullet \circ}_{123} = (K_{123} )^{\Delta_{1,0}} \, (K_{213} )^{\Delta_{2,0}} \, \frac{\, \vint  \, q_{1} \, q_{2} \, e^{- \phi_3 u} 
 \frac{du}{ 2 \pi i u}
 }{\sqrt{ \vint q_1 \, {q}_1  \, \frac{du}{2\pi i u} } \ \,\sqrt{ \vint q_2 \, {q}_2  \, \frac{du}{2\pi i u} } 
 } , 
\eeq
where the constants $K_{123}$, $K_{213}$ are defined as in (\ref{eq:K123}) by permutation of the indices.
Again, it simplifies further in terms of the bracket $\br{\cdot}$ defined in \eq{eq:thebracket}
\beq\la{correlator_2}
\boxed{
C^{\bullet \bullet \circ}_{123} = 
\, \frac{\, \br{  q_{1} \, q_{2}\, e^{-\phi_3 u} }
}{\sqrt{ \br{ q_1^2}\br{ q_2^2}}  
}} \ \ .
\eeq
In this form it is clear that the final expression is explicitly symmetric for $1 \leftrightarrow 2$,
even though for the derivation we treated cusp $x_1$ differently from $x_2$.

This strikingly compact expression is one of our main results. Notice that it also covers the HLL case, namely if we send one of the effective couplings $\hat g_1, \hat g_2$ to zero we recover
\eq{CHLL1} as for zero coupling $\br{ q^2 } = 1$.

\section{Excited states}\label{sec:excited}

In this section we explore the meaning of the excited states and give them a QFT interpretation
as insertions at the cusps. We will also extend our result for the structure constant to the excited states.

\begin{figure}
    \centering
    \includegraphics[scale=2.0]{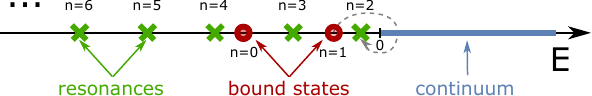}
    \caption{Structure of the spectrum of the Schrodinger operator.
    For finite coupling there are finitely many bound states. When the coupling is decreased, eventually the top bound state touches the continuum and goes to another sheet, becoming a resonance.
    There are infinitely many resonances for any value of the coupling. The spectrum of dimensions is related to the energy of the Schrodinger equation by $\Delta = -\sqrt{-E}$. This map resolves the branch cut of the continuum spectrum making the bound states and the resonances indistinguishable and equally important.}
    \label{fig:speccut}
\end{figure}

\subsection{Excited states and insertions}
\label{sec:insert}

First, let us discuss the structure of the spectrum of the Schr\"odinger equation.
When we increase the coupling we find more and more bound states in the spectrum at $E<0$.
If we analytically continue the bound state energy by slowly decreasing the coupling we will find
that the level approaches the continuum at $E=0$  and then reflects back. After that point
the state will strictly speaking disappear from the spectrum of the bound states as the wave function
will no longer be normalizable. However, if we define the bound state as a pole of the resolvent,
it will continue to be a pole, just not on the physical sheet, but under the cut of the continuum part of the spectrum.

\begin{figure}
    \centering
    \includegraphics[scale=2.0]{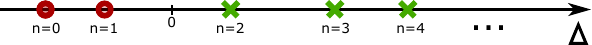}
    \caption{Structure of the spectrum of the QSC.
    The map $\Delta = -\sqrt{-E}$, which relates the spectrum obtained from QSC to the Schrodinger equation, resolves the cut of the continuum spectrum, revealing an infinite set of states.}
    \label{fig:speccut2}
\end{figure}

At the same time,  from the expression for $G(\Lambda_1,\Lambda_2,\Lambda_3,\Lambda_4)$
in \eq{G1234} we see that the natural variable is not $E$ but rather $\Delta = -\sqrt{-E}$.
In the $\Delta$-plane the branch cut of the continuum spectrum will open revealing all the infinite number of the resonances bringing them back into the physical spectrum (see Fig.~\ref{fig:speccut2}).

In order to give the field theory   interpretation of those bound states we build projectors,
which acting on our main object $G(\Lambda_1,\Lambda_2,\Lambda_3,\Lambda_4)$ will project on the excited states $\Delta_n$ in the large $\Lambda_i$ limit. First let us rewrite \eq{G1234} in terms of $\Delta_n$'s\footnote{To obtain \eq{G1234delta} rigorously from \eq{G1234}, one should take the coupling very large bringing many bound states into the spectrum and neglect the continuum part of the spectrum, which will get exponentially suppressed
w.r.t. the bound states with $\Delta_n<0$. After that one can continue in the coupling to smaller values. Alternatively, one can open the integral over the continuum part of the spectrum into the next sheet picking the poles at the resonances.}
\beq\la{G1234delta}
G(\Lambda_1,\Lambda_2,\Lambda_3,\Lambda_4)
\simeq\sum_n \frac{2F_n(\Lambda_1-\Lambda_2)F_n(\Lambda_4-\Lambda_3)}{||F_n||^2(-\Delta_n)}
\exp\left(-\Delta_n \frac{\Lambda_1+\Lambda_2+\Lambda_3+\Lambda_4}{2}\right) \;.
\eeq
Since $G$ has an interpretation as a $4$-BPS correlator, one can think about \eq{G1234delta}
as an OPE expansion in the $t$-channel. 
We will also see soon that the coefficients appearing there are the HLL structure constants with excited states. We will come back to this point in section \ref{sec:ope}.

When $\Lambda$'s tend to infinity the sum is saturated by the smallest $\Delta_n$.
To suppress the lowest states we define the following differential operators:
\beq\la{Odef}
{\cal O}_{2m}=\prod_{i=0}^{m-1}\frac{\d_++\Delta_{2i}}{-\Delta_{2m}+\Delta_{2i}}\;\;,\;\;
{\cal O}_{2m+1}=\prod_{i=0}^{m-1}\frac{\d_++\Delta_{2i+1}}{-\Delta_{2m+1}+\Delta_{2i+1}}\times\frac{1}{2}\d_-
\eeq
where 
$
\d_\pm\equiv \d_{\Lambda_1}\pm \d_{\Lambda_2}\;\;,\;\;\bar\d_\pm \equiv \d_{\Lambda_4}\pm \d_{\Lambda_3}
$.
With the help of these operators we define
\beqa
W_{n}&\equiv& \left.{\cal O}_n \bar {\cal O}_n   \, G(\Lambda_1,\Lambda_2,\Lambda_3,\Lambda_4)\right|_{\Lambda_1=\Lambda,\;\Lambda_2=\Lambda,\;\Lambda_3=\Lambda,\;\Lambda_4=\Lambda}\; ,
\eeqa
which at large $\Lambda$ scales as $e^{-2\Delta_n \Lambda}$ since all terms with $k < n$ are projected out!
Notice that, as discussed in Sec.  \ref{sec:2ptcusp}, $G( \Lambda, \Lambda, \Lambda,  \Lambda )$ can be used to describe a regularized two-point function, where the cutoff is identified with $ x_{12}e^{-\Lambda}= \epsilon$, similarly we get 
\beq\label{2ptn}
W_{2m}\simeq \left(\frac{\epsilon}{x_{12}}\right)^{2\Delta_{2m}}\frac{-2[F_{2m}(0)]^2}{ ||F_{2m }||^2 \, \Delta_{2m}}\;\;,\;\;
W_{2m+1}\simeq \left(\frac{\epsilon}{x_{12}}\right)^{2\Delta_{2m+1}}\frac{-2[F'_{2m+1}(0)]^2}{ ||F_{2m + 1 }||^2 \,  \Delta_{2m+1}} ,
\eeq
which indeed has the structure of the two point function of operators with dimension $\Delta_n$! These are the two point functions of the cusps with extra insertions due to the action of ${\cal O}_n$.
The specific form of the operator insertion in general depends on the  regularization scheme. The operators ${\cal O}_n$ give an explicit form of these insertions for the point-splitting regularization\footnote{We expect that for the finite $\theta$ case, i.e. away from the ladder limit, one should simply replace $\partial_\pm$ with the corresponding covariant derivatives at least at weak coupling.}.  
For instance, the first two operators ${\cal O}_1=\frac{1}{2}\d_-$ and ${\cal O}_2=\frac{\d_++\Delta_0}{\Delta_0-\Delta_2}$ will produce the following insertions\footnote{In \eq{O1phi} and \eq{O2phi} the scalar coupled to $n_1$ is located at position $-\Lambda_1$ on the contour, and the scalar coupled to $n_2$ is at $\Lambda_1$.  } 
\beqa\label{O1phi}
{\cal O}_1\;\;&\leftrightarrow&\;\;\frac{1}{2}
\(
-\Phi^a n_2^a |\dot x(\Lambda_2)|+\Phi^a n_1^a |\dot x(-\Lambda_1)|
\)=
\(
\Phi^a n_1^a-\Phi^a n_2^a 
\)\frac{\epsilon}{2}\;,\\ 
\label{O2phi}
{\cal O}_2\;\;&\leftrightarrow&\;\;\frac{(\Phi^a n_2^a+\Phi^a n_1^a)\epsilon+\Delta_0}{\Delta_0-\Delta_2}\;.\label{eq:O2def}
\eeqa
Naively, the interpretation of the operators corresponding to the excited states is only valid for large enough coupling
when $\Delta_n<0$. In the next section we verify that it remains true at weak coupling at one loop level.

Below, we also extend our result for the 3-cusp correlator to  excited states. For this, we will need to know the long-time asymptotics of $\tilde G_{\Lambda_1, \Lambda_2}(x, y)$ computed with the new type of boundary conditions described by the action of the projector $\mathcal O_n$.  
 We have, for $y \rightarrow \infty$, 
\beq\la{largetimen}
\mathcal O_n \, \tilde G_{\Lambda_1, \Lambda_2}(x, y) 
\simeq c_n F_n(x) \;  e^{-\Delta_n y } ,
\eeq
where
\beq
c_{2m}=-\frac{2 F_{2m}(0)}{ ||F_{2m }||^2 \, \Delta_{2m}}\;\;,\;\;c_{2m+1}=-\frac{2 F'_{2m+1}(0)}{ ||F_{2m+1}||^2 \, \Delta_{2m+1}}\;. \label{eq:cn}
\eeq

Finally, from the $2$-point correlator \eq{2ptn} we extract the normalization coefficients
\beq\la{Nn2}
{\cal N}_{\Delta_n}=\epsilon^{\Delta_n} \, c_{n} \, \sqrt{\frac{-\Delta_n \, ||F_{n}||^2}{2} }  ,
\eeq
which we will need  to normalize the structure constant in the next section. 
\subsection{Correlator with excited states}\label{sec:HLLexc}
We will redo the calculation of the HLL correlator for the case
when the heavy state is excited. We mostly notice that all the steps are essentially the same as in the case of the ground state.
We begin by applying the projector operator ${\cal O}_n$, defined in \eq{Odef} to the cusp at $x_1$  
and use that in the small $\epsilon$ limit we simply use the leading asymptotics \eq{largetimen} to obtain, very similarly to the ground state \eq{Yground} 
\beq
{\cal O}_n Y_{\vec{x}_1, \epsilon}(\vec{x}_2 , \vec{x}_3 ) =c_n
\, F_{n}(-\delta x_1) \, \epsilon ^{\Delta _{n}} \, \frac{
   ( L_{123}  )^{ {\Delta _{n}} } }{ x_{12}^{\Delta _{n}} x_{13}^{\Delta _{n}} x_{23}^{-\Delta _{n}}} ,
\eeq
with $c_n$ defined in (\ref{eq:cn}). 
Normalizing the result with \eq{Nn2} to get a finite result for the structure constant we get
\beq
C_{123}^{\bullet_{n}\circ\circ}=
\sqrt{\frac{2}{-\Delta_{n}||F_{n}||^2}} \, 
{F_{n}(\delta x_1)} \, ( L_{123} )^{ {\Delta_{n}}} \; .
\eeq
rewriting it in terms of q-functions exactly as for the ground state we obtain
\beq
\boxed{
C^{\bullet_{n} \circ \circ}_{123}= \frac{ \br{q_{1,n} e^{\phi_2 u-\phi_3 u}} }{\sqrt{(-1)^n\br{q_{1,n}^2}}}
}\;,
\eeq
where $q_{1,n}$ denotes the solution of the QSC corresponding to the $n$-th excited state, with parameters $\hat g=\hat g_1$, $\phi = \phi_1$. The $(-1)^n$ appears from the corresponding factor in the relation for the norm of the wavefunction in \eq{eq:normQ}, it is needed to ensure the denominator is real at large couplings.

Similarly for the HHL correlator we simply replace q-functions and the corresponding dimensions, but the expression stays the same! 
\beq
\boxed{
C^{\bullet_{n}\bullet_{m} \circ}_{123}= \frac{(-1)^m \br{q_{1,n}q_{2,m} e^{-\phi_3 u}} }{\sqrt{(-1)^{n+m}\br{q_{1,n}^2}\br{q_{2,m}^2}}}
}\;.
\eeq

\subsection{Excited states at weak coupling from QSC}

\label{sec:qscexc}

\begin{figure}[t]
    \centering
    \includegraphics[scale=0.8]{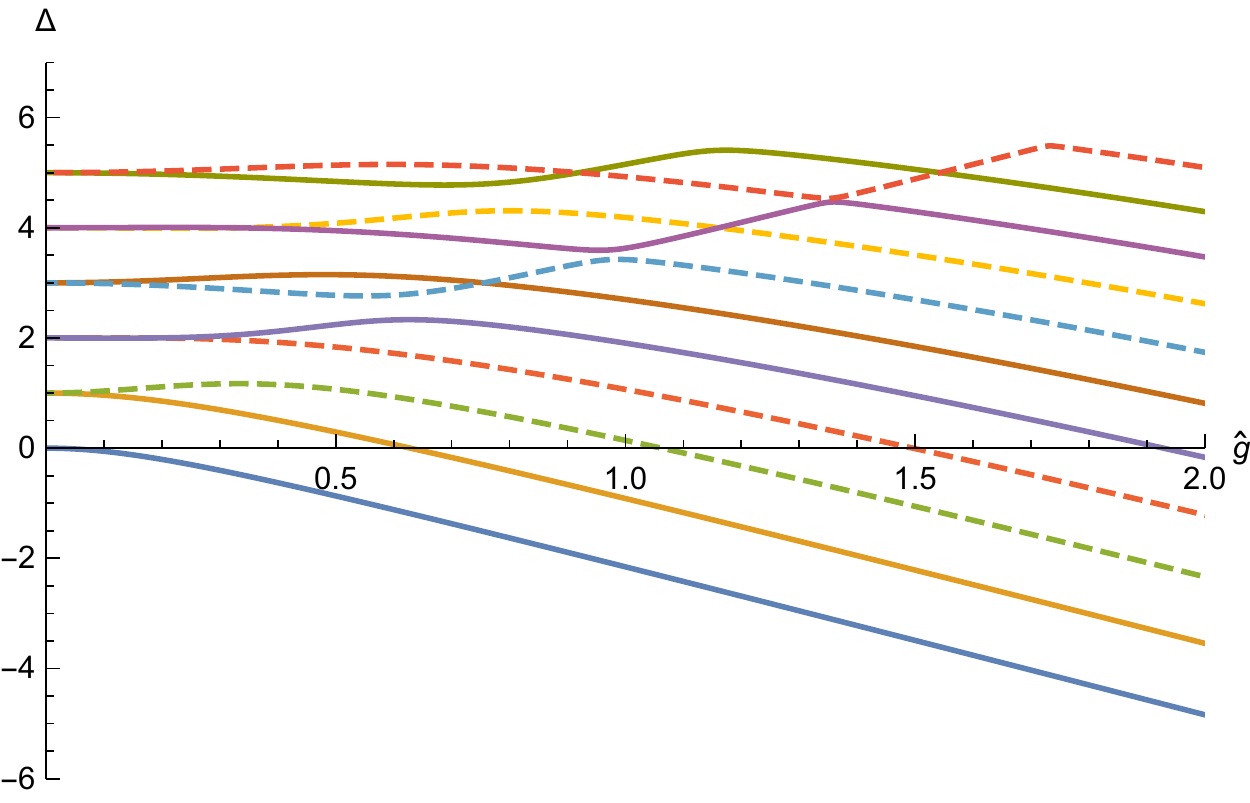}
    \caption{The first  few states for $\phi=1.5\,\,$. We show numerical data for $\Delta$ as a function of $\hat g$, obtained from the Baxter equation.
    We see that all the states, except the ground state, are paired together at weak coupling. 
    \label{fig:spectrum2}}
\end{figure}

\begin{figure}[t]
\centering
    \includegraphics[scale=0.8]{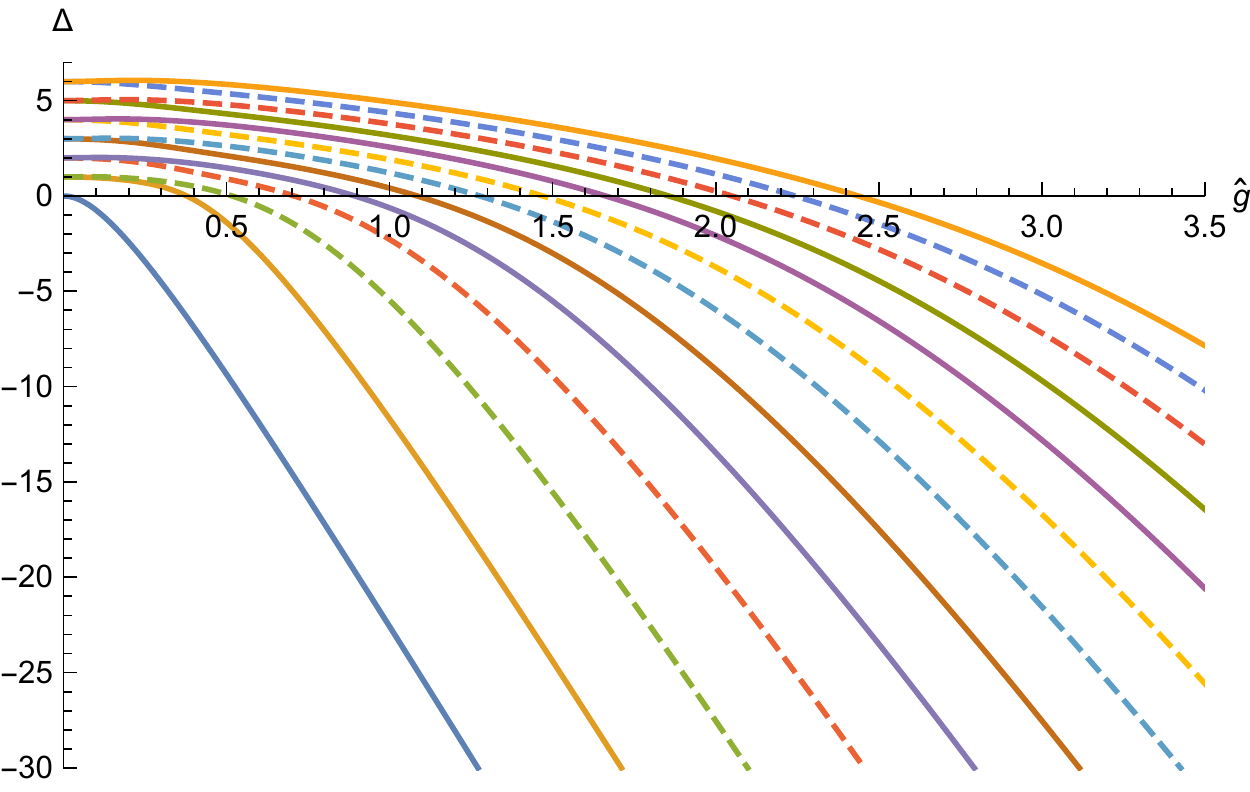}
    \caption{The first  few states for  $\phi=3.0\,\,$. We plot $\Delta$ as a function of the coupling $\hat g$ similarly to Fig.~\ref{fig:spectrum2}. \label{fig:spectrum2p2}}
\end{figure}

As we discussed above (see section \ref{sec:schr}), for large coupling the Schr\"odinger equation has several bound states while for small coupling all of them except the ground state disappear. Nevertheless the excited states have remnants at weak coupling which are not immediately apparent in the Schr\"odinger equation but are directly visible in the QSC. By solving the Baxter equation \eq{Bax2p} and the gluing condition \eq{qquant} numerically, we can follow any excited state from large to small coupling and we find that $\Delta$ has a perfectly smooth dependence on $\hat g$. The first several states are shown on Fig.~\ref{fig:spectrum2} and Fig.~\ref{fig:spectrum2p2} which also demonstrate an intricate pattern of level crossings that we will discuss below. For $\hat g\to 0$ we moreover observe that $\Delta$ becomes a positive integer $L$,
\beq
\label{dexp}
    \Delta=L+\Delta^{(1)}\hat g^2+\Delta^{(2)}\hat g^4+\dots, \ \ \ \ 
    L=1,2,\dots
    \ \ .
\eeq
Remarkably, for each $L>0$ we have two states which become degenerate at zero coupling.  In contrast, the ground state (corresponding to $L=0$) does not merge with any other state. 
This pattern is consistent with our proposal for the insertions \eq{Odef} -- the states with $n=2m$ and $n=2m-1$ have the same number of derivatives and thus should have the same bare dimension.

We can  explicitly compute $\Delta$ for these states at weak coupling from the Baxter equation.  
We solve it perturbatively using the efficient iterative method of \cite{Gromov:2015vua} and the Mathematica package provided with \cite{Gromov:2015dfa}. We start from the solution at $\hat g=0$ and improve it order by order in $\hat g$. 
At $\hat g=0$ the solution for any $L\equiv \left.\Delta\right|_{\hat g=0}$ has the form of a polynomial of 
degree $L$ multiplied by $e^{u\phi}$. At the next order we already encounter nontrivial pole structures.
This procedure gives $q$-functions 
written in terms of generalized $\eta$-functions \cite{Marboe:2014gma, Gromov:2015dfa} defined as
\beq
\label{defeta}
\eta_{s_1,\dots,s_k}^{z_1,\dots,z_k}(u)\equiv \sum_{n_1 > n_2 > \dots > n_k \geq 0}\frac{z_1^{n_1}\dots z_k^{n_k}}{(u+i n_1)^{s_1}\dots (u+i n_k)^{s_k}} \ .
\eeq
As an example, for $L=1$ we find \beq
    q=e^{u\phi }\[u+\hat g^2
    \(-i\Delta^{(1)}u\;\eta_1^1 -\frac{2}{\sin\phi} 
    +\frac{\Delta^{(1)}}{2} (-2 u e^{2 i \phi }+\cot \phi+i)\)
    \]+\cO(\hat g^4)
\eeq
where $\Delta^{(1)}$ is the 1-loop coefficient in \eq{dexp}. 
 The second solution $q_-$ is more complicated and already involves twisted $\eta$-functions such as $\eta_1^{e^{2i\phi}}$, but fortunately we only need $q_+$ to close the equations.
The quantization condition \eq{qquant} then gives a quadratic equation on $\Delta^{(1)}$ which fixes  \beq
    \Delta^{(1)}=\pm 4\ \ \ \text{for} \ \ \ L=1 \ .
\eeq
Thus as expected from the numerical analytsis we find two separate states, which become degenerate at zero coupling.

For comparison, for the ground state ($L=0$) we have
\beq
\label{qgs}
q = e^{u\phi } \, \left[ 1 + \hat{g}^2 \, \frac{2 i }{\sin\phi} \,  \left( 2 \, \phi \, (\eta^1_1-  \eta ^{e^{2 i \phi }}_1 )  -  (\eta^1_2-\eta ^{e^{2 i \phi }}_2 )\right) \right] +\cO(\hat g^4) \ .
\eeq

Repeating this calculation for $L=2,3,4,5$ we were able to guess a simple closed formula for the 1-loop correction,
\beq
\label{Dpm}
    \Delta_{L,\pm}=L\pm \frac{4}{L}\frac{\sin L\phi}{\sin\phi}\hat g^2+\dots, \ \ \ \ 
    L=1,2,\dots\;.
\eeq
For the ground state ($L\to 0$) this formula also gives the correct result although only the minus sign is admissible.

For the first several states we also computed $\Delta$ to two loops, e.g. for $L=1$
\beqa
\label{dn12}
    \Delta_{1,-}&=&1-4\hat g^2
   + 16 \left(\phi  \cot \frac{\phi }{2}-1\right)\hat g^4
    +\dots 
    \\
    \label{dp12}
    \Delta_{1,+}
    &=&1+4\hat g^2-16 \left(\phi  \tan \frac{\phi }{2}+1\right)\hat g^4
    +\dots\;.
\eeqa
The two-loop results for $L=2,3$ are given in\footnote{ Notice that for $\phi=\pi/L$ the two states with $\Delta=L$ at zero coupling are degenerate at one loop but not at two loops, at least for $L=1,2,3$.} Appendix \ref{app:exc}. All these results are also in excellent agreement with QSC numerics.  For completeness, the ground state anomalous dimension to two loops is \cite{Makeenko:2006ds,Drukker:2011za}\footnote{See \cite{Correa:2012nk,Henn:2013wfa,Henn:2012qz} for higher-loop results.}
\beqa
\label{eq:gs2loops}
   \Delta_0&=&0 -4\frac{\phi}{\sin\phi}\hat g^2
   \\ \nn &&
   +\frac{4}{\sin^2\phi} \left[2 i \phi  \left(\text{Li}_2(e^{2 i \phi
   })-\text{Li}_2(e^{-2 i \phi })\right)-2 \left(\text{Li}_3(e^{-2 i \phi
   })+\text{Li}_3(e^{2 i \phi })\right)+4 \zeta_3\right]\hat g^4
   +\dots\;. 
\eeqa

Let us note that for the ground state the leading weak coupling solution $q=e^{\phi u}$ immediately provides the 1-loop anomalous dimension via the quantization condition \eq{qquant}. However for excited states the leading order $q$-function is not enough because it vanishes at $u=0$, leading to a singularity in the quantization condition (resolved at higher order in $\hat g$).

\bigskip

\begin{table}[h]
\begin{tabular}{ | l| l |l |l |l | l| l| l| l | l | l  |}
 strong coupling &  $\Delta_0$ & $\Delta_1$ & $\Delta_2$ & $\Delta_3$ & $\Delta_4$ & $\Delta_5$ &  $\Delta_6$ &  $\Delta_7$ & $\Delta_8$ & $\Delta_9$ \\
 \hline 
 weak coupling & $\Delta_{0,-}$ & $\Delta_{1, -}$ & $\Delta_{1,+} $& $\Delta_{2,+} $& $\Delta_{2,-}$ & $\Delta_{3,-}$ & $\Delta_{3,+}$ & $\Delta_{4,+}$ & $\Delta_{4,-}$ & $\Delta_{5,-}$ 
\end{tabular}
\caption{The table shows the correspondence between the weak  and strong coupling behaviour of the first few excited states. The notation $\Delta_n$ denotes the ordering of the states at strong coupling (in particular see (\ref{eq:strongcoupling})), while the notation $\Delta_{L, \pm}$ is related to the form of the one-loop correction, see (\ref{Dpm}). The pattern evident from the table continues for all excited states.\label{tab:states}}
\end{table}

\paragraph{Comments on level crossing.} Let us now discuss another curious feature of the spectrum, namely the presence of level crossings for $\Delta>0$ which is evident from Fig.~\ref{fig:spectrum2}. Level crossings are of course forbidden in 1d quantum mechanics, but there is no contradiction as our states only correspond to energies of the Schr\"odinger problem when  $\Delta<0$. As we increase the coupling, for any state $\Delta$ eventually becomes negative and the levels get cleanly separated.
At the same time the odd (even) levels do seem to repel from each other.

\begin{figure}[t]
    \centering
    \includegraphics[scale=0.7]{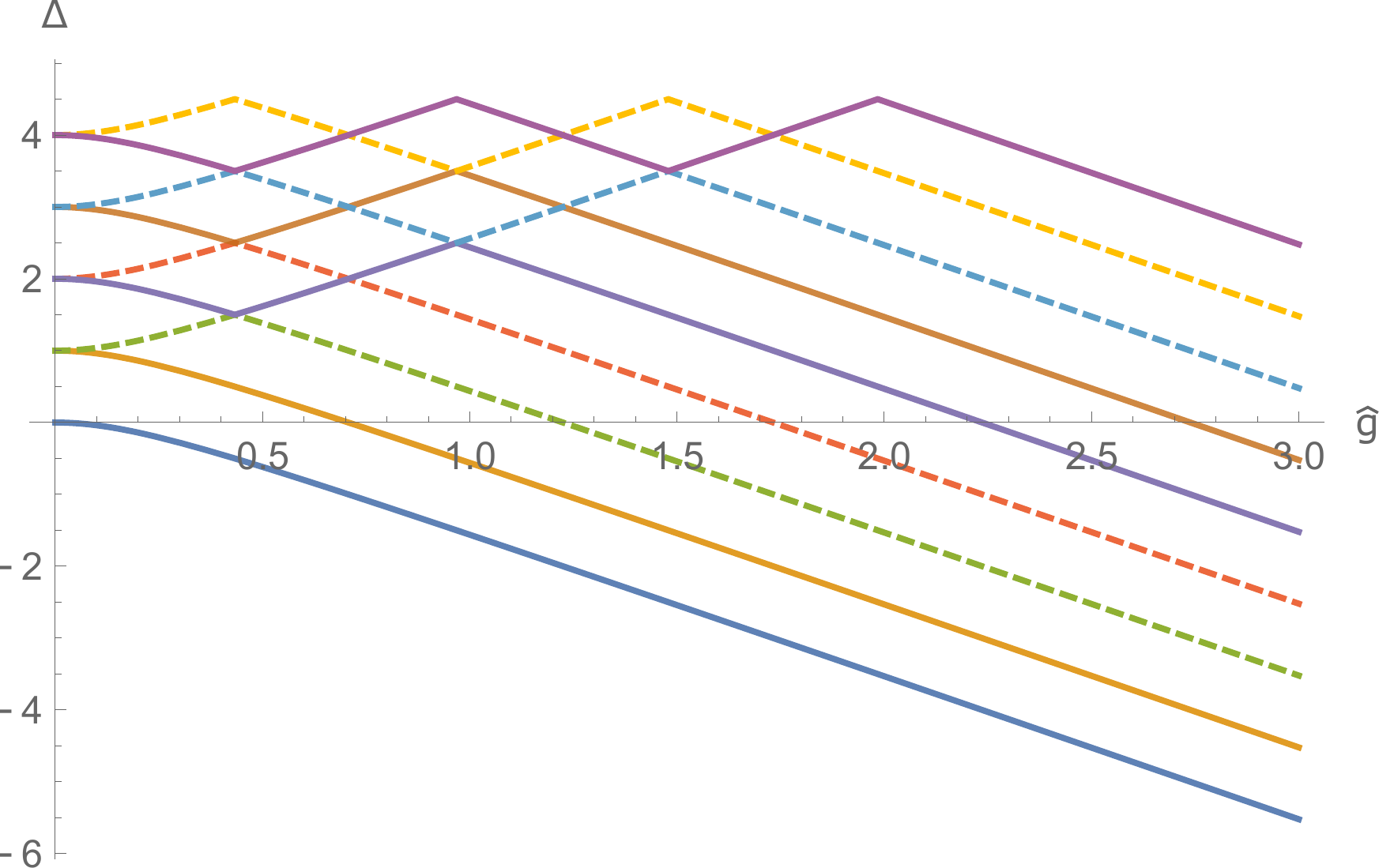}
    \caption{The first several states at $\phi=0$. For each level the dependence of $\Delta$ on the coupling alternates between \eq{Dn0} and \eq{Dm0} before taking the form \eq{Dn0} at large coupling.}
    \label{fig:spectrumzero2}
\end{figure}

At large coupling it is natural to label the states by $n=0,1,2,\dots$ starting from the ground state.
However the reshuffling of levels makes it a priori nontrivial to say what is the weak coupling behavior of a state with given $n$. First, we observe that $\Delta$ at zero coupling is given by $L= n/2$ (rounded up).  Moreover we found a nice relationship between $n$ and the signs plus or minus in \eq{Dpm} determining the 1-loop anomalous dimension. Namely, the levels with $n=0,1,2,\dots$ correspond to the following sequence of signs:
\beq
\label{MMPP}
    --++--++--++\ \dots
\eeq
In order to understand this pattern it is helpful to consider the analytically solvable case when $\phi=0$. We plot the states for this case on Fig.~\ref{fig:spectrumzero2}.  The spectrum of the Schr\"odinger problem for $\phi=0$ is known exactly \cite{Correa:2012nk},
\beq
\label{Dn0}
    \Delta_n=\frac{1}{2}\[1-\sqrt{16\hat g^2+1}\]+n\ , \ \ n=0,1,2,\dots
\eeq
Here only the values of $n$ for which $\Delta_n<0$ actually correspond to bound states.  One may try to analytically continue $\Delta_n$ in $\hat g$ starting from large coupling where it is negative, and arrive to weak coupling. However this would not be correct, as we know that half the levels should have positive slope at weak coupling, corresponding to the choice of the plus sign in the 1-loop correction\footnote{Clearly,  \eq{Dn0} would instead give a negative 1-loop coefficient with $\Delta=n-4\hat g^2+\dots$. Also note that for $\phi=0$ the 1-loop correction \eq{Dpm} becomes equal to $\pm 4 \hat g^2$ and does not depend on $n$.} \eq{Dpm}. The true levels instead are shown on Fig.~\ref{fig:spectrumzero2}.  At weak coupling half of them are given by an expression of the same form \eq{Dn0} but with opposite sign of the square root,
\beq
\label{Dm0}
    \Delta_m'=\frac{1}{2}\[1+\sqrt{16\hat g^2+1}\]+m\ , \ \ m=0,1,2,\dots
\eeq
At large coupling the levels are 
given by \eq{Dn0}, so dependence on the coupling switches from \eq{Dn0} to \eq{Dm0} (where $m$ and $n$ may be different) at the point where these two curves intersect. Moreover, at this point two levels meet, and they correspond to adjacent values of $n$ of the same parity. In this way e.g. the levels with even $n$ `bounce' off each other, and the same is true for odd $n$. That explains the pattern of signs in \eq{MMPP}.

In fact as we see in Fig.~\ref{fig:spectrumzero2} the behavior of $\Delta$ can switch multiple times between forms \eq{Dn0} and \eq{Dm0}, before finally becoming the expected curve \eq{Dn0} at large coupling.
The derivative $\d \Delta/\d\hat g$ is discontinuous at these switching points. However when  $\phi$ becomes nonzero the picture smoothes out and the level crossing at the intersection point is also avoided (though some other level crossings truly remain) as can be see on Fig.~\ref{fig:spectrum2}. 

Having $\Delta$ as a piecewise-defined function made up of parts given by \eq{Dn0} and \eq{Dm0} reminds somewhat the spectrum of local twist-2 operators at zero coupling, where the anomalous dimension becomes a piecewise linear function of the spin (with different regions corresponding e.g. to the BFKL limit \cite{bfkl2, bfklint} or to usual perturbation theory\footnote{ See e.g. \cite{Costa:2012cb} for a discussion and \cite{Gromov:2015wca} for some finite coupling plots.}).

One may regard \eq{Dm0} as an analytic continuation of \eq{Dn0} around the branch point at $\hat g=i/4$. There are more branch points at complex values of $\hat g$ where curves of the form \eq{Dn0} and \eq{Dm0} intersect, and  we expect all the levels to be obtained from each other by analytic continuation in $\hat g$, even for generic $\phi$. Again this situation is reminiscent of the twist operator spectrum.

\subsection{Excited states at weak coupling from Feynman diagrams}

\begin{figure}
    \centering
    \def\svgwidth{200pt}
    \input{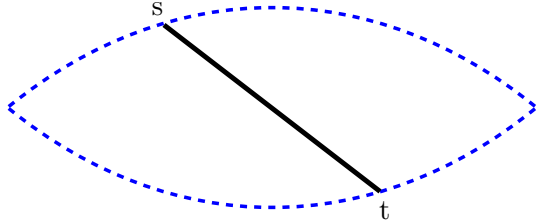}
    \caption{One loop diagram, contributing to the ground state anomalous dimension.}
    \label{fig:groundstate}
\end{figure}

In this section we compute the diagrams contributing to the anomalous dimensions of the lowest excited states.
First let us reproduce the one loop correction to the ground state. For that case there is only one diagram, shown on Fig.~\ref{fig:groundstate},
\beq
D_0=\int_{-\Lambda}^\Lambda dt\int_{-\Lambda}^\Lambda ds \frac{2\hat{g}^2}{\cosh (s-t)+\cos (\phi )}\label{eq:D0def} \ .
\eeq
It can be computed exactly for any $\Lambda$,
\beq
D_0= \frac{4\hat g^2}{\sin\phi} \left(2 \Lambda  \phi -i \text{Li}_2\left(-e^{-2 \Lambda -i \phi }\right)+i \text{Li}_2\left(-e^{i \phi -2 \Lambda }\right)+i
   \text{Li}_2\left(-e^{-i \phi }\right)-i \text{Li}_2\left(-e^{i \phi }\right)\right)\label{eq:D0eval}
\eeq
and at large $\Lambda$ it diverges linearly as $D_0=8\hat g^2 \frac{\phi}{\sin\phi}\Lambda+{\cal O}(\Lambda^0)$. Recalling that $\Lambda=\log\frac{x_{12}}{\epsilon}$
we read-off the anomalous dimension $\gamma_0=-4\hat g^2 \frac{\phi}{\sin\phi}$ in agreement with \eq{eq:gs2loops}.

\begin{figure}
    \centering
    \def\svgwidth{300pt}
    \input{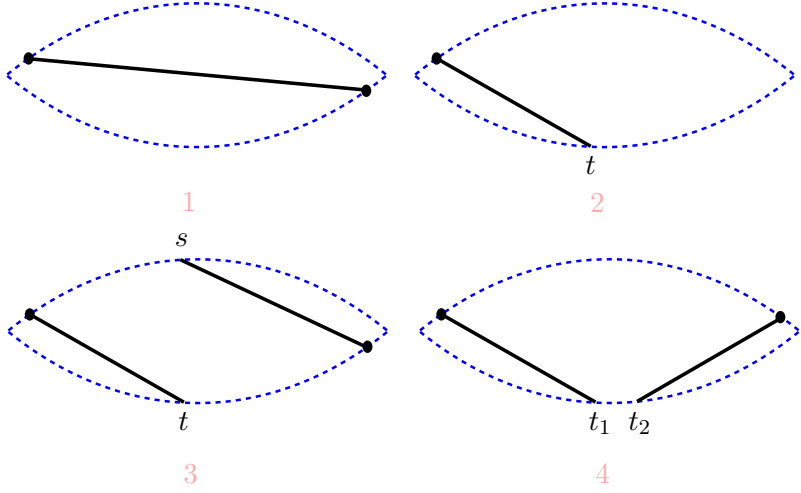}
    \caption{Four diagrams contributing to the mixing matrix of the cusps with insertions of a scalar operator.}
    \label{fig:diagrams}
\end{figure}

For the lowest excited states we have $4$ diagrams (see Fig.~\ref{fig:diagrams}). 
For example, the 4th diagram $D_4$ is given by the double integral
\beq
D_4=\int_{-\Lambda}^{\Lambda} dt_1 \int_{t_1}^{\Lambda} dt_2 \, \frac{ 4 \, \hat g^4}{( \cosh(-\Lambda - t_1 ) + \cos\phi ) \, (\cosh(\Lambda - t_2 ) + \cos\phi ) } ,
\eeq
 and corresponds to the following differentiation of the four point function:
\beq
\partial_{\Lambda_1} \partial_{\Lambda_3} \left. G(\Lambda_1, \Lambda_2 , \Lambda_3 , \Lambda_4 ) \right|_{\Lambda_i=\Lambda} = D_4 + O( \hat g^6 )\  .
\eeq
Below we give the result for these diagrams for large $\Lambda$, keeping $e^{-2\Lambda}$ terms:
\beqa
D_1&=&4\hat g^2 e^{-2\Lambda} \ ,\\ \nn
D_2&=&2 \hat{g}^2 \phi  \csc \phi -4 \hat{g}^2 e^{-2 \Lambda }+{\cal O}( e^{-4 \Lambda }) \ ,\\ \nn
D_3&=&(D_2)^2=4 \hat{g}^4 \phi ^2 \csc ^2\phi -16 \hat{g}^4 e^{-2 \Lambda } \phi  \csc \phi +{\cal O}( e^{-4 \Lambda })\ ,\\ \nn
D_4&=&4 \hat{g}^4 \phi ^2 \csc ^2\phi  + 16 \hat{g}^4 e^{-2 \Lambda } (-2 \Lambda +\phi  \cot \phi +\log (\cos \phi +1)-1+\log 2) \ .
\eeqa
Combining these diagrams we can construct the   operators described in section \ref{sec:insert}, in particular here we consider operators obtained with the insertion of one scalar at the cusp\footnote{The operators with more scalar insertions built this way may include derivatives acting on the scalars.}. We have\footnote{In the r.h.s. of \eq{O11G} and \eq{O22G} we omit an overall irrelevant prefactor.} 
\beq
\label{O11G}
2 \,  \mathcal{O}_1 \,  \mathcal{\bar{O}}_1 \left. G(\Lambda_1, \Lambda_2 , \Lambda_3 , \Lambda_4 ) \right|_{\Lambda_i=\Lambda} = D_1+D_3 -D_4 + O ( \hat g^6 ) \eeq
 and from the diagrams computed above we find
 \beq
2 \,  \mathcal{O}_1 \,  \mathcal{\bar{O}}_1 \left. G(\Lambda_1, \Lambda_2 , \Lambda_3 , \Lambda_4 ) \right|_{\Lambda_i=\Lambda} ={4 \hat{g}^2}e^{-2\Lambda}\left(1+8 \hat{g}^2 \Lambda \right) + \dots 
\eeq
Again identifying the cutoff  with $\Lambda=\log\frac{x_{12}}{\epsilon}$,  we read off the  one-loop dimension $\Delta_1=1-4\hat g^2$. Remarkably, it perfectly matches the analytic continuation to weak coupling of the first excited state energy, computed from the QSC above in  \eq{dn12}. This state corresponds to the second line from below on Fig.~\ref{fig:spectrum2}.

Another operator one can build is obtained from the following combination of derivatives: 
\beqa
\label{O22G}
&& (\Delta_0 - \Delta_2)^2 \, \mathcal{O}_2 \,  \mathcal{\bar{O}}_2 \left. G(\Lambda_1, \Lambda_2,\Lambda_3 , \Lambda_4 ) \right|_{\Lambda_i=\Lambda}  
 \\ \nn &&
 =\left[ 2 \, (\partial_{\Lambda_1} \partial_{\Lambda_3} + \partial_{\Lambda_1} \partial_{\Lambda_4} ) + \Delta_0^2 + 4 \, \Delta_0 \, \partial_{\Lambda_1} \right]  \,\left. G(\Lambda_1,\Lambda_2,\Lambda_3 , \Lambda_4 ) \right|_{\Lambda_i=\Lambda}  \ \ \ .
\eeqa 
The r.h.s. here can be written in terms of the diagrams we have computed and is equal to
\beqa 
\gamma_0^2 + 4 \gamma_0 \, D_2 + 2 (D_1 + D_3 + D_4 ) + O ( \hat g^6 ) = {4 \hat{g}^2}e^{-2\Lambda}\left(1-8 \hat{g}^2 \Lambda \right) + \dots \ \ ,\label{eq:secondO2}
\eeqa
where $\gamma_0 = -4 \hat g^2 \phi/\sin\phi $ is the  one-loop scaling dimension for the ground state. 
The logarithmic divergence in (\ref{eq:secondO2}) correctly reproduces the energy of the analytic continuation of the second excited state at one loop $\Delta_2=1+4\hat g^2$, matching the QSC result  \eq{Dpm}. 
This state corresponds to the third line from below in  Fig.~\ref{fig:spectrum2}. The one-loop result agrees with the one obtained in \cite{Alday:2007he,Bruser:2018jnc} at $\theta=0$ (we expect in the ladders limit this result should be the same).

\section{Simplifying limit}\label{sec:smallPhi}
In this section we consider the limit when $\phi_1+\phi_2\to \phi_3$. 
Geometrically this limit, which lies at the boundary of the regime of parameters considered in the rest of the paper (\ref{eq:ineq}), describes the situation where the cusp point $\vec{x}_3$ belongs to the circle defined by the extension of the arc $(12)$. In this situation, the points $A$ and $B$ shown in Fig. \ref{fig:triangle} both coincide with the cusp point $\vec{x}_3$. 
A special case of this limit is the situation when all angles are zero and the triangle reduces to a straight line.

The main simplification comes from the most important part of the result
\beq
\vint\frac{du}{u} q_{1}q_{2}e^{-\phi_3 u}
\eeq
which now can be evaluated explicitly.
When $\phi_1+\phi_2\to \phi_3$ we can deform the integration contour to infinity and notice that only the large $u$ asymptotic of the integrand contributes. This is clear from the following integral
\beq
\frac{1}{2 i \pi }\vint {du}\frac{e^{\beta u}}{u^\alpha} =\frac{ \beta ^{\alpha -1}}{\Gamma (\alpha )}\;\;\label{eq:gint}
\eeq
where in our case $\beta=\phi_1+\phi_2-\phi_3$ is small and positive. We see that the integral \eq{eq:gint} allows us to convert the large $u$ expansion into small $\beta$ series.
The large $u$ expansion of the integrand is very easy to deduce from the Baxter equation \eq{Bax2p}, one just has to plug into the Baxter equation \eq{Bax2p} the ansatz
\beq
q=e^{\phi u}u^{\Delta}\(1+\frac{k_1}{u}+\frac{k_2}{u^2}+\dots\) \label{eq:largeuq}
\eeq
to get a simple linear system for the coefficients $k_i$, which gives
\beqa
k_1\sin\phi &=& \frac{1}{2} (\Delta -1) \Delta  \cos \phi -2 \hat g^2\\
\nn k_2\sin^2\phi &=& 
\frac{1}{48} (\Delta -2) (\Delta -1) \Delta  ((3 \Delta -1) \cos (2 \phi )+3 \Delta -5)-(\Delta -1)^2 \hat g^2 \cos \phi+2 \hat g^4 , \\
&\dots& \label{eq:an}
\eeqa
which allows us to compute explicitly
\beqa
&&\vint\frac{du}{u} q_{1}q_{2}e^{-\phi_3 u}=\frac{2 i \pi  \beta ^{-\Delta _1-\Delta _2}}{\Gamma \left(-\Delta _1-\Delta _2+1\right)}\\
\nn &&-\frac{i \pi  \beta ^{-\Delta _1-\Delta _2+1} \left(-\left(\Delta
   _1-1\right) \Delta _1 \cot \phi _1-\left(\Delta _2-1\right) \Delta _2 \cot \phi _2+4 \left(g_1^2 \csc \phi _1+ g_2^2\csc \phi
   _2\right)\right)}{\Gamma \left(-\Delta _1-\Delta _2+2\right)} + \dots 
\eeqa
In this way we get the following small-$\beta$ expansion for the bracket in the numerator of structure constant with insertions at $1$ and $2$: 
\beqa
\label{eq:CsmallPhi}
\br{ q_1 q_2 e^{- \phi_3 u } }
&=& \frac{1 \,}{\Gamma (-\Delta_1
   -\Delta_2+1)}\\
   & + &\frac{\left((\Delta_1 -1) \Delta_1 
   \cot \phi_1 +(\Delta_2-1)
   \Delta_2 \cot
   \phi_2-4 \left( \hat g_1^2 \csc
   \phi_1 +\hat g_2^2 \csc
   \phi_2\right)\right)}{2 \, \Gamma
   (-\Delta_1 -\Delta_2+2)} \, \beta \nn \\ \nn
   &+& \dots  .
\eeqa
In principle, the expansion can be performed to an arbitrary order in $\beta = \phi_1 + \phi_2 - \phi_3$.

Similarly, the norm factors appearing in the denominator of the structure constants simplify when $\phi_i \rightarrow 0$ for one of the cusps $i=1$ or $i=2$. 
This limit describes the situation where the cusp angle disappears. As we reviewed in Sec. \ref{sec:qscexc}, at $\phi=0$ the Schr\"odinger equation becomes exactly solvable and the spectrum is explicitly known \cite{Correa:2012nk}.  

The main ingredient for the computation of the norm is the integral  (\ref{eq:normQ}), and it is clear that for small $\phi$ it simplifies for the very same mechanism we have just described. In particular,  every term in the $1/u$ expansion of the integrand gives an integral of the kind (\ref{eq:gint}), which allow us to organize the result in powers of $\phi$. Naturally we should also take into account the scaling of the coefficients $k_i$ appearing in (\ref{eq:largeuq}) for $\phi \sim 0$. Notice that the expressions (\ref{eq:an}) are apparently  singular at $\phi \sim 0$. 
However, a nice feature of this limit is that most of these divergences are cancelled systematically due to the fact that the scaling dimension too depends on $\phi$ in a nontrivial way. 
 In particular, we found numerically that, for the QSC solution corresponding to the ground state, the coefficients $k_n$ have the following scaling for $\phi \rightarrow 0$:
 \beq
\left\{ k_1 , k_2, k_3, k_4,k_5, \dots \right\} \sim \left\{0 \, ,\,  O(1)\,, \,0\,, \, O(1) \, , \, 0\,,\, O(1) \,, \dots \right\}. \label{eq:gsScale}
 \eeq
 This observation is quite powerful. 
 Indeed, combined with the parametric form of the coefficients (\ref{eq:an}), the requirement that they scale as (\ref{eq:gsScale}) fixes all terms\footnote{A very similar observation was made in the context of the fishnet models at strong coupling  in \cite{Gromov:2017cja}.}  in the expansion of $\Delta$ for small  $\phi$ !
 
More precisely, we find that the scaling (\ref{eq:gsScale}) corresponds to two solutions for $\Delta( \phi)$: one is the ground state, for which we reproduce the results of \cite{Correa:2012nk} obtained using perturbation theory of the Schr\"odinger equation, namely, for the first two orders,
 \beqa
\Delta_0  &=&  \frac{1}{2}\left(1  - \sqrt{1 + 16 \, \hat{g}^2 }\right) + \frac{\hat g^2 \left(-16 \hat g^2+\sqrt{16 \hat g^2+1}+1\right)}{\left(16
  \hat  g^2-3\right) \sqrt{16 \hat g^2+1}}\, \phi^2 + \dots . \label{eq:DeltagsPhi}
  \eeqa
The other solution describes one of the excited states trajectories\footnote{As explained in Sec. \ref{sec:qscexc}, this trajectory strictly speaking is formed patching together pieces of infinitely many  levels, which are separate for finite $\phi$, see Fig. \ref{fig:spectrumzero2}.  }
 \beq
\Delta_0' =  \frac{1}{2} \left(1+\sqrt{16 \hat g^2+1}\right)+\frac{\hat g^2 \left(16 \hat g^2+\sqrt{16 \hat g^2+1}-1\right)}{\left(16
  \hat  g^2-3\right) \sqrt{16 \hat g^2+1}}\, \phi^2 + \dots 
 \eeq
 It is straightforward to generate higher orders in $\phi$ with this method. The remaining infinitely many states can be described allowing for a more general scaling of the coefficients $k_m$, see  Appendix~\ref{app:smallPhi} for details and some results. 
 
 Plugging in the scaling of coefficients (\ref{eq:gsScale}), for the solution corresponding to the ground state we find 
\beq
\br{ q^2 } = \frac{ 1 }{\Gamma( 1 - 2 \Delta ) }  +   O(\phi^2 )   ,
\eeq
which combined with (\ref{eq:CsmallPhi}) gives a finite result for the OPE coefficient at $\phi_1  =\phi_2 = \phi_3 = 0$:
\beq
\left. C^{\bullet \bullet \circ}_{123} \right|_{\phi_i=0} = \left.\frac{ \sqrt{  \Gamma( 1 - 2 \Delta_1 ) \, \Gamma( 1 - 2 \Delta_2 ) }}{\Gamma( 1- \Delta_1 - \Delta_2 ) } \right|_{\phi_i=0} = \frac{  \sqrt{\Gamma
   \left(\sqrt{16 \hat g_1^2+1}\right)}
   \sqrt{\Gamma
   \left(\sqrt{16
    \hat g_2^2+1}\right)} }{ \Gamma
   \left(\frac{1}{2} \left(\sqrt{16
   \hat g_1^2+1}+\sqrt{16
   \hat g_2^2+1}\right)\right)}, \label{eq:Cphi0}
\eeq
where we used (\ref{eq:DeltagsPhi}) in the last step. 
This is in perfect agreement with the result of \cite{Kim:2017sju}. It is simple to obtain further orders in a small angle expansion, the next-to leading order in all angles is reported in Appendix~\ref{app:smallPhi}.

\section{Numerical evaluation}
\label{sec:num}

\begin{figure}
    \centering
    \includegraphics[scale=0.6]{HHL1p0.pdf}
    \includegraphics[scale=0.6]{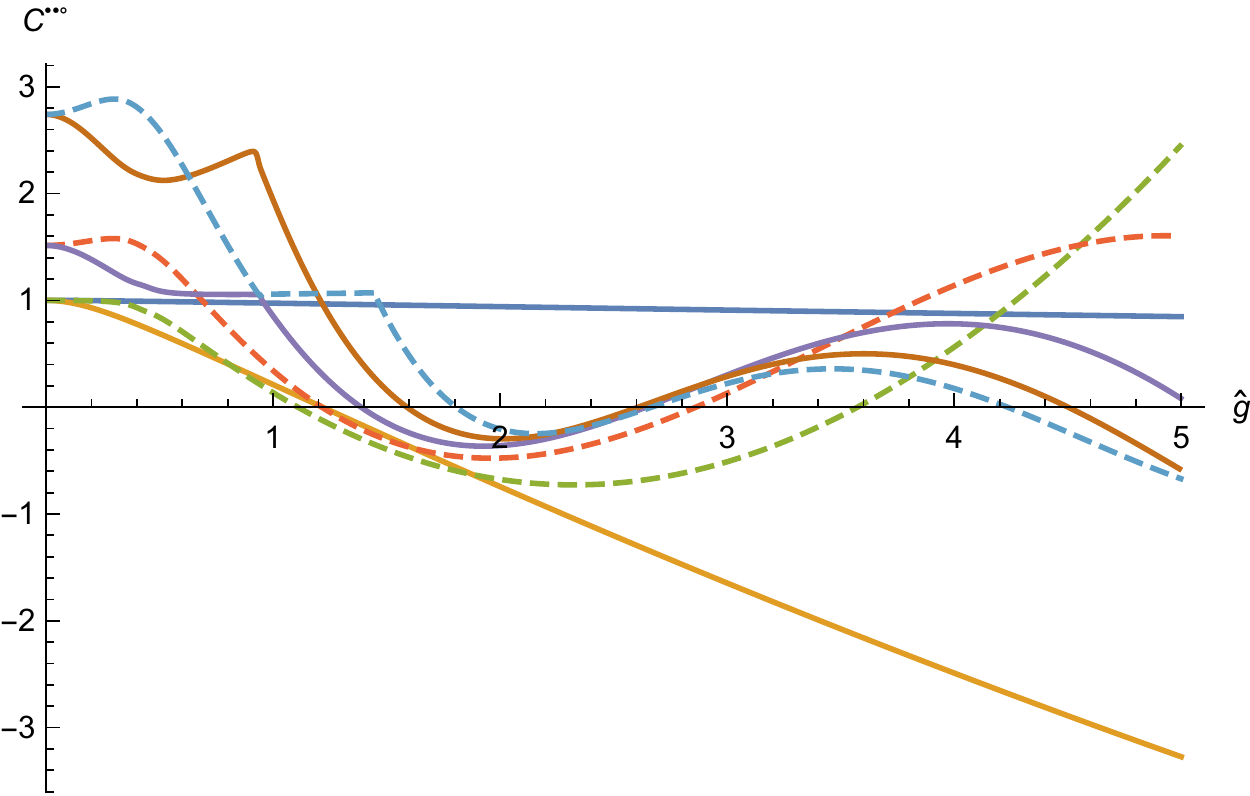}
    \caption{Diagonal HHL correlator for several first excited states ($n=0,1,\dots,7$) with all angles equal to $\phi=1 $ (left) or $\phi=1/3 $ (right). Colors are the same as on Figure \ref{fig:spectrum2}. }
    \label{fig:HHLnum}
\end{figure}
The expression for the 3-cusp correlator we found has the form of an integral $\vint q_{\Delta_1}q_{\Delta_2}e^{-u\phi_3}\frac{du}{2\pi i u}$ which is guaranteed to converge for large enough coupling as the q-functions behave as $e^{\phi u}u^\Delta$ where $\Delta$ decreases linearly with $\hat g$ and reaches  arbitrarily large negative values. However, we would like to be able to use these expressions at small coupling too, where the convergence of the integral is only guaranteed when both states are ground states, but for the excited states the integral is formally not defined.

To define the integrals we introduce the following $\zeta$-type of regularization. We multiply the integrand by some negative power $u^\alpha$, compute the integral for large negative enough $\alpha$ and then analytically continue it to zero value. The key integral is \eq{eq:gint}
where the r.h.s. gives the ananlytic continuation to all values of $\alpha$.

We see that for large negative $\alpha$ the expression decays factorially.
This fact is crucial for our numerical evaluation of the correlation function. 
Once the value of the energy is known numerically it is very easy to get an asymptotic expansion of the q-functions at large $u$ to essentially any order. However, since the poles of the q-functions accumulate at infinity, this expansion is doomed to have zero convergence radios. Nevertheless if we expand the integrand at large $u$ and then integrate each term of the expansion using
\eq{eq:gint} we enhance the convergence of this series by a factorially decaying factor making it a very efficient tool for the numerical evaluation.

We applied this method to compute the correlation function for several excited states 
(see Fig. \ref{fig:HHLnum}). The method allows one to compute the correlator even faster than the spectrum.
We checked that it  works very well for $\phi\sim 1$ giving $10$ digits precision easily, but seems to diverge for $\phi=1.5$.
To cross check our precision we also used the $d\Delta/dg$ correlator \eq{eq:dddg}, which is given by the same type of integrals.

\section{Correlation functions at weak coupling}
\label{sec:weak}
In this section we present some explicit results for the structure constants at weak coupling. 

Our all-loop expression for the structure constants \eq{correlator} is rather straightforward to evaluate perturbatively. First one should find the Q-function $q$ at weak coupling, which can be done by iteratively solving the Baxter equation  as discussed in section \ref{sec:qscexc}. The result at each order is given as a linear combination of twisted $\eta$-functions (see \eq{defeta}) multiplied by exponentials $e^{\phi u}$ and rational functions of $u$, as in e.g. \eq{qgs}. Then the integrals appearing in the numerator and denominator of  \eq{correlator} can be easily done by closing the integration contour to encircle the poles of $q(u)$ in the lower half-plane, giving an infinite sum of residues\footnote{For excited states the integral in the lhs of \eq{intres} may be divergent. We still replace it by the (convergent) sum of residues, which corresponds to the $\zeta$-type regularization discussed in section \ref{sec:num}. }:
\beq
\label{intres}
\frac{1}{2 \pi i} \, \vint f(u) \, du = \sum_{n=0}^{\infty} \left. \text{Res} \, f(u) \right|_{u=-i n} .
\eeq
The residues come from poles of the $\eta$-functions, e.g.
\beq
\eta^z_n= \frac{z^m}{(u + i m )^n} + O(1) , \;\;\; u \to -i m , \;\;\; m=0,1,\dots ,
\eeq
To get the residue one may need more coefficients of this Laurent expansion, which are given by zeta values or polylogarithms. Finally one should take the infinite sum in \eq{intres} which again may give  polylogs. 

In this way we have computed the first 1-2 orders of the weak coupling expansions, as a demonstration (going to higher orders is in principle straightforward, limited by computer time and the need to simplify the resulting multiple polylogarithms).
The integrals giving the norm of $q$-functions are especially simple. 
Below, we assume that $q(u)$ is normalized\footnote{Notice that, while the brackets in the numerator and denominator of (\ref{correlator}) depend on this normalization, the structure constants are clearly invariant. }  such that the leading coefficient in the large $u$ expansion is $1$, so $q(u) \simeq u^{\Delta} \, e^{\phi u}$.  
For the ground state ($L=0$) we find 
\beq
\br{ q^2 }_{L=0} = 1+8 \, \hat g^2 \, \frac{ \phi}{\sin\phi } \, \gamma_{\text{E}} + \cO(\hat g^4 ) ,
\eeq
where $\gamma_{\text{E}}$ is the Euler-Mascheroni constant. For the excited states $(L, \pm)$\footnote{This notation for the excited states is explained in Section \ref{sec:qscexc}, see also Table \ref{tab:states}.} corresponding to insertion of $L$ scalars, we have 
\beqa
\br{ q^2 }_{1,\pm}& =& \pm 8 \hat g^2  +\dots \\ 
\br{ q^2 }_{2,\pm} &=& \pm 16 \,   \cos\phi \;\hat g^2 +\dots \label{eq:2pm}
\eeqa
The $L=3$ result is given in \eq{norm3w}. Notice here that for the states $2^+$ and $2^-$ the signs of $\br{ q^2 }$ are different at weak and strong coupling. Indeed, at strong coupling the relation with the wavefunctions \eq{eq:normQ} implies that $\br{q^2}$ is positive/negative for even/odd states, respectively. Since the even state is $2^-$ (see Table~\ref{tab:states}),  in (\ref{eq:2pm}) we see explicitly that these signs can change at weak coupling. 

The structure constants are more involved. For the HHL correlator without scalar insertions we have to 1-loop order
\beq
   ( C^{\bullet \bullet \circ})_{L=0}=1+\hat g_1^2 F_{123}+\hat g_2^2 F_{213}+\dots
\eeq
where
\beqa\la{F123}
    F_{123}=&&
    \frac{1}{\sin\phi_1}
    \[2 i \left(\text{Li}_2(e^{-2 i \phi_1})-\text{Li}_2(e^{-i\phi_1-i\phi_2+i\phi_3})+\text{Li}_2(e^{i \phi_1-i\phi_2+i\phi_3})\right)-\frac{i \pi
   ^2}{3}
   \right. \\ \nn 
   && \left.
   +2 \left(\phi _1-\phi _2+\phi _3\right) \log
   \left(\frac{1-e^{-i \phi _1-i\phi _2+i\phi _3}}{1-e^{i \phi _1-i\phi
   _2+i\phi _3}}\right)-4 \phi _1 \log \left(\frac{\sin \frac{1}{2} \left(\phi
   _1+\phi _2-\phi _3\right)}{\sin\phi_1} \right)\] \ .
\eeqa
For the correlators with excited states both the numerator and the denominator in the expression \eq{correlator} for $C^{\bullet \bullet o}$ vanish at weak coupling. Due to this even the leading order in the expansion is nontrivial and requires using $q(u)$ computed to $\hat g^2$ accuracy. For the correlators with two $L=1$ states we find  
\beqa
( C^{ \bullet \bullet \circ} )_{L=1}  &=&\frac{1}{2} \, \left(\frac{\hat g_1 }{ \hat g_2 } \pm \frac{\hat g_2 }{ \hat g_1 } \right) +\dots \ ,
\eeqa
while for $L=2$ we get a nontrivial dependence on the angles,
\beqa
( C^{ \bullet \bullet \circ} )_{L=2} =&& \frac{1}{2} \, \left( \sqrt{ \frac{ \hat g_1^2 \, \cos\phi_1 }{\hat g_2^2 \, \cos\phi_2} } \pm \sqrt{ \frac{ \hat g_2^2 \, \cos\phi_2 }{\hat g_1^2 \, \cos\phi_1} } \right) 
\\ \nn && \times \left(- \frac{\cos\phi_3 }{\sin\phi_1 \, \sin\phi_2 }+\cot\phi_1  \, \cot
   \phi_2 +2 \right) + \dots \ \ .
\eeqa
Here we have the plus sign for correlators corresponding to  $(L^+,L^+)$ or $(L^-,L^-)$ states, and the minus sign for the $(L^+,L^-)$ correlator.

Curiously, the HHL results do not have a smooth limit when one of the couplings goes to zero corresponding to the HLL case (this is related to a singularity in the 2-pt function normalization). This means we have to compute the HLL correlators separately. For H${}_n$LL with the excited state being $\Delta_{1,+}$ we get
\beq\la{C1p}
(C^{\bullet \circ \circ} )_{1^+,0,0}= -\sqrt{2   \, \hat g^2} \,  \frac{\cos(\frac{1}{2}(\phi_2 - \phi_3 ))}{\cos{\frac{1}{2}\phi_1}} \ \ ,
\eeq
while for $\Delta_{1,-}$ we have
\beq\la{C1m}
(C^{\bullet \circ \circ} )_{1^-,0,0}= -\sqrt{2   \, \hat g^2} \, \frac{\sin(\frac{1}{2}(\phi_2 - \phi_3 ))}{\sin{\frac{1}{2}\phi_1}} \ \  .
\eeq
For the $L=2$ states we find
\beqa\la{C2p}
(C^{\bullet \circ \circ} )_{2^+,0,0}&=& -\hat g\,i \, \frac{\sin(\phi_2 - \phi_3 )}{\sin{\phi_1} } \, \sqrt{\cos\phi_1} \ \ , \\
\la{C2m}
(C^{\bullet \circ \circ} )_{2^-,0,0}&=& \hat g\,i \, \frac{ \sqrt{\cos\phi_1} }{\sin^2\phi_1} \, ( \cos\phi_1 \, \cos(\phi_2 - \phi_3 ) - 1)\ \ .
\eeqa
These two structure constants are purely imaginary due to the sign of $\br{q^2 }$ at weak coupling. We also present the results for the $L=3$ states in Appendix \ref{app:exc}.

\section{The 4-point function and twisted OPE}
\label{sec:ope}

In this section we examine more closely the expression for the 4-point function which we obtained in \eq{G1234delta}.
We interpret it as an OPE expansion and cross-test it at weak coupling against our perturbative data for the correlation functions. 
 We also present some conjectures on the generalization of this OPE expansion and its applications to the computation of more general correlators. 

\subsection{The 4-cusp correlation function}\label{sec:4cusp}
\begin{figure}
    \centering
    \includegraphics[scale=1.5]{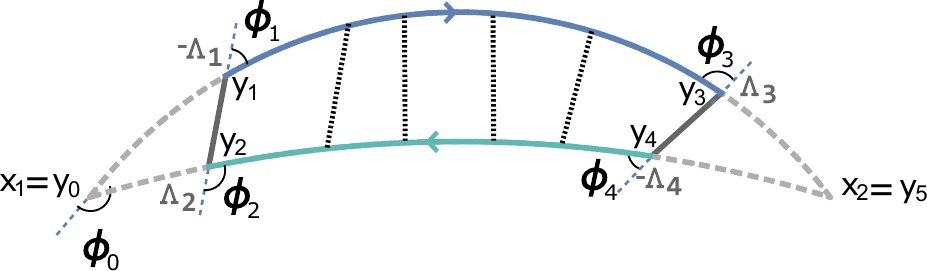}
    \caption{The 4-cusp correlator. Its OPE-like expansion \eq{G1234delta4} provides predictions for the HLL structure constants.}
    \label{fig:4cusp}
\end{figure}
Our starting point is an OPE-like formula \eq{G1234delta} for the 4-cusp correlator. It is based on the 2-pt function of cusps with angle $\phi_0$, but the four cutoffs $\Lambda_1,\dots,\Lambda_4$ give it the structure of a 4-point function with four cusp angles $\phi_a$ determined by $\Lambda$'s as shown on
Fig.~\ref{fig:4cusp}. To make the analogy more clear we notice that we can get rid of the wavefunctions in \eq{G1234delta} entirely 
and rewrite it in terms of the structure constants as follows
\beqa\la{G1234delta4}
G(\Lambda_1,\Lambda_2,\Lambda_3,\Lambda_4)
=\sum_{n=0}^\infty &&{C_{012}^{\bullet_{n}\circ\circ}}
{C^{\bullet_{n}\circ\circ}_{043}}{
}\(\frac{e^{-2\Lambda}}{
L_{043}
L_{012}
}\)^{\Delta_n} \;,
\eeqa
where $\Lambda\equiv \frac{\Lambda_1+\Lambda_2+\Lambda_3+\Lambda_4}{4}$, while the angles $\phi_1,\dots,\phi_4$ at the cusps $y_a$ (see Fig.~\ref{fig:4cusp}) can be found from $w_{\phi_0}(\Lambda_a-\Lambda_b)=\phi_b-\phi_a$ with $w$  defined by \eq{defw}. More explicitly,
\beq\la{tophi}
e^{-i \phi_{12}}=\frac{e^{\Lambda _{12}}+e^{i \phi_0 }}{1+e^{\Lambda _{12}+i \phi_0 }}\;\;,\;\;
e^{-i \phi_{43}}=\frac{e^{\Lambda _{43}}+e^{i \phi_0 }}{1+e^{\Lambda _{43}+i \phi_0 }}\;,
\eeq
where we denoted $\phi_{ab}=\phi_a-\phi_b$ and
\beq
\label{L12def}
\Lambda_{12}=\Lambda_1-\Lambda_2, \ \  \Lambda_{43}=\Lambda_4-\Lambda_3 \ .
\eeq
The factor $L_{abc}$ as before is defined by
\beq\label{eq:defL}
L_{abc} = \frac{ \sqrt{ \sin\frac{1}{2}( \phi_a + \phi_{b}-\phi_c) \, \sin\frac{1}{2}( \phi_a - \phi_{b}+\phi_c)}
}{\sin\phi_a} .
\eeq
We can view  equation \eq{G1234delta4} as defining the 4-cusp correlator in terms of the structure constants, opening an easy way for computing this quantity in various regimes including numerically at finite coupling. This equation suggests a natural interpretation in terms of an OPE expansion for pairs of cusps. To understand this point, let us first investigate the space-time dependence of the  4pt function (\ref{G1234delta4}), 
which comes through the factors
\beq\label{eq:4ptfactors}
\left( \frac{e^{-2 \Lambda} }{L_{012} \, L_{034} } \right)^{\Delta_n} . 
\eeq
To decode the  dependence of  (\ref{eq:4ptfactors}) on the cusp positions, it is convenient to introduce six complex parameters:  four space-time positions $y_i$, $i=1, \dots, 4$, defined as
\beq
y_1 = \zeta_+( -\Lambda_1 ) , \;\;\; y_2 = \zeta_-(\Lambda_2 ) , \;\;\; y_3 = \zeta_+(\Lambda_3) , \;\;\; y_4 =\zeta_-(-\Lambda_4) ,
\eeq
(where $\zeta_\pm$ is the parameterization defined by \eq{eq:zazb}) together with the intersection points of the two arcs $x_1$, $x_2$ (see Fig.  \ref{fig:4cusp}), which we denote as $y_0 \equiv x_1$, $y_5 \equiv x_2$. 
 These six points are not all independent as we can express $y_5$ in terms of the other five complex  coordinates through the solution of the equations\footnote{These equations express the fact that four points lying on the same line or circle have a real cross ratio.}
 \beq
\frac{ y_{53} \, y_{10} }{y_{31} \, y_{50}  } = \frac{ y_{53}^* \, y_{10}^* }{y_{31}^* \, y_{50}^*  }  , \;\;\;\;\; \frac{ y_{54} \, y_{20} }{y_{42} \, y_{50}  } = \frac{ y_{54}^* \, y_{20}^* }{y_{42}^* \, y_{50}^*  } ,
 \eeq
where $y_{ab} = y_a-y_b$. From these two relations we can obtain $y_5$ as a rational function of $y_i$, $i=0,\dots,4$ and their complex conjugates.\footnote{We have also found  nice explicit parameterizations of the spacetime dependence in terms of crossratios of these points and we present them in Appendix \ref{sec:4ptparam}. }

\begin{figure}
	    \centering
	\includegraphics[scale=0.75]{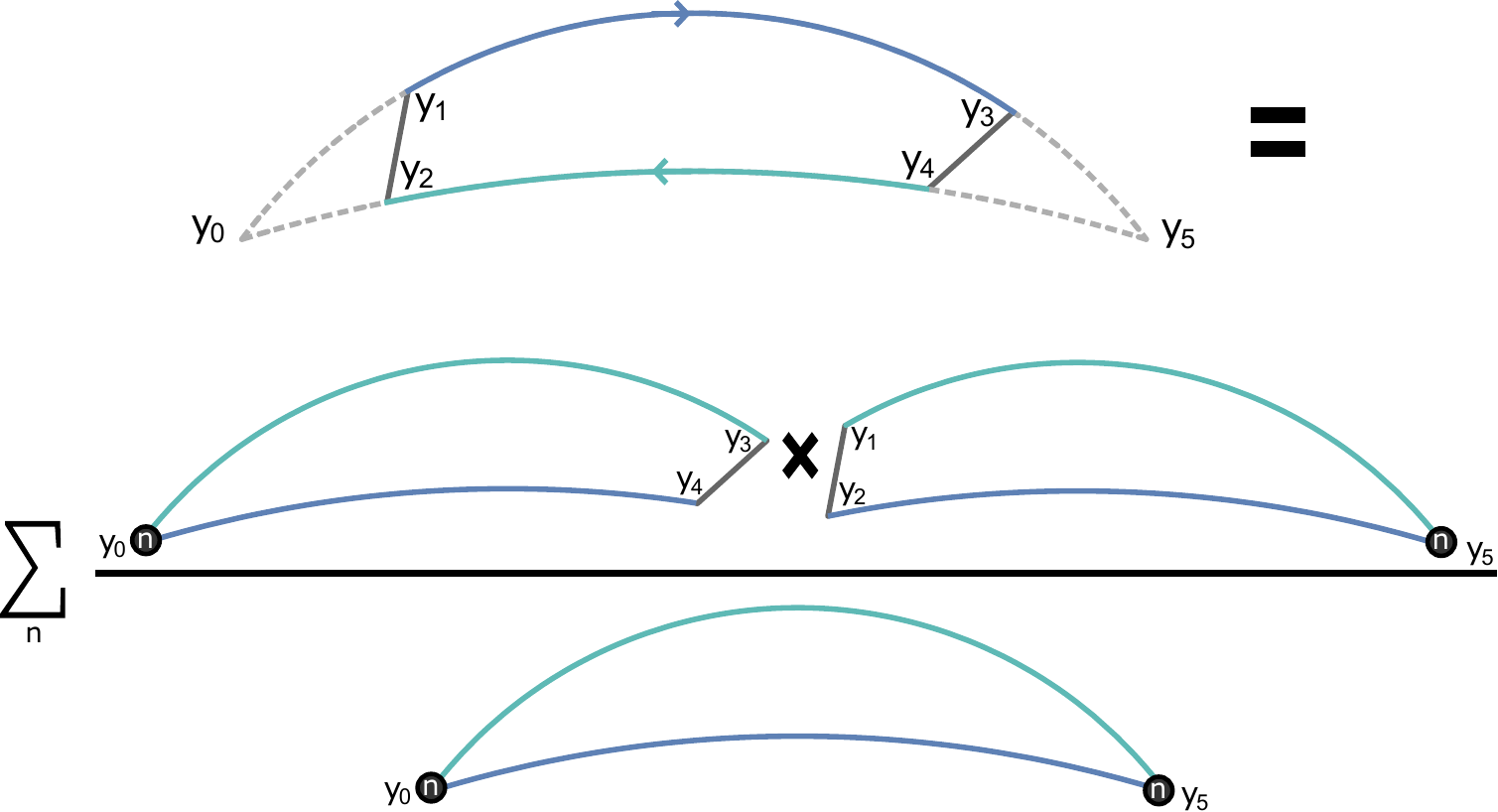}
	\caption{The OPE decomposition of the four-point function, illustrating equation \eq{eq:finalG}.}
	\label{fig:archesOPE}
\end{figure}

Eliminating the parameters $\Lambda_i$ in favour of the $y_i$  coordinates, we find that the term  (\ref{eq:4ptfactors}) appearing in the 4pt function can be written as 
\beq
\left( \frac{e^{-2 \Lambda} }{L_{012} \, L_{034} } \right)^{\Delta_n} = | y_{05}^2 |^{\Delta_n} \, \frac{ | y_{12} |^{\Delta_n} }{| y_{15} \, y_{25} \,  |^{\Delta_n}}\, \frac{ | y_{34} |^{\Delta_n} }{| y_{30} \, y_{40} \,  |^{\Delta_n}} . 
\eeq
Notice that this is the space-time dependence of the product of two 3pt functions, divided by a 2pt function,  and (\ref{G1234delta4}) can be rewritten  suggestively as
\beq\label{eq:finalG}
G = \sum_n  \frac{ C_{512}^{\bullet_n \circ \circ}  }{| y_{15} \, y_{25} \,  |^{\Delta_n} \, | y_{12} |^{-\Delta_n}}\, \frac{ C_{043}^{\bullet_n \circ \circ}  }{| y_{30} \, y_{40} \,  |^{\Delta_n} \, | y_{34} |^{-\Delta_n}} \, \left( \frac{1}{| y_{05} |^{2\Delta_n} } \right)^{-1} 
.
\eeq
This relation is illustrated in Fig. \ref{fig:archesOPE} and it strongly reminds the usual OPE decomposition of a 4pt function in terms of 3pt correlators. In the next subsection we provide an interpretation of this relation on the operator level.

\subsection{The cusp OPE}
\begin{figure}
	    \centering
	\includegraphics[scale=0.8]{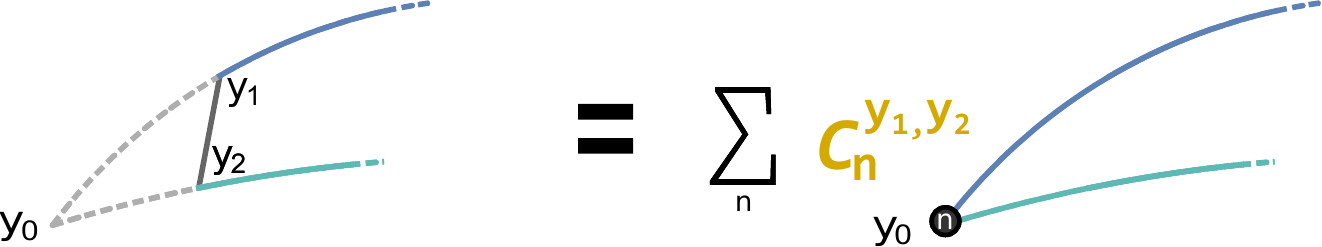}
	\caption{Expansion of the Wilson lines  starting at points $y_1$, $y_2$ in terms of Wilson arcs emanating from $y_0$, as written in equation \eq{eq:OPEG}.}
	\label{fig:archesOPE2}
\end{figure}

\begin{figure}
	    \centering
	\includegraphics[scale=0.63]{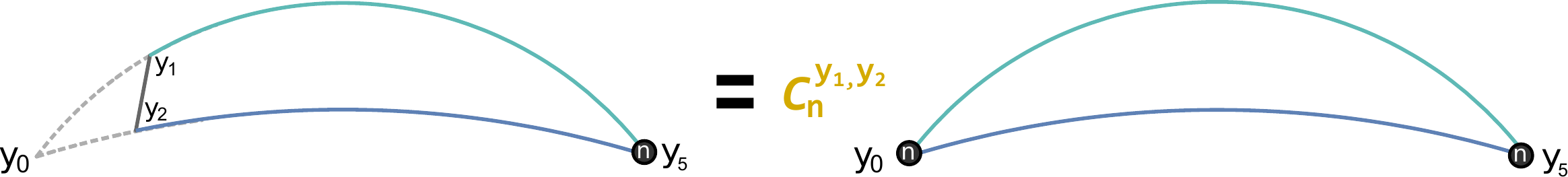}
	\caption{Graphical representation of equation \eq{eqCnope} determining the value of $\mathcal{C}_n^{y_1,y_2}$.}
	\label{fig:OPEexp2}
\end{figure}

Let us now rederive the decomposition \eq{eq:finalG} of the 4pt function  from first principles using the logic inspired by the usual OPE. 
The idea, illustrated in Fig.  \ref{fig:archesOPE2}, is to express the cusps at $y_1$, $y_2$ as a combination of cusp operators inserted at $y_0$: 
\beq\label{eq:OPEG}
W_{y_1}^{y_3} \, W_{y_2}^{y_4}
=\sum_n \, \mathcal{C}^{y_1, y_2}_n  \;   \left[\mathcal{O}_n \, \left(  W_{y_4}^{y_0} \, W_{y_0}^{y_3} \right) \right] ,
\eeq
where $\mathcal{C}^{y_1, y_2}_n $ are some coefficients, $W_x^y$ are the Wilson line operators defined in (\ref{eq:Wxyop}), and $\mathcal{O}_n$ represent  projector operators on the $n$-th excitation of the cusp at $y_0$. To make sense of the  rhs of (\ref{eq:OPEG}), we need to specify a regularization scheme; we assume that the $\epsilon$ regularization defined in the rest of the paper is used, and the projectors $\mathcal{O}_n$ are  the ones defined explicitly in  section \ref{sec:insert}. 
 Notice that the expansion corresponds to a change in the limit of integration of the Wilson lines. Derivatives of the Wilson line with respect to its endpoints produce the scalar insertions described in Sec. \ref{sec:insert}. For this reason, at least in the ladder limit considered here, we expect that only these excitations are involved in the OPE. 
To determine the coefficients $\mathcal{C}_n^{y_1,y_2}$, we proceed in the standard logic of the OPE and place equation (\ref{eq:OPEG}) inside an expectation value.  
Considering the limit where $y_3 $,$y_4$ converge towards $y_5$ (with the usual point-splitting regulator $\epsilon$), and projecting on the $n$-th state, we have
\beq
\bar{\mathcal{O}}_n \langle W_{y_1}^{y_5} \, W_{y_2}^{y_5} \rangle = \mathcal{N}_{\Delta_n, \epsilon}  \,  \frac{C_{512}^{\bullet_n \circ \circ} }{| y_{15} \, y_{25} |^{\Delta_n} \, | y_{12}|^{-\Delta_n} } , 
\eeq
where we noticed that in this limit the configuration reduces to an HLL 3pt function, which we related to the structure constant as in Sec. \ref{sec:HLLexc}. Here, the constant $\mathcal{N}_{\Delta_n, \epsilon}$ is the square root of the normalization of the 2pt function, explicitly defined in (\ref{Nn2}). 
 On the other hand from the rhs of  (\ref{eq:OPEG}) we obtain (see Fig. \ref{fig:OPEexp2}):
\beq
\label{eqCnope}
\bar{\mathcal{O}}_n \, \left( \sum_m  \, \mathcal{O}_m  \; \langle \, W_{y_0 }^{y_5 } \; W_{y_0  }^{y_5 } \, \rangle \right) = \mathcal{C}_{n}^{y_1 , y_2} \, \frac{\mathcal{N}_{\Delta_n, \epsilon}^2 }{|y_{05} |^{2 \Delta_n} } ,
\eeq
therefore we find the coefficients: 
\beq
\mathcal{C}_n^{y_1,y_2} = C_{512}^{\bullet_n \circ \circ } \, \left(\frac{ |y_{12} \, y_{05}^2 | }{|y_{15} \, y_{25} |} \right)^{\Delta_n}  \, \mathcal{N}_{\Delta_n, \epsilon}^{-1} .
\eeq
 Taking the expectation value of (\ref{eq:OPEG}) now fixes the 4pt function precisely to the form (\ref{eq:finalG}). 
 
 In the next subsection we will discuss how to apply similar logic to higher-point correlators.
 
 \subsection{OPE expansion of more general correlators}
 




\begin{figure}
	    \centering
	\includegraphics[scale=0.75]{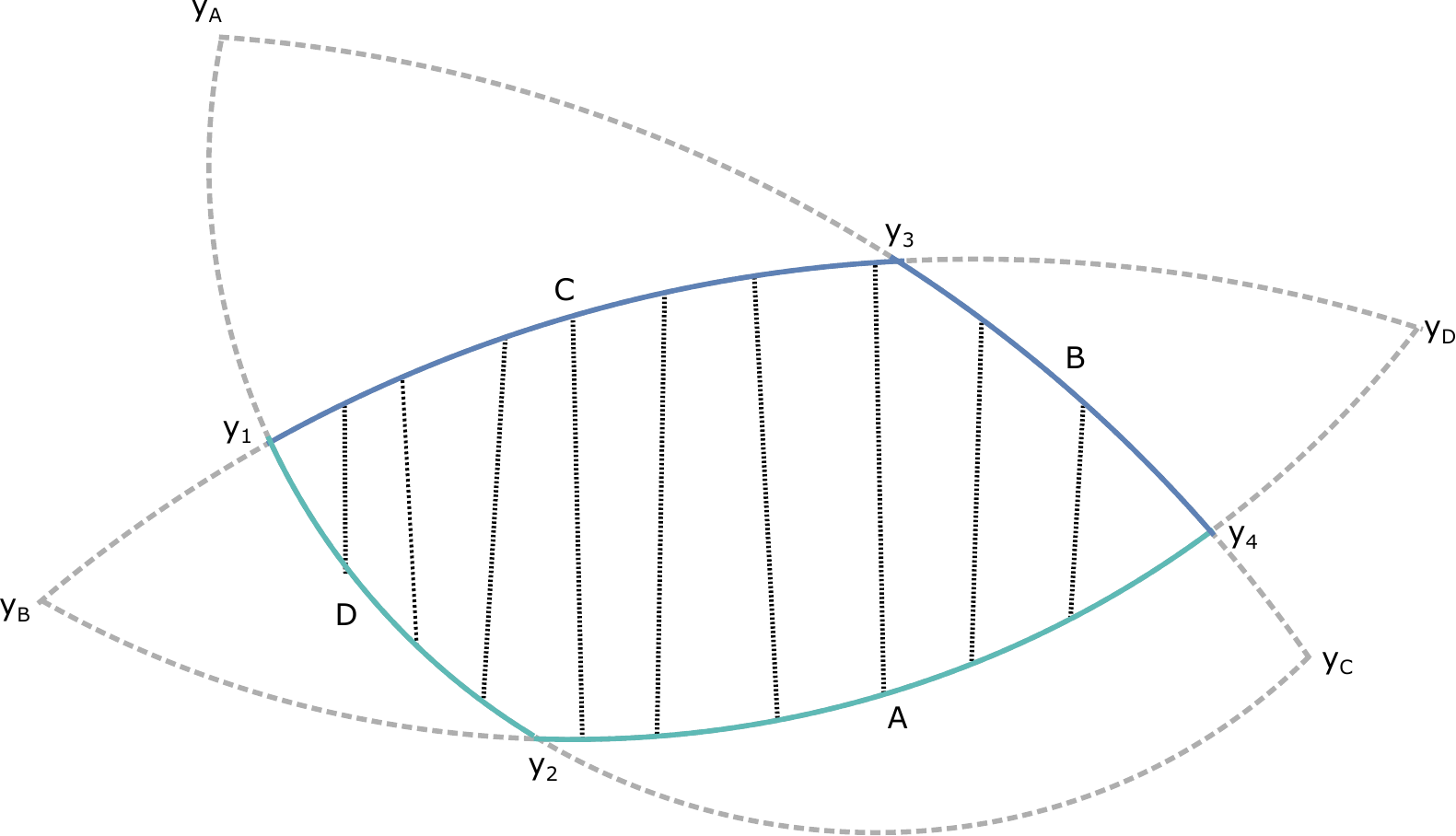}
	\caption{The 4pt function $G_{1243}^{\bullet \circ \bullet \circ}$ of two protected and two unprotected cusps. We assume that only two scalar polarizations are involved:  $\vec{n}$ on the arcs $B$,$C$ and $\vec{m}$ on the arcs $A$,$D$, so that the configuration depends on a single effective coupling. }
	\label{fig:OPE4pt}
\end{figure}

The OPE approach we presented above can also be applied to more general correlation functions. As one of the possible generalizations\footnote{One could also consider correlators with more than four protected cusps.  In particular, the 4pt function considered in this section can naturally be viewed as a limit of the correlator of six protected cusps, which is obtained by introducing a finite $\epsilon$ cutoff around $y_1$ and $y_4$. This six point function can also be decomposed using the OPE. }, let us consider the four point function shown in Figure \ref{fig:OPE4pt}. For simplicity of notation, we assume that the same scalar polarization $\vec{n}$ is chosen for the Wilson lines denoted as $C$ and $B$, while on lines $A$ and $D$ we have a different polarization vector $\vec{m}$. This defines a configuration where the two cusps at $y_1$ and $y_4$ are not protected, while the remaining two are. Explicitly, we are considering the expectation value:
\beq
G_{1243}^{\bullet \circ \bullet \circ} =   \frac{ \langle W_{y_1 }^{y_2}(\vec{m} ) \; W_{y_2}^{y_4  }(\vec{m} ) \; W_{y_4 }^{y_3}(\vec{n} ) \; W_{y_3}^{y_1  }(\vec{n} ) \;   \rangle }{  \mathcal{N}_{1}\, \mathcal{N}_{4} } ,
\eeq
 where we divided by the usual 2pt function normalization factors $\mathcal{N}_1$, $\mathcal{N}_4$ for the unprotected cusps (defined explicitly in (\ref{Nn2})) in order to get a finite result\footnote{As usual we assume the point-splitting $\epsilon$-regularization close to the cusps. }. 
 
   Our conjecture for this quantity  is based on the assumption that we can use the same type of OPE expansion as in the previous section. 
   This allows us to replace each pair of consecutive cusps  with a sum over excitations of a single cusp, whose position is defined by the geometry. 
   For instance, the two cusps at $y_3$ and $y_4$, which are defined by the consecutive sides $A$ $B$ $C$ of the Wilson loop, are traded for a sum over excitations of a single cusp at the point $D$, defined by the extension of the lines $A$ and $C$.  

\begin{figure}
	    \centering
	\includegraphics[scale=0.3]{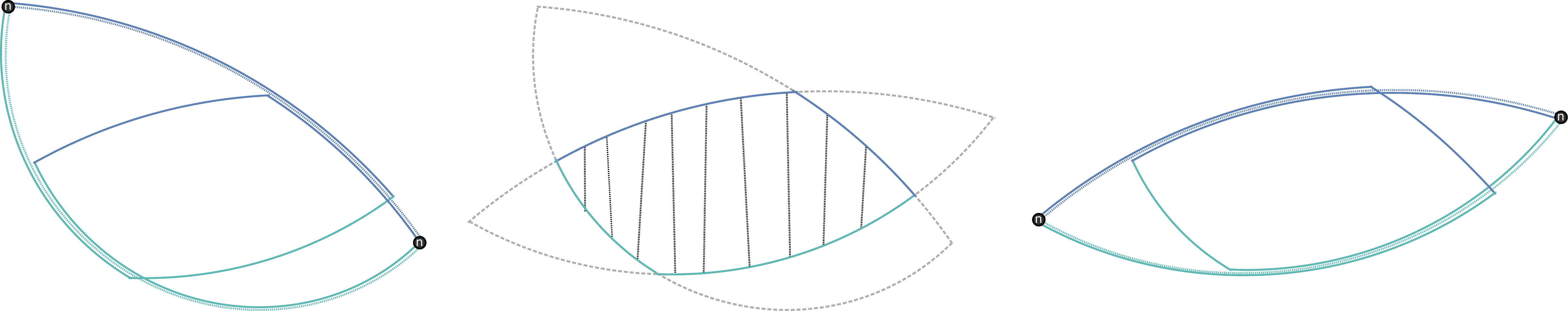}
	\caption{The two alternative OPE decompositions of the 4pt function $G_{1243}^{\bullet \circ \bullet \circ }$. }
	\label{fig:OPE4pttwo}
\end{figure}

As expected, the OPE expansion gives rise to nontrivial crossing equations. Let us see this explicitly here.
Taking into account the space-time dependence as in the previous section, from the contraction of $y_3$ and $y_4$ we obtain (see Fig. \ref{fig:OPE4pttwo} on the right):
  \beq
 G_{1243}^{\bullet \circ \bullet \circ} = \sum_n \, \frac{ C_{D12}^{\bullet_n \bullet \circ} }{ |y_{1D} |^{\Delta_n + \Delta_0} \, |y_{2D} |^{\Delta_n - \Delta_0} \, |y_{12} |^{\Delta_0 - \Delta_n} } \, \frac{ C_{B43}^{\bullet_n \bullet \circ} }{ |y_{B4} |^{\Delta_n + \Delta_0} \, |y_{B3} |^{\Delta_n - \Delta_0} \, |y_{34} |^{\Delta_0 - \Delta_n} } \, |y_{BD} |^{2\Delta_n } ,\label{eq:OPE4pt1}
  \eeq
  which now involves HHL structure constants\footnote{Here we assume that the excited states studied in the rest of this paper constitute a full enough basis which makes possible this decomposition. This point requires further investigation. If that is not the case one will have to add a sum over some additional states as well.}. Performing the OPE decomposition in the crossed channel, which corresponds to contracting $y_1$ and $y_3$ (see Fig. \ref{fig:OPE4pttwo} on the left), yields a different expansion:
   \beq
 G_{1243}^{\bullet \circ \bullet \circ} = \sum_{n} \, \frac{ C_{A42}^{\bullet_n \bullet \circ} }{ |y_{4A} |^{\Delta_n + \Delta_0} \, |y_{2A} |^{\Delta_n - \Delta_0} \, |y_{24} |^{\Delta_0 - \Delta_n} } \, \frac{ C_{C13}^{\bullet_n \bullet \circ} }{ |y_{C1} |^{\Delta_n + \Delta_0} \, |y_{C3} |^{\Delta_n - \Delta_0} \, |y_{13} |^{\Delta_0 - \Delta_n} } \, |y_{AC} |^{2\Delta_n }.\label{eq:OPE4pt2}
  \eeq
  Notice that we left the dependence on all angles implicit; however, we point out that the sums in (\ref{eq:OPE4pt1}) and (\ref{eq:OPE4pt2}) are over different spectra, characterized by the same coupling but different cusp angles. 
 Proving the equivalence between (\ref{eq:OPE4pt1}) and (\ref{eq:OPE4pt2}) would be an important test of these expressions, and more generally of the OPE expansion on which they are based\footnote{A somewhat related OPE approach was discussed in \cite{Kim:2017sju} for the $\phi=0$ case. It would be interesting to clarify possible connections with the OPE that we discuss here, which seems to be not a completely trivial task. We thank S.~Komatsu for discussions of this point.}. We leave this nontrivial task for the future.  Crossing relations such as the one presented above could perhaps also be used to gain information on the HHH structure constants, which would appear in one of the two channels in the OPE expansion of correlators of the form $G_{1234}^{\bullet \bullet \circ \circ}$. 

\subsection{Checks at weak coupling}
In this section, we present some tests of the 4pt OPE expansion \eq{eq:finalG} at weak coupling. We will show that perturbative expansion of the 4pt function reproduces our results for HLL structure constants. In Appendix \ref{HLLspace} we also verify at 1 loop that when two of the four points collide, the 4pt function reduces precisely to a 3pt HLL correlator, including the expected spacetime dependence. This provides an important test of our results for the structure constants and also of the OPE expression for the 4pt function.


At one loop it is very easy to compute the 4pt function, and we find
\beq
G(\Lambda_1,\Lambda_2,\Lambda_3,\Lambda_4)=1+
\int_{-\Lambda_4}^{\Lambda_2} ds\int_{-\Lambda_1}^{\Lambda_3} dt \frac{2\hat{g}^2}{\cosh (s+t)+\cos (\phi_0 )} \ , 
\eeq
resulting in
\beqa\nn
G&=&1+\frac{2 i \hat g^2}{\sin\phi_0}\Big[
\text{Li}_2\left(-e^{-i \phi_0 +\Lambda _{12}}\right)-\text{Li}_2\left(-e^{i
   \phi_0 +\Lambda _{12}}\right)
   -\text{Li}_2\left(-e^{-i \phi_0 -\Lambda
   _{23}}\right)+\text{Li}_2\left(-e^{i \phi_0 -\Lambda _{23}}\right)
 \\&-&\text{Li}_2\left(-e^{-i \phi_0 +\Lambda _{14}}\right)+\text{Li}_2\left(-e^{i \phi_0 +\Lambda _{14}}\right)+\text{Li}_2\left(-e^{-i \phi_0 +\Lambda
   _{43}}\right)-\text{Li}_2\left(-e^{i \phi_0 +\Lambda _{43}}\right)\Big] \ , \label{G41loop}
\eeqa
where we denoted (note the difference with \eq{L12def})
\beq
\Lambda_{23}=\Lambda_2+\Lambda_3, \ \  \Lambda_{14}=\Lambda_1+\Lambda_4\ .
\eeq
Expanding this expression at large $\Lambda$ we get:
\beqa
G&=&g_0+\Lambda h_0+e^{-2\Lambda}g_1+e^{-4\Lambda}g_2+e^{-6\Lambda}g_3+{\cal O}(e^{-8\Lambda})
\ ,
\eeqa
where the first coefficient is rather involved, 
\beqa
g_0&=&\nn 2
   \frac{\hat g^2}{\sin\phi_0} \left(i
   \text{Li}_2\left(-e^{\Lambda _{12}-i \phi_0
   }\right)-i \text{Li}_2\left(-e^{i \phi_0
   +\Lambda _{12}}\right)+i
   \text{Li}_2\left(-e^{\Lambda _{43}-i \phi_0
   }\right)-i \text{Li}_2\left(-e^{i \phi_0
   +\Lambda _{43}}\right)\right)\\
 &+&2 \, {\hat g^2}{}\frac{\Lambda _{12} \phi_0
   +\Lambda _{43} \phi_0}{\sin\phi_0}
+1 \ , \label{resg0}
\eeqa
while the rest are simpler,
\beqa
h_0&=&8 \frac{\hat g^2 \phi_0}{\sin\phi_0} \ , \ \ \ \ \ \ \ \ \ \ \ \ 
\ \ \ \ \ \ \ \ \ 
\ \ \ \ \ \ \ \ 
\ \ \ \ \ 
g_1=
8 \hat g^2 \cosh \left(\frac{\Lambda
   _{12}+\Lambda _{43}}{2} \right)\ ,\\
\nn g_2&=&- 
4 \hat g^2 \cosh \left(\Lambda _{12}+\Lambda
   _{43}\right) \cos (\phi_0 )\ , \ \ \ \ \ g_3=
   \frac{8\hat g^2}{9}  \cosh \left(\frac{3(\Lambda _{12}+\Lambda
   _{43})}{2}
   \right) (2 \cos (2 \phi_0 )+1) \ .\nn
\eeqa
Rewriting this in terms of the angles using \eq{tophi} we obtain
\beq
L_{012}L_{043}\;g_1=
2 \hat g^2 \left(\frac{\cos \frac{\phi _{12}}{2} \cos \frac{\phi
   _{43}}{2}}{\cos ^2\frac{\phi _0}{2}}+\frac{\sin \frac{\phi
   _{12}}{2} \sin \frac{\phi _{43}}{2}}{\sin ^2\frac{\phi
   _0}{2}}\right)=C^{\bullet_1 \circ \circ}_{012} C^{\bullet_1 \circ \circ}_{043} +C^{\bullet_2 \circ \circ}_{012} C^{\bullet_2 \circ \circ}_{043} \ ,
\eeq
where we used that there are only two states $n=1,2$ which converge to $\Delta=1$ at weak coupling.
Furthermore, we can identify precisely $n=1$ and $n=2$, by using the fact that the $n=1$ state is associated with an odd state
and thus should give an odd function in $\phi_{12}$. This results in
\beq
C^{\bullet_1 \circ \circ}_{012} =\pm \sqrt{2\hat g^2}\frac{\sin\frac{\phi_{12}}{2}}{\sin\frac{\phi_0}{2}}\;\;,\;\;
C^{\bullet_2 \circ \circ}_{012} =\pm \sqrt{2\hat g^2}\frac{\cos\frac{\phi_{12}}{2}}{\cos\frac{\phi_0}{2}} \ ,
\eeq
in complete agreement with our perturbative results \eq{C1m} and \eq{C1p} ! In the same way we find for the $L=2$ states
\beqa
C^{\bullet_3 \circ \circ}_{012} &=&\pm i \hat g \frac{\sin \phi _{12}}{\sin\phi_0} \sqrt{\cos \phi _0} \ ,\\
C^{\bullet_4 \circ \circ}_{012} &=&\pm  i \hat g \frac{\sqrt{\cos \phi _0}}{\sin^2\phi _0} \left(\cos \phi _0 \cos
   \phi _{12}-1\right) \ , 
\eeqa
in agreement with \eq{C2m} and \eq{C2p}. We also verified the $L=3$ states and reproduced expressions \eq{HLL3pw}, \eq{HLL3mw} given in Appendix \ref{app:exc}.

We also notice that the term $h_0$ is indeed equal to $2\Delta_0^{(1)}$ i.e. the ground state energy at 1 loop. Finally, 
the expression $g_0$ can be compared with the HLL structure constant of three ground states, which reads at weak coupling
\beq
    (C^{\bullet o o})_{L=0}=1+\hat g^2 F_{123}+\dots
\eeq
where $F_{123}$ is given explicitly by the lengthy formula \eq{F123}. From the OPE (\ref{G1234delta4}) we expect that
\beq
g_0 = 1 + \hat g^2 \, \left(- \Delta_0^{(1)} \, \log{( L_{043} \, L_{012} )}  + F_{012} + F_{043} \right)\ ,
\eeq
and indeed our result \eq{resg0} for $g_0$ precisely matches this complicated expression! This is a nontrivial check of the OPE as well as the HLL structure constant at 1 loop. 



\section{Conclusions}

\label{sec:concl}

Our main result is the all-loop computation  of the expectation value of a Wilson line with three cusps with particular class of insertions at the cusps in the ladders limit. We demonstrated that in terms of the q-functions it takes 
a very simple form, reminiscent of the SoV scalar product. 
The key ingredient in the construction is the bracket $\br\cdot$, which allows to wrote the 
result in a very compact form \eq{correlator}.
We also found a similar representation for the diagonal correlator of two cusps and the Lagrangian \eq{Cinsert}.
This gives a clear indication that the Quantum Spectral Curve and the SoV approach can be able to provide an all-loop description of  3-point correlators.

In order to generalise our results one could consider correlators with more complicated insertions which should help to reveal more generally the structure of the SoV-type scalar product. We expect in this case that the bracket $\br\cdot$ will involve product of several Q-functions:
\beq\la{measureg}
\br{q_1q_2}=\int \mu(u_1,\dots,u_L)q_1(u_1)\dots q_1(u_L)
q_2(u_1)\dots q_2(u_L)du_1\dots du_L
\eeq
for some universal measure function $\mu$, which should not depend on the states, but could be a non-trivial function of coupling\footnote{In fact $L$ itself may be nontrivial to define at finite coupling as states with different values of the charges can be linked by analytic continuation.}. 
It would also be important to extend the results obtained in this paper to the more general  HHH configuration where all three effective couplings  are nonzero. The form of our result \eq{correlator}, where the BPS cusp always appears with a different sign for the rapidity, suggests that in the most general case one of the Q-functions may need to be treated on a different footing as the other two. Therefore, the generalization to the HHH case may be nontrivial and reveal new important elements. 

Going away from the ladders limit (see e.g. \cite{Bykov:2012sc, Henn:2012qz}) could also give some hints about the measure in the complete ${\cal N}=4$ SYM theory and eventually lead to the solution of the planar theory.
Potentially a simpler  problem is the fishnet theory \cite{Gurdogan:2015csr,Gromov:2017cja,Grabner:2017pgm}, where some $3-$ and $4-$point correlators were found explicitly and have a very similar form to the $\phi\to 0$ limit of our correlator. As they involve only conventional local operators this is another natural  setting for further developing our approach. It would be also interesting to consider the cusp in ABJM theory for which the ladders limit was recently elucidated in \cite{Bonini:2016fnc}.
It would be also useful to utilize the perturbative data from other approaches \cite{Escobedo:2010xs,Gromov:2012uv,Caetano:2014gwa,Basso:2015zoa,Basso:2017muf,Eden:2016xvg,Fleury:2016ykk} in order to guess the measure factor.

Let us mention that our result  incorporates all finite size corrections (in particular the 2-point functions are given exclusively by wrapping contributions). These  corrections are rather nontrivial to deal with in the hexagon~\cite{Basso:2015zoa} approach to computation of correlators  (see also \cite{Eden:2016xvg,Fleury:2016ykk,Bargheer:2017nne,Eden:2017ozn}). The diagonal correlators, which we studied numerically in this paper at any value of coupling, are proven to be particularly hard in the hexagon formulation which is known to be incomplete in this situation. Nevertheless, it would be interesting to draw parallels 
between the two approaches. 
The hexagon techniques could be especially helpful in generalisation of our results for the longer states, where the wrapping corrections are suppressed by powers of `t Hooft coupling. 

Another possible limit which would be interesting to consider is near-BPS.
This could be either the small spin limit of twist-2 local operators or the  $\phi\simeq\theta$
limit of the cusps. In both cases the analytic solutions of the QSC are known explicitly~\cite{Gromov:2013qga,Gromov:2014bva} (see also \cite{Gromov:2012eu}), which could be helpful in fixing the measure factor. In particular, at the leading order,
the Q-functions $q(u)$  describing the excited states of a cusp are orthogonal on $[-2g,2g]$ with the  measure $\mu(u)=
\sinh(2\pi u)$~\cite{Gromov:2012eu,Gromov:2013qga,Sizov:2013joa}. It is not clear how this measure is related to our result yet, but there are some promising signs which we discuss in the Appendix~\ref{app:bps}.
Let us point out that the naive guess that this is the measure we need is not consistent 
in an obvious way with the structure expected from SoV \eq{measureg}, where we expect multiple interactions for the insertions of such scalars. 
It would be really interesting to compare with localisation methods, which are applicable in the near-BPS limit. Some preliminary results were reported
recently~\cite{Giombi:2018qox} (see also \cite{Bonini:2015fng} for partial results for the spectrum). Let us also mention that often the measure can be bootstrapped from the orthogonality requirement, see \cite{Nsl2} for a higher-loop result in the $sl(2)$ sector. One could try this strategy too in order to find the measure in ${\cal N}=4$ SYM.

As another new result, we understood the meaning of the bound states of the Schr\"odinger problem resulting from Bethe-Salpeter resummation of ladder diagrams. 
They correspond to  insertion of scalar operators of the same type as those on the Wilson lines\footnote{and of their derivatives, when there is more than one scalar inserted}, see \cite{Klebanov:2006jj} for a string theory interpretation. 
From the point of view of the Bethe-Salpeter equation the excited states can be interpreted as resonances -- poles of the resolvent on the non-physical sheet, which can be reached by analytic continuation under the branch cut of the continuum. 
As such they are hard to study analytically or numerically. In the QSC approach there is no continuum spectrum and the bound states can be studied on completely equal footing with the vacuum state. Moreover they can be easily tracked away from the ladders limit and should still correspond 
to scalar insertions. In addition, we showed that our results for the 3-cusp correlators immediately generalize to the case with these scalar insertions.

Our result opens the way to efficiently study the cusp with scalar insertions at arbitrary values of $\theta$ using the powerful QSC methods, both analytically and numerically. 
We already found the first few orders in the weak and strong coupling expansions of the  energies of  excited states in the ladders limit. The result at $1$ loop for the first excited state matches the known 1-loop prediction \cite{Alday:2007he} (assuming it is not changed in the ladders limit).

It would be also important to further investigate the OPE picture we presented in section \ref{sec:ope}. In order to reveal more structure for higher point correlators it would be very useful to find a compact way to perform the spectral sums appearing in the OPE. Recent results of \cite{Gross:2017aos} for the SYK model suggest that this could be feasible at least in the ladder limit. One could also explore the applicability of modern conformal bootstrap techniques \cite{Rattazzi:2008pe,ElShowk:2012ht} for the OPE expansion we considered.
 Finally, the structure of our OPE expansion  is very reminiscent of the one for null polygonal Wilson loops \cite{Alday:2010ku}, and it could be useful to explore this analogy.

\section*{Acknowledgements}

We thank N.~Drukker,
D.~Grabner, V.~Kazakov, E.~Sobko, A.~Sever, A.~Tseytlin, A.~Tumanov and K.~Zarembo for related discussions.
We are especially grateful to A.~Pushnitsky and F.~Smirnov for inspirational comments, and to
S.~Komatsu for sharing the manuscript of \cite{Kim:2017sju} before publication. F.~L.-M. was supported by LabEX ENS-ICFP: ANR-10-LABX-0010/ANR-10-IDEX-0001-02 PSL*.
A.C. was supported by the STFC grant (ST/P000258/1) “Fundamental Physics from the Planck Scale to the LHC”. N.G. wishes to thank STFC for support from Consolidated grant number ST/J002798/1.

\appendix

\newpage
\section{Technical details on the QSC}

\la{app:qsc}

Here we provide details concerning the formulation of the QSC for the cusp anomalous dimension at generic values of the coupling $g$ and the angles $\phi,\theta$ \cite{Gromov:2015dfa}.

The $\bP$-functions of the QSC can be written in a compact form as
\beqa
\label{cuspas}
\bP_1(u)&=&+\epsilon\;u^{1/2}\; e^{+\theta u}\; {\bf f}(+u)\ ,\\
\nn
\bP_2(u)&=&-\epsilon\;u^{1/2}\; e^{-\theta u}\; {\bf f}(-u)\ ,\\
\nn
\bP_3(u)&=&+\epsilon\;u^{1/2}\; e^{+\theta u}\; {\bf g}(+u)\ ,\\
\nn
\bP_4(u)&=&+\epsilon\;u^{1/2}\; e^{-\theta u}\; {\bf g}(-u)\ .
\eeqa
where the functions ${\bf f}(u)$ and ${\bf g}(u)$ have powerlike asymptotics at large $u$ with ${\bf f}\simeq 1/u$ and ${\bf g}\simeq u$. The prefactor $\epsilon$ in this normalization reads
\beq
    \epsilon=\sqrt{\frac i2}\;
\frac{\cos \theta -\cos \phi}{\sin \theta}
\;.
\eeq

The functions ${\bf f}(u)$ and ${\bf g}(u)$ are regular outside of the cut $[-2g,2g]$, which can be resolved using the Zhukovsky variable $x(u)$,
\beq
\label{defx}
    x(u)=\frac{u+\sqrt{u-2g}\sqrt{u+2g}}{2g}\;\;,\;\;u=(x+1/x)g\;
\eeq
where we choose the solution with $|x|>1$. In terms of $x$ these functions simply become power series,
\beqa
\label{fgAB}
{\bf f}(u)=
\frac{1}{gx}+\sum_{n=1}^\infty \frac{g^{n-1}A_n}{x^{n+1}}
\;\;,\;\;
{\bf g}(u)=
\frac{u^2+B_0u}{gx}+\sum_{n=1}^\infty  \frac{g^{n-1}B_n}{x^{n+1}}\;.
\eeqa
The coefficients $A_n$ and $B_n$ encode nontrivial information about the AdS conserved charges including $\Delta$. In particular, for the first few of them we have
\beqa
\label{epsab0}
A_1g^2-B_0&=&-\frac{ 2 \cos \theta  \cos \phi+\cos (2 \theta) -3}{2 \sin\theta (\cos \theta -\cos \phi)}\;,\\
\nn
	\Delta^2&=&
	\frac{(\cos\theta-\cos\phi)^3}{\sin\theta\sin^2\phi}\[
A_3 g^6+\frac{A_1^2 g^4 (1-\cos\theta \cos\phi)}{
\sin\theta(\cos \theta -\cos\phi)}-A_2 g^4 \cot\theta
\right.\\
\label{deltaas0}&&   \left.
-g^2 \left(B_0+B_1+\cot \theta\right)
-A_1 g^2
   \left(A_2 g^4-2 g^2+\frac{1}{\sin^2\theta}\right)
\]\;.
\eeqa

The fourth order Baxter type equation \eq{bax5} on $\bQ_i$ is written in terms of 
several determinants involving the $\bP$-functions. They are given by:
\beqa
D_0&=&{\rm det}
\(
\bea{llll}
\bP^{1[+2]}&\bP^{2[+2]}&\bP^{3[+2]}&\bP^{4[+2]}\\
\bP^{1}&\bP^{2}&\bP^{3}&\bP^{4}\\
\bP^{1[-2]}&\bP^{2[-2]}&\bP^{3[-2]}&\bP^{4[-2]}\\
\bP^{1[-4]}&\bP^{2[-4]}&\bP^{3[-4]}&\bP^{4[-4]}
\eea
\)\;,\\
D_1&=&{\rm det}
\(
\bea{llll}
\bP^{1[+4]}&\bP^{2[+4]}&\bP^{3[+4]}&\bP^{4[+4]}\\
\bP^{1}&\bP^{2}&\bP^{3}&\bP^{4}\\
\bP^{1[-2]}&\bP^{2[-2]}&\bP^{3[-2]}&\bP^{4[-2]}\\
\bP^{1[-4]}&\bP^{2[-4]}&\bP^{3[-4]}&\bP^{4[-4]}
\eea
\)\;,\\
D_2&=&{\rm det}
\(
\bea{llll}
\bP^{1[+4]}&\bP^{2[+4]}&\bP^{3[+4]}&\bP^{4[+4]}\\
\bP^{1[+2]}&\bP^{2[+2]}&\bP^{3[+2]}&\bP^{4[+2]}\\
\bP^{1[-2]}&\bP^{2[-2]}&\bP^{3[-2]}&\bP^{4[-2]}\\
\bP^{1[-4]}&\bP^{2[-4]}&\bP^{3[-4]}&\bP^{4[-4]}
\eea
\)\;,\\
\bar D_1&=&{\rm det}
\(
\bea{llll}
\bP^{1[-4]}&\bP^{2[-4]}&\bP^{3[-4]}&\bP^{4[-4]}\\
\bP^{1}&\bP^{2}&\bP^{3}&\bP^{4}\\
\bP^{1[+2]}&\bP^{2[+2]}&\bP^{3[+2]}&\bP^{4[+2]}\\
\bP^{1[+4]}&\bP^{2[+4]}&\bP^{3[+4]}&\bP^{4[+4]}
\eea
\)\;,\\
\bar D_0&=&{\rm det}
\(
\bea{llll}
\bP^{1[-2]}&\bP^{2[-2]}&\bP^{3[-2]}&\bP^{4[-2]}\\
\bP^{1}&\bP^{2}&\bP^{3}&\bP^{4}\\
\bP^{1[+2]}&\bP^{2[+2]}&\bP^{3[+2]}&\bP^{4[+2]}\\
\bP^{1[+4]}&\bP^{2[+4]}&\bP^{3[+4]}&\bP^{4[+4]}
\eea
\)\;.
\eeqa

\subsection{Derivation of the quantization condition}\label{sec:gluingomega}

Let us explain the derivation of \eq{qquant} in detail.
For consistency with standard QSC notation \cite{Gromov:2015dfa} we denote in this section the two solutions of the Baxter equation \eq{Bax2p} as $q_1$ and $q_4$ which in the notation of section \ref{sec:baxter} corresponds to
\beq
    q_+=q_1, \ \ \ q_-=q_4 \ \ ,\label{eq:notationapp}
\eeq
with large $u$ asymptotics $q_1\sim e^{u\phi}u^\Delta,\ \ q_4\sim e^{-u\phi}u^{-\Delta}$.

First we notice that the Baxter equation \eq{Bax2p} is invariant under complex conjugation, so $\bar q_1$ and $\bar q_4$ are linear combination of the two solutions $q_1$ and $q_4$ with $i$-periodic coefficients that we denote $\Omega_i^j$,
\beqa
\label{q1O}
	\bar q_1&=&\Omega_1^1q_1+\Omega_1^4q_4 \\ \label{q4O}
	\bar q_4&=&\Omega_4^1q_1+\Omega_4^4q_4 \ .
\eeqa
Our strategy is to constrain as much as possible the form of $\Omega$'s and then fix them completely using the gluing conditions from the QSC.

The analytic properties of $q$'s already impose strong restrictions on $\Omega_i^j$. Both $q_1(u)$ and $q_4(u)$ are analytic in the upper half-plane, but the Baxter equation implies that they can have second order poles at $u=-in, \; n=1,2,\dots$ in the lower half-plane.
 Accordingly, $\bar q_1, \bar q_4$ will have second order poles in the upper half plane which can only originate from $\Omega$'s in the r.h.s.\! of \eq{q1O} and \eq{q4O}. Therefore these $\Omega$'s can have at most 2nd order poles. Their rate of growth at $u\to +\infty$ and $u\to -\infty$ is moreover constrained by the known asymptotics of $q_1,q_4$. To fix normalization we impose for $u\to +\infty$  
\beq
\label{qbas}
	q_1\sim e^{u \phi}u^\Delta, \ \ \    q_4\sim \frac{i}{8t^4\Delta \sin^2\phi }e^{-u \phi}u^{-\Delta}
\eeq
where the constant prefactor for $q_4$ is determined by the canonical normalisation of Q-functions\footnote{At finite angles we should have $q_1q_4\simeq i\frac{(\cos\theta-\cos\phi)^2}{2\Delta\sin^2\phi}$ at large $u$, see \cite{Gromov:2015dfa}.}). Assuming $\phi>0$ we see that $q_1$ is the dominant solution at $u\to +\infty$ and therefore e.g. $\Omega_4^1$ must vanish for large positive $u$ (though not necessarily for $u\to -\infty)$. By arguments of this type we can write all the components of $\Omega$ in terms of just a few parameters, namely 
\beqa
\label{q1b}
	\bar q_1&=&q_1\(\frac{a_1}{(e^{2\pi u}-1)^2}+\frac{a_2}{e^{2\pi u}-1}+1\)
	+q_4\(\frac{a_3}{(e^{2\pi u}-1)^2}+\frac{a_4}{e^{2\pi u}-1}-a_3+a_4\) \\
\label{q4b}
	\bar q_4&=&q_1\(\frac{b_1}{(e^{2\pi u}-1)^2}+\frac{b_2}{e^{2\pi u}-1}\)
	+q_4\(\frac{b_3}{(e^{2\pi u}-1)^2}+\frac{b_4}{e^{2\pi u}-1}-1\)
\eeqa
Moreover, we can use the trick suggested in \cite{Gromov:2017cja} to express these parameters $a_n,b_n$ in terms of $q$'s. As in \cite{Gromov:2016rrp} we will focus on $\Omega_1^4$, which as we see from \eq{q1b} is given by
\beq
\label{om14d}
   \Omega_1^4=\frac{a_3}{(e^{2\pi u}-1)^2}+\frac{a_4}{e^{2\pi u}-1}-a_3+a_4 \ \ . 
\eeq
Shifting $u\to u+i$ in \eq{q1O}, \eq{q4O} and using $i$-periodicity of $\Omega$ we find
\beqa
\label{q1O2}
	\bar q_1(u+i)&=&\Omega_1^1(u)q_1(u+i)
	+\Omega_1^4(u)q_4(u+i) \\ 
	\label{q4O2}
	\bar q_4(u+i)&=&\Omega_4^1(u)q_1(u+i)
	+\Omega_4^4(u)q_4(u+i) \ .
\eeqa
Now we can view the four equations \eq{q1O}, \eq{q4O}, \eq{q1O2}, \eq{q4O2} as a linear system on the four components of $\Omega$. Solving it we can we find $\Omega_1^4$,
\beq
\label{om14q1}
    \Omega_1^4=\frac{q_1(u+i)  \bar q_1(u)-q_1(u)\bar q_1(u+i)}{q_1(u+i)q_4(u)-q_1(u)q_4(u+i)} \ \ .
\eeq
Nicely, the denominator of \eq{om14q1} is precisely the Wronskian of the Baxter equation, which is a constant we denote by $C_W$. Its precise value is not important here but can be found from the asymptotics \eq{qbas},
\beq
    C_W\equiv q_1(u+i)q_4(u)-q_1(u)q_4(u+i)=-{\Delta t^4\sin\phi} \ .
\eeq
Thus we have
\beq
\label{omq}
    \Omega_1^4(u)=\frac{1}{C_W}
    \[q_1(u+i)  \bar q_1(u)-q_1(u)\bar q_1(u+i)\] \ .
\eeq
We expect that $\Omega_1^4$ has a singularity at $u=0$, which in this expression can only come from $\bar q_1(u+i)$. Using the fact that $\bar q_1$ satisfies the original Baxter equation \eq{Bax2p}, we find
\beq
    \bar q_1(u+i)=-\frac{4\hat g^2
   \bar q_1(0)}{u^2}-\frac{4 \hat g^2 \bar q_1'(0)+2 \Delta  \bar q_1(0) \sin \phi }{u}+\cO(1), \ \ u\to 0
\eeq
Plugging this into \eq{omq} gives
\beq
\label{ome1}
    \Omega_1^4= \frac{4 \hat g^2 q_1(0) \bar q_1(0)}{C_W u^2}+\frac{4 \hat g^2
   \left(\bar q_1(0) q_1'(0)+q_1(0) \bar q_1'(0)\right)+2 \Delta  q_1(0)
   \bar q_1(0) \sin \phi}{C_W u}+\cO(1) , \ \ u\to 0
\eeq
At the same time, expanding the expression for $\Omega_1^4$ from \eq{om14d} we find
\beq
\label{ome2}
    \Omega_1^4=\frac{a_3}{4 \pi ^2 u^2}+\frac{a_4-a_3}{2 \pi  u}+\cO(1), \ \ u\to 0
\eeq
Comparing \eq{ome1} with \eq{ome2} we can express $a_3$ and $a_4$ in terms of $q_1(0)$ and $q_1'(0)$, in particular\footnote{In a similar way we can express all parameters $a_n,b_n$ appearing in \eq{q1b}, \eq{q4b} in terms of the values of $q_1$, $q_4$ and their derivatives at $u=0$.}
\beq
\label{adif}
    a_3-a_4=
    -\frac{4 \pi  \left[2\hat g^2 \left(\bar q_1(0) q_1'(0)+q_1(0)
   \bar q_1'(0)\right)+\Delta  q_1(0) \bar q_1(0) \sin \phi
   \right]}{C_W} \ \ .
\eeq
So far we have not used any relations from the QSC involving analytic continuation around the branch points. Now we will apply one of such relations, which was derived in \cite{Gromov:2016rrp} using the gluing condition for $\tilde q_1$ given in \eq{qtil}. It reads
\beqa
\label{om14disc}
	\tilde\Omega_1^4-\Omega_1^4&=&u\bar q_1(u) q_1(u)-u\bar q_1(-u) q_1(-u)  \ \ .
\eeqa
In fact we will only use that as a consequence of this relation $\Omega_1^4$ must be even, which gives
\beq
	a_3=a_4 \ , \ \ \ \ \Omega_1^4=\frac{a_3}{4\sinh^2\pi u}\ \ .
\eeq
Combining the first relation with \eq{adif} we get precisely the quantization condition \eq{qquant} presented above.

\subsection{Quantization condition from asymptotics of the $\Omega$ functions}

There is also an alternative way to arrive at the quantization condition, which though just an observation at the moment is very instructive for the discussion that will follow in section \ref{sec:schr}. In this alternative approach we start from the same Baxter equation \eq{Bax2p} but never use any relations from the QSC involving tilde, i.e. analytic continuation around the branch points such as in \eq{qtil}. Instead we observed that it is sufficient to demand that $\Omega_1^4$ vanishes at $u\to +\infty$. This immediately fixes $a_3=a_4$ and thus leads via \eq{adif} (which as we showed above follows from the Baxter equation) to the same quantization condition \eq{qquant}. The importance of this observation will become apparent in section \ref{sec:schr}, where we will see that the vanishing asymptotics of $\Omega_1^4$ ensures finiteness of various scalar products that play a key role in our construction.

Curiously, in the fishnet theory \cite{Gurdogan:2015csr,Grabner:2017pgm} it is also possible to derive the quantization condition solely from asymptotics of $\Omega$ as was recently found in \cite{Grtoapp}.  It would be interesting to better understand the underlying reason behind this.

\section{Quantization condition and square-integrability of the wave function}
\la{app:quant}

In Sec. \ref{sec:BaxtoSchrod}, we introduced an explicit map  between the Q-function and a solution of the stationary  Schr\"odinger equation:
\beq
\frac{F(z)}{2\pi} = \, e^{-\Delta z/2 }  \vint \, q(u) \, e^{w_{\phi}(z) \, u} \, \frac{du}{2\pi i u} . \label{eq:qToFapp} 
\eeq
 As we showed there, the fact that $q(u)$ satisfies the Baxter equation implies that  $F(z)$ solves the Schr\"odinger equation. This statement does not require that the quantization conditions are satisfied, and is valid for any value of the parameter $\Delta$\footnote{ Notice that, strictly speaking, the integral transform in (\ref{eq:qToF}) requires $-1 < \Delta < 0$ for convergence. In this section we restrict consideration to this range of parameters, and then extend the result by analytic continuation. 
 }. 
 In this Appendix we show that, for $\Delta < 0$, the quantization conditions are equivalent to the square-integrability of $F(z)$. 
 In particular, notice that, since the potential in the Schr\"odinger equation is vanishing at infinity, any solution to (\ref{eq:Schrodinger}) 
 can have one of the two behaviours $ \sim e^{\pm \Delta z/2 }$ at large $z$, therefore it can either decay or grow exponentially. We will show that $F(z)$ is always decaying at $z \rightarrow + \infty$,  while it is decaying at $z \rightarrow -\infty$ if and only if $q(u)$ satisfies the quantization conditions. 

 We will use the same convention as in Sec. \ref{app:qsc} and denote the two independent solution of the Baxter equation as $q_1$ and $q_4$, see (\ref{eq:notationapp}), where $q(u)=q_1(u)$.

 They are characterized by the following asymptotics in the upper half plane
 \beq
 q_1(u) \sim e^{\phi u} \, u^{\Delta} ,  \;\;\;\;\; q_4(u) \sim   e^{-\phi u} \, u^{-\Delta}.\label{asyappq}
 \eeq
 In preparation for the following argument, we will need to determine the asymptotics of $q_1(u)$  also along the part of the integration contour in (\ref{eq:qToFapp}) which  extends in the lower half plane. 
 To determine the asymptotics along this line, we reflect it to the upper half plane using complex conjugation, and then use the exact relation (\ref{q1b}) between $q$ and $\bar{q}$. This leads to
 \beqa
  ( q(c-is) )^* &=& \bar{q}(c+is) = \Omega_1^1(c + i s) \, q_1(c+is)  +\Omega_1^4(c + i s) \, q_4(c + is) \simeq \Omega_1^4( c + i s) \, q_4( c + i s) \nn\\ 
 &\sim & e^{ - \phi  \,(c + i s)  } \,(c + i s)^{-\Delta} \, \left( a_4 - a_3 + \frac{a_4}{ e^{ 2 \pi ( c + i s ) }-1} + \frac{a_3}{\left( e^{ 2 \pi ( c + i s ) } -1 \right)^2} \right) ,\nn\\
\label{eq:qexpl}
 \eeqa
where the constants $a_3$, $a_4$ are  defined in (\ref{om14d}).  Notice that in (\ref{eq:qexpl}) we dropped the terms proportional to $\Omega_1^1$, since they give a subdominant contribution  suppressed as $ \sim u^{\Delta}$ (in this appendix we assume $\Delta < 0$ throughout). 
 Equation (\ref{eq:qexpl}) shows that $q(u)$ grows for large $| \text{Im}(u) |$ in the lower half plane. Despite this fact, notice that the integral (\ref{eq:qToF}) still converges as long as $-1 < \Delta < 0$, since, for any finite $z$, the integrand is oscillatory. 

Let us now come to the core of the argument. To determine the behaviour of $F(z)$ for $z \rightarrow +\infty$, we study the following limit
\beq
\lim_{z \rightarrow +\infty} e^{ \frac{ \Delta z}{2} } \, F(z) ,  \;\;\; ( \Delta < 0 ) ,
\eeq
which vanishes if and only if $F(z)$ is a decaying solution of the Schr\"odinger equation. From (\ref{eq:qToF}), we find
\beq
\lim_{z \rightarrow +\infty} e^{ \frac{ \Delta z}{2} } \, F(z) = \lim_{z \rightarrow +\infty} \vint \,\frac{ q(u) \, e^{-\phi u}  }{u}\,  e^{ +2 \sin\phi \, e^{-z} u } \, du = \vint \,\frac{ q(u) \, e^{-\phi u}  }{u}\, du  = 0 , \label{eq:vanint}
\eeq
 where the last term in (\ref{eq:vanint}) is zero due to the fact that the  integrand is suppressed at least as $\sim u^{\Delta - 1}$ at large $u$. Therefore, we found that $F(z)$ is always decaying for $z \rightarrow \infty$. 
 
 To analyse the situation at $z \sim -\infty$ we now look at the limit
 \beqa
\lim_{z \rightarrow - \infty} e^{ - \frac{\Delta  z }{2} } \, F(z) &=& \lim_{z \rightarrow -\infty} -i \,  \left( \vint \,\frac{ q(u) }{u}\, e^{ + \phi u } \, e^{ -2 \sin\phi e^z \, u } du \right) \label{eq:Fminus}\\
&=&  \vint \,\frac{ q(u) }{u}\, e^{ + \phi u } \, du  .\label{eq:testint}
\eeqa
Notice that by definition this limit is finite if and only if $F(z)$ is decaying at $z \sim -\infty$. Accordingly, we find that, for a generic value of $\Delta$, the last integral in (\ref{eq:testint})  is not convergent. To understand why, notice that, as a consequence of (\ref{eq:qexpl}), the integrand in (\ref{eq:testint}) behaves as
 \beq
 \frac{ q(u) \, e^{ \phi u} }{u} \sim (a_3 - a_4 )  \, u^{-1-\Delta}  , \;\;\;\; u \sim -i \infty \label{eq:qasysing}
 \eeq
 along the part of the contour extending in the lower-half plane. Therefore, the integral is clearly divergent. 
 
However, the quantization conditions coming from the QSC correspond precisely to $a_3 = a_4$ (see (\ref{adif}) )! 
When they are satisfied, the most singular part of the asymptotics (\ref{eq:qasysing}) is cancelled and the integral (\ref{eq:testint}) is still convergent, which implies that $F(z)$  is a square-integrable function.  
Therefore we have just shown that the (negative) scaling dimensions described by the QSC are associated with the spectrum of bound states of the Schr\"odinger equation (\ref{eq:Schrodinger}). While we derived this relation for $\Delta$ in a specified range $-1 < \Delta < 0$,  this correspondence can be extended beyond this regime by analytic continuation in the coupling constant. This analytic continuation is such that, for small enough coupling,  $\Delta_n$ becomes positive for almost all levels except for the ground state. In this regime, the scaling dimensions no longer correspond to bound states in terms of the Schr\"odinger potential problem, but can be understood as resonances.

\section{Perturbative results}
\label{app:exc}

Here we list our weak coupling results supplementing the main text.

First we present The perturbative results for $\Delta$ corresponding to the excited states with $L=2,3$, complementing the result for $L=1$ given in \eq{dn12}:
\beqa
    \Delta_{2,-}&=&2-2\frac{\sin2\phi}{\sin\phi}\hat g^2
    +(-8 \cos ^2\phi +16 \phi  \cos ^2\phi  \cot \phi +8)\hat g^4
    +\dots
    \\
    \Delta_{2,+}&=&2+2\frac{\sin2\phi}{\sin\phi}\hat g^2
   +(-8 \cos ^2\phi -16 \phi  \sin \phi  \cos \phi -8)\hat g^4
   +\dots
    \\
   \Delta_{3,-}&=&3-\frac{4\sin3\phi}{3\sin\phi}\hat g^2
   \\ \nn &&
   +\[\frac{16}{9} \phi  (2 \cos \phi -1)^3 (2 \cos \phi +1) \cot \frac{\phi
   }{2}-\frac{16}{27} \left((2 \cos 2 \phi +1)^2-18 \cos \phi \right)\]\hat g^4+\dots
   \\
   \Delta_{3,+}&=&3+\frac{4\sin3\phi}{3\sin\phi}\hat g^2
   \\ \nn &&
   +\[
   -\frac{16}{9} \phi 
   (2 \cos \phi -1) (2 \cos \phi +1)^3 \tan \frac{\phi }{2}
   -\frac{16}{27} \left((2 \cos 2 \phi +1)^2+18 \cos \phi \right)
   \]\hat g^4+\dots
\eeqa
For the $L=3^{\pm}$ excited states we also have\footnote{in the normalization where $q(u)\simeq  e^{u\phi}u^\Delta$ at large $u$ } 
\beq
\label{norm3w}
\br{ q^2 }= \pm 32 \, \hat g^2 \, (2 \cos (2 \phi )+1) +\dots \ \ \ .
\eeq

 Now let us present further results for the structure constants. For the HLL correlator with the $3^+$ state we have to leading order in the coupling
\beqa
\label{HLL3pw}
\left. C^{\bullet \circ \circ} \right|_{3^+,0,0} &=& \frac{\hat g }{6} \;\left(\cos (2 \phi_1 )+\frac{1}{2} \right)^{\frac{1}{2} } \,  \csc \left(\frac{\phi_1 }{2}\right) \cot(\phi_1 ) \sec ^2\left(\frac{\phi_1 }{2}\right)
   \cos \left(\frac{\phi_2-\phi_3}{2}\right) \\ \nn
   &\times &   ((\sec (\phi_1 )+2) \cos (\phi_2-\phi_3)-2 \sec (\phi_1 )-1) +\dots
\eeqa
and for the $3^-$ state:
\beqa
\label{HLL3mw}
\left. C^{\bullet \circ \circ} \right|_{3^-,0,0} &=& -\frac{\hat g }{6} \;\left(\cos (2 \phi_1 )+\frac{1}{2} \right)^{\frac{1}{2} } \, \, \csc^2\left({\frac{\phi_1 }{2}}\right) \csc({\phi_1 }) \, \sec \left({\frac{\phi_1 }{2}}\right)
    \, \sin \left({\frac{\phi_2-\phi_3}{2}}\right) \\ \nn
   &\times &  \left( ( 2 \cos( {\phi_1})-1) \, \cos( \phi_2-\phi_3)+\cos( {\phi_1})-2\right) +\dots
\eeqa
For the HHL structure constant with $(3^+,3^+)$ or $(3^-,3^-)$ states we find at leading order
\beqa
( C^{ \bullet \bullet \circ} )_{3^{\pm},3^{\pm}} &=& \frac{\cot (\phi_2 ) \csc (2 \phi_2 ) \cot (\phi_1 ) \csc (2 \phi_1 ) \, \left(2
   \hat g_2 ^2 \cos (2 \phi_2 )+\hat g_2 ^2+2 \hat g_1^2 \cos (2 \phi_1 )+ \hat g_1^2\right) }{4 \, \hat g_2  \, \hat g_1  \, \sqrt{2
   \cos (2 \phi_2 )+1} \sqrt{2 \cos (2 \phi )+1}} \nn\\   \label{HHL3ss}
   &\times&
  { \Big ( } 12 \cos\phi_3 \, (3 \sin\phi_1 \sin \phi_2+\cos\phi_1 \cos\phi_2 )-10 \cos(2 (\phi_1- \phi_2))\\ \nn &&-\cos (2 (\phi_1+\phi_2)) +8 \cos (2 \phi_1)+8 \cos (2 \phi_2)-3
   \cos (2 \phi_3)-14 { \Big ) } \\ \nn &+& \dots
\eeqa

\section{Results for the small-$\phi$ expansion}\label{app:smallPhi}
At $\phi=0$ the spectrum and resonances are described by the following trajectories \footnote{ Except for the ground state $\Delta_0$, each $\Delta_{n, \phi=0}$, $\Delta_{n, \phi=0}'$ corresponds to a patchwork of different excited states levels, which split at finite $\phi$, see Sec. \ref{sec:qscexc}. }
\beq
\Delta_{n, {\phi=0}} = \frac{1}{2} \, \left((2 n + 1) - \sqrt{ 1 + 16 \, \hat{g}^2 }  \right), \;\;\;\; \Delta_{n,\phi=0}'  = \frac{1}{2} \, \left((2 n + 1) + \sqrt{ 1 + 16 \, \hat{g}^2 }  \right),
\label{eq:Deltapm}
 \eeq
for $n=0,1,2,3,\dots$. 
Inspecting the numerical solution of the QSC equations for a few states, we observe a clear pattern in the scaling of the coefficients in the large-$u$ expansion (\ref{eq:largeuq}) for small $\phi$. 
 For a state converging to one of the trajectories $\Delta =\Delta_{n, {\phi=0}}  + O(\phi)^2$ or $\Delta = \Delta_{n, {\phi=0}}' + O(\phi^2)$, the coefficients scale as follows as $\phi \sim 0$: for even $n$,
\beq
\left\{ k_1,k_2, k_3, k_4 , \dots  \right\} \sim \left\{ \phi, \, 1, \, \phi,  \, 1, \, \phi,  \,1, \dots   \right\} \, \phi^{-n}, \;\;\; n=0, 2, 4, 6 , \dots , \label{eq:scale1}
\eeq
while for odd $n$:
\beq
\left\{ k_1, k_2, k_3, k_4 , \dots  \right\} \sim \left\{ \, 1, \,\phi, \, 1, \, \phi , \, 1, \, \phi,\, 1, \dots   \right\} \, \phi^{-n}, \;\;\; n=1, 3, 5 , 7,\dots .\label{eq:scale2}
\eeq
Notice that this means that the large-$u$ expansion becomes approximately even or odd for even or odd $n$, respectively.
Imposing the validity of a given scaling behaviour such as (\ref{eq:scale1}) or (\ref{eq:scale2}) generates all terms in the small-$\phi$ expansion of $\Delta$. 
 
 In particular, from the inspection of a few trajectories we conjecture a general formula for the expansion up to order $\phi^2$:

\beq\la{specsmallphi}
 \Delta = \frac{1}{2} \left((2 n + 1) \pm \sqrt{16 \hat g^2+1}\right)  \pm \frac{ \hat g^2 \left(  16 \hat g^2 \pm (2 n + 1)\, \sqrt{16 \hat g^2+1} + (2 n (n+1) -1 ) \right)}{\left(16
   \hat g^2-3\right) \sqrt{16 \hat g^2+1}} \, \phi^2 + \dots .
\eeq
We cross-checked this result at finite $\phi$ but large $g$ in section \eq{app:strong}.
Higher orders in $\phi$ are straightforward to obtain, even though the expressions become cumbersome. We report the result only for the ground state:
\beqa
\Delta_{0} &=& \frac{1}{2}\left(1  - \sqrt{1 + 16 \, \hat{g}^2 }\right) + \frac{\hat g^2 \left(-16 \hat g^2+\sqrt{16 \hat g^2+1}+1\right)}{\left(16
 \hat  g^2-3\right) \sqrt{16 \hat g^2+1}}\, \phi^2+ \\
 &+& \left( -\frac{(\tau-1) (\tau+1)^2 \left(5 \tau^5+40
   \tau^4+97 \tau^3+68 \tau^2+18
   \tau+24\right)}{768 \tau^3 (\tau+2)^3
   (\tau+4)} \right) \, \phi^4 + \dots ,
\eeqa
where we set $\tau = \sqrt{1 + 16 \, \hat{g}^2 }$. 

 As explained in Sec. \ref{sec:smallPhi}, one can also obtain a systematic expansion of the structure constants in the limit where $\phi_1 \sim \phi_2 \sim \phi_3 \sim 0$. In the case where the ground state is inserted at every cusp we obtain, up to next-to-leading order:
\beqa
&&\frac{C^{ \bullet \bullet \circ }_{123} }{\left. C^{ \bullet \bullet \circ }_{123} \right|_{\phi_1=\phi_2=\phi_3=0} }= 1 -2 (\phi_1 + \phi_2 - \phi_3 ) \, \left( \frac{ \left(\hat g_1^2 \phi_1 + \hat g_2^2
   \phi_2\right)}{\sqrt{16
   \hat g_1^2+1}+\sqrt{16 \hat g_2^2+1}} \right) \\
   &+& \phi_1^2 \,\left( \frac{\hat g_1^2 \left(-48 \hat g_1^2+\sqrt{16
   \hat g_1^2+1}-2 \left(-16 \hat g_1^2+\sqrt{16
   \hat g_1^2+1}+1\right) \psi
   ^{(0)}\left(\sqrt{16
   \hat g_1^2+1}\right)+7\right)}{2 \left(16
   \hat g_1^2-3\right) \sqrt{16 \hat g_1^2+1}} \right) \nn\\
   &+& \phi_2^2 \,\left( \frac{\hat g_2^2 \left(-48 \hat g_2^2+\sqrt{16
   \hat g_2^2+1}-2 \left(-16 \hat g_2^2+\sqrt{16
   \hat g_2^2+1}+1\right) \psi
   ^{(0)}\left(\sqrt{16
   \hat g_2^2+1}\right)+7\right)}{2 \left(16
   \hat g_2^2-3\right) \sqrt{16 \hat g_2^2+1}} \right) ,\nn
\eeqa
where $\psi^{(0)}(z) = \Gamma'(z)/\Gamma(z)$ and $ \left.  C^{ \bullet \bullet \circ }_{123} \right|_{\phi_1=\phi_2=\phi_3=0} $ is given in (\ref{eq:Cphi0}). 

For the norm of excited states at small $\phi$ we get
, in proximity of the trajectories (\ref{eq:Deltapm}),
\beqa\label{eq:normappDpm}
\br{ q^2_{\Delta_{n, \phi=0} } } &=&  \frac{(-1)^n \,(n!)}{\Gamma( 1 + n - 2  \, \Delta_{n, \phi=0} ) }  +   \dots = \frac{(-1)^n \,(n!)}{\Gamma( - n + \sqrt{ 1+ 16 \, \hat{g}^2 } ) }  +   \dots,\\
\br{ q^2_{\Delta'_{n, \phi=0} } } &=&  \frac{(-1)^n \,(n!)}{\Gamma( 1 + n - 2  \, \Delta'_{n, \phi=0} ) }  + \dots = \frac{(-1)^n \,(n!)}{\Gamma( - n - \sqrt{ 1+ 16 \, \hat{g}^2 } ) }  +   \dots 
\eeqa
 In the case of excited states, the small-angles limit for the numerator of structure constants depends on the relative scaling of the three angles. For example, for the HHL structure constants involving two $n=1$ trajectories, assuming $\phi_3=0$ and $\phi_1=\phi_2= \phi \sim 0$ small, we get
\beqa
&&\left( \left. C^{ \bullet \bullet \circ }_{123} \right|_{\phi_1=\phi_2= \phi \, , \, \phi_3=0} \right)_{n_1 = 1, n_2=1}\\
&=&-\frac{-\Delta_1^2+2 \Delta_1 \, \Delta_2 +\Delta_1 -\Delta_2^2+\Delta_2-2}{\Gamma (-\Delta_1 -\Delta_2+3)} \, \sqrt{  \Gamma( 2 - 2 \Delta_{1} ) \, \Gamma( 2 - 2 \Delta_{2} ) } +  O( \phi^2) ,\nn
\eeqa
while in the scaling $\phi_2 << \phi_1 \sim \phi_3 \sim 0$ we get
\beq
\left( \left. C^{ \bullet \bullet \circ }_{123} \right|_{\phi_1=\phi_3= \phi \, , \, \phi_2=0} \right)_{n_1 = 1, n_2=1}=-\frac{\sqrt{  \Gamma( 2 - 2 \Delta_{1} ) \, \Gamma( 2 - 2 \Delta_{2} ) }}{\Gamma (-\Delta_1 -\Delta_2+1)}  +  O( \phi^2) .
\eeq

\section{Strong coupling expansion}\la{app:strong}

Here we will describe the large $\hat g$ expansion of the spectrum. We will apply the WKB method used in the Fishnet theory in \cite{Gromov:2017cja}.
One should replace the Q-function in the Baxter equation by its semiclassical expression in terms of the quasi-momenta
$q(u)=\exp\({\hat g\int^{u/\hat g}p(x)dx}\)$, while also rescaling the spectral parameter to $x=u/\hat g$ and defining $d=-\Delta/\hat g$. After that we get
\beq
0=\left(-\frac{2 d \sin (\phi )}{x}+2 \cos
   (p(x))+\frac{4}{x^2}-2 \cos (\phi
   )\right)-\frac{p'(x) \cos
   (p(x))}{\hat g}+{\cal O}\left(\hat g^{-2}\right)
\eeq
Now we can solve for $p(x)$ at each order in $\hat g$:
\beqa
P\equiv e^{i p^{(0)}(x)}&=&\frac{2 x^2}{\sqrt{4 \left(d x \sin
   (\phi )+x^2 \cos (\phi )-2\right)^2-4
   x^4}+2 d x \sin (\phi )+2 x^2 \cos
   (\phi )-4}\\
p^{(1)}(x)&=&-\frac{i \left(P^2+1\right)
   p'^{(0)}(x)}{2
   \left(P^2-1\right)}\\
 p^{(2)}(x)&=&
 -\frac{\left(P^4+10 P^2+1\right)
   p''^{(0)}(x)}{12
   \left(P^2-1\right)^2}+\frac{3 i P^2
   \left(P^2+1\right)
   \left(p'^{(0)}(x)\right)^2}{2
   \left(P^2-1\right)^3}
\eeqa
where $p(x)=p^{(0)}(x)+\frac{1}{\hat g}p^{(1)}(x)+\frac{1}{\hat g^2}p^{(2)}(x)+\dots$.
Finally we impose
\beq\la{intp}
\frac{i}{2\pi}\oint p(x)dx=\frac{1}{\hat g}(n+1)\;.
\eeq
For $n\sim \hat g$ we get rather complicated elliptic integrals. However, for $n\sim 1$
the integral \eq{intp} can be computed easily by poles and the equation \eq{intp} gives the 
quantization condition for $\Delta_n$,
\beqa
{\Delta_n}\cos \left(\frac{\phi }{2}\right)
&=&-2\hat g+\(n+\frac{1}{2}\)+\frac{1}{\hat g}\(\frac{1}{16} (-2 n (n+1)-1)
   s^2-\frac{1}{16}\)\\
\nn &+& \frac{1}{\hat g^2}\(
 \frac{3}{64} (2 n+1) s^2-\frac{1}{128}
   (2 n+1) \left(n^2+n+1\right) s^4
 \)
+{\cal O}(1/\hat g^3)
\label{eq:strongcoupling}
\eeqa
where $s=\sin\frac{\phi}{2}$.
Re-expanding these relations at small $\phi$ 
we reproduce the large $\hat g$ expansion of \eq{specsmallphi}.
It would be interesting to compute the strong coupling asymptotics of the correlation functions using the WKB expansion presented in this appendix.

\section{The near-BPS limit}
\label{app:bps}

In this section we show that a formula very similar to the one we presented in \eq{Cinsert} in the ladders limit captures  $\d\Delta/\d\phi$ in a completely different regime -- namely in the near-BPS limit when $\phi\to\theta$. We will consider the generalized cusp dimension corresponding to $L$ scalars inserted at the cusp, which should however be independent from those coupling to the lines.\footnote{This observable is simpler than the one with insertions discussed in  section \ref{sec:excited} and corresponds from that perspective to the ground state, not an excited one.}

The QSC solution in this case was presented in \cite{Gromov:2013qga, Gromov:2015dfa} where the details can be found. The Q-function which we will use is $q=\bQ_1/\sqrt{u}$ which to leading order in $\phi-\theta$ is given by (up to irrelevant normalization)
\beq
    q_L=P_L(x)e^{g\phi(x-1/x)} \ ,
\eeq
where $L=0,1,2,\dots$ labels the R-charge of the inserted scalar operator and $x$ is the usual Zhukovsky variable \eq{defx} such that $x+1/x=u/g, \ |x|>1$.  Here $P_L(x)$ is given by
\beq
	P_L(x)=\frac{1}{\det {\cal M}_{2L}}\left|\begin{matrix}
	I_1^{\phi}& I_0^{\phi}& \cdots & I_{2-2L}^{\phi}  &I_{1-2L}^{\phi}\\
	I_2^{\phi}& I_1^{\phi}& \cdots & I_{3-2L}^{\phi} &I_{2-2L}^{\phi}\\
	\vdots      &  \vdots     &\ddots & \vdots            &\vdots           \\
	I_{2L}^{\phi}& I_{2L-1}^{\phi}& \cdots & I_{1}^{\phi}  &I_{0}^{\phi}\\
	x^{-L}& x^{1-L}& \cdots & x^{L-1} &x^{L}\\
	\end{matrix}\right|
\eeq
where
\beq
	{\cal M}_{N}=\begin{pmatrix}
	I_1^{\phi}& I_0^{\phi}& \cdots & I_{2-N}^{\phi}  &I_{1-N}^{\phi}\\
	I_2^{\phi}& I_1^{\phi}& \cdots & I_{3-N}^{\phi} &I_{2-N}^{\phi}\\
	\vdots      &  \vdots     &\ddots & \vdots            &\vdots           \\
	I_{N}^{\phi}& I_{N-1}^{\phi}& \cdots & I_{1}^{\phi}  &I_{0}^{\phi}\\
	I_{N+1}^{\phi}& I_{N}^{\phi}& \cdots & I_{2}^{\phi} &I_{1}^{\phi}
	\end{pmatrix}
\eeq
and the twisted Bessel functions are defined as
\beq
	I_n^\phi=\frac{1}{2}I_{n}\(4\pi g\sqrt{1-\frac{\phi^2}{\pi^2}}\)\[
	\(\sqrt{\frac{\pi+\phi}{\pi-\phi}}\)^{n}-
	(-1)^n\(\sqrt{\frac{\pi-\phi}{\pi+\phi}}\)^{n}
	\]\;.
\eeq
Notice a useful property
\beq
    P_L(x)=P_L(-1/x) \ .
\eeq

The key point is that for $P_L(x)$ we have a natural scalar product with respect to which they are orthogonal\footnote{It is also natural from their interpretation in matrix model terms, see \cite{Sizov:2013joa} and \cite{Gromov:2012eu}.}. For Q-functions it translates into orthogonality with respect to the scalar product
\beq
    \br{q_a\; q_{b}}_{\rm guess}\equiv \(2\sin\frac{\beta}{2}\)^\alpha\oint dx  \sinh(2\pi u) \;q_a\;q_b 
\eeq
where $q_aq_b\sim e^{\beta u}u^\alpha$ and the integral goes along the unit circle (which in the $u$ variable would correspond to going around the cut $[-2g,2g]$ \footnote{ Notice that this integration contour is consistent with the vertical one used in the main text of the paper. Indeed, our vertical integration contour can be bent and closed to the left; in general, we would need to take into account an infinite sequence of cuts of the Q functions at $[-2g, 2g] - i n$, but in the near-BPS limit only the cut at $[-2g,2g]$ remains.  }), i.e. we have
\beq
      \br{q_L\; q_{L'}}_{\rm guess} \propto \delta_{LL'} \ .
\eeq
The prefactor in the scalar product is defined in the same way as for the bracket \eq{eq:thebracket} we use in the main text. The full meaning of this scalar product and its precise relation with the bracket we used in the ladders limit are not completely clear yet. However it allows us to write $\d\Delta/\d\phi$ in almost exactly the same way as in the ladders limit where according to \eq{Cinsert} it corresponds to an insertion of $u$ in the integral:
\beq
\label{derlad}
   -2\frac{\d(\sin\phi\Delta)}{\d\phi}=\frac{\br{q^2 u}}{\br{q^2}}
     \ \ \ \ \ \ \ \  \text{(ladders limit)}
\eeq
Remarkably we find that in the near-BPS case this derivative again corresponds to an insertion of $u$ ! That is,
\beq
\label{derbps}
    2\left.\frac{\d(\sin\phi\Delta)}{\d\phi}\right|_{\phi=\theta}=
    \frac{\br{q^2u}_{\rm guess}}{\br{q^2}_{\rm guess}} \ \ \ \ \ \ \ \text{(near-BPS limit)}
\eeq
so the only difference with the ladders limit is the overall sign (whose interpretation remains to be understood).
Concretely,  in the near-BPS limit we have
\beq
    \Delta=(\phi-\theta)\Delta^{(1)}(g,\phi)+\cO((\phi-\theta)^2)
\eeq
so that
\beq
    \left.\frac{\d\Delta}{\d\phi}\right|_{\phi=\theta}=\Delta^{(1)}(g,\phi)
\eeq
and our formula \eq{derbps} precisely reproduces the complicated all-loop result from \cite{Gromov:2013qga} which reads
\beq
    \Delta^{(1)}(g,\phi)=(-1)^{L+1}(\phi-\theta)g
    \frac{{\rm det}{\cal M}_{2L+1}^{(1,2L+2)}}{{\rm det}{\cal M}_{2L}}
\eeq
where ${\cal M}_N^{(a,b)}$ is the matrix ${\cal M}_N$ with row $a$ and column $b$ deleted.

 Regardless, it is rather nontrivial that \eq{derbps} provides the correct non-perturbative result. This may be viewed as a hint towards the existence of an underlying structure capturing the exact result at all values of the parameters.
 As an important testing ground, it would be very interesting to see whether replacing $\br{} \rightarrow \br{}_{\text{guess}}$ in our main result \eq{correlator} yields the structure constants in the near-BPS limit, which should also be accessible with localization \cite{Giombi:2018qox}.

\section{More details on the space-time dependence of 4pt functions}

\label{app:4pt}

Here we give a few more details on the space-time dependence of the basic 4pt function \eq{G4WW} (given in OPE terms in \eq{G1234delta4}). First we discuss some alternative parameterization of the spacetime dependence in terms of the angles and crossratios. Then we show that when two points collide the spacetime dependence matches the one for a 3pt correlator as expected.

\subsection{Parameterization of the four points}

\label{sec:4ptparam}

 Let us first show how to eliminate the two coordinates $y_0$, $y_5$, defined in Sec. \ref{sec:4cusp}, in favour of the angles $\phi$, $\phi_{12}\equiv\phi_1-\phi_2$, $\phi_{43}\equiv\phi_4-\phi_3$ (defined by \eq{tophi})\footnote{Notice that  the angles can be seen as parameters specifying the configuration, i.e. the four operators corresponding to the four points. In particular the structure constants depend on these angles. }. 
 We will see that the result depends only on the cross ratio $r_{1234}$ of the four insertion points, together with the angles $\phi$, $\phi_{43}$, $\phi_{21}$. 
 Translating between the $\Lambda$ parametrization and the space-time coordinates, we find
\beq
\frac{y_{12} \, y_{34} }{y_{13} \, y_{24}} =
-\frac{\cosh(\frac{\Lambda_3 - \Lambda_4 - i \phi }{2}) \, \cosh(\frac{\Lambda_1 - \Lambda_2 + i \phi }{2})}{\sinh(\frac{\Lambda_2 + \Lambda_4}{2} ) \, \sinh(\frac{\Lambda_1 + \Lambda_3}{2} )},
\eeq
which, together with (\ref{tophi}),  implies
\beq\label{eq:quadratic}
\frac{1}{r_{1234} } =  L_{034} \, L_{012} \, \left( e^{ 2 \Lambda  } + e^{-2 \Lambda} \right)  - K_{043} \, K_{021} - K_{012} \, K_{034}  ,
\eeq
where $\Lambda = \frac{1}{4} ( \Lambda_1 + \Lambda_2 + \Lambda_3 + \Lambda_4 )$, 
\beq
r_{abcd} = \frac{ | y_{ab} \, y_{cd } | }{ | y_{ac} \, y_{bd} | },
\eeq
and we recall that $L_{abc}$ and $K_{abc}$ are defined as
\beq
K_{abc} = \frac{ \sin\frac{1}{2}( \phi_a + \phi_b - \phi_c ) }{\sin\phi_a }, \;\;\;\;\; L_{abc} = \sqrt{ K_{abc} \, K_{acb}  }.
\eeq
Solving (\ref{eq:quadratic})  for $e^{-2 \Lambda}$, and plugging it back in the four point function, we see that the terms  (\ref{eq:4ptfactors}) appearing in the OPE expansion of the correlator are simple algebraic functions of the cross ratio $r_{1234}$ . 

Finally, let us mention that the factors $K_{0ab}$ can be interpreted as particular cross ratios involving the points $y_0$ and $y_5$. 
In fact from (\ref{tophi}), converting from  $\Lambda_i$'s to  space-time points we find 
\beqa
e^{-i \phi_{43} } &=&
e^{i \phi} + 2 i \sin\phi \, \left(\frac{y_{40} \, y_{35}}{ y_{34} \, y_{05} } \right) = e^{-i \phi} + 2 i \sin\phi \, \left(\frac{y_{45} \, y_{30}}{ y_{34} \, y_{05} } \right) , \\
e^{-i \phi_{12} } &=& e^{i \phi} + 2 i \sin\phi \, \left(\frac{y_{20} \, y_{15}}{ y_{12} \, y_{05} } \right) = e^{-i \phi} + 2 i \sin\phi \, \left(\frac{y_{25} \, y_{10}}{ y_{12} \, y_{05} } \right) , 
\eeqa
from which we see that 
\begin{align}
&r_{3045} = \frac{\sin\frac{1}{2}(\phi + \phi_3 - \phi_4 ) }{\sin\phi} = K_{034} , \;\;\;\;\;\; 
&r_{3540} = \frac{\sin\frac{1}{2}(\phi + \phi_4 - \phi_3 ) }{\sin\phi} = K_{043}, \\ 
&r_{1520} = \frac{\sin\frac{1}{2}(\phi + \phi_1 - \phi_2 ) }{\sin\phi} = K_{012} , \;\;\;\; \;\; &r_{1025} = \frac{\sin\frac{1}{2}(\phi + \phi_2 - \phi_1 ) }{\sin\phi} = K_{021}.
\end{align}

\subsection{HLL correlator from the 4-point function}
\label{HLLspace}
Let us verify explicitly that taking the limit of two coincident points in our 4-point function reproduces the correct spacetime dependence of the 3-point HLL correlator. The general proof of this was given in Section~\ref{sec:3pt1}, 
here we will check this at 1 loop (testing also the 1-loop HLL structure constant).  

 We will consider the limit when
\beq
    \Lambda_1=\Lambda_2 \equiv \Lambda\to\infty
\eeq
but $\Lambda_3,\Lambda_4$ are finite.
Then the left ends of the two arcs in Fig.~\ref{fig:4cusp} will approach the first cusp point. The four arc endpoints  correspond to $y_{1},\dots, y_4$\ , and for large $\Lambda$ the two left endpoints are at equal small distance $\epsilon$ from the cusp,
\beq
    |y_1-x_1|=|y_2-x_1|=\epsilon, \ \epsilon\to 0
\eeq
so that $\Lambda$ is related to the distance as (see \eq{eq:lambdacutoff0})
\beq
    \Lambda=\log\frac{|x_1-x_2|}{\epsilon} \ .
\eeq
The perturbative expression for the 4-pt function \eq{G41loop} reduces in this limit to
\beqa
    G=&&1-\frac{2i \hat g^2}{\sin\phi} 
    \left[2 i \Lambda _3 \phi -2 i \phi  \log \left(\frac{1}{\Lambda
   }\right)-\text{Li}_2\left(-e^{-i \phi +\Lambda _3-\Lambda _4}\right)+\text{Li}_2\left(-e^{i
   \phi +\Lambda _3-\Lambda _4}\right)
   \right. \nn \\  && \left.
   -\text{Li}_2\left(-e^{-i \phi
   }\right)+\text{Li}_2\left(-e^{i \phi }\right)\right]\label{GLL}
\eeqa
It is far from obvious that the dependence on the 3 endpoint positions here (two are parameterized by $\Lambda_3,\Lambda_4$ while the last one is $x_1$) is the one expected for a CFT 3-pt correlator. In the notation given on Fig.~\ref{fig:4cusp} this dependence should be of the form
\beq
\label{Gcft3}
    { G}_{\rm CFT}=\frac{1}{|y_3-y_4|^{-\Delta_0}|x_1-y_3|^{\Delta_0}|x_1-y_4|^{\Delta_0}}
\eeq
corresponding to a HLL correlator of 3 cusps without insertions, with $\Delta_0$ being the ground state anomalous dimension. In order to compare this expression with \eq{GLL} we plug into \eq{Gcft3} the coordinates $y_3=\zeta_+(\Lambda_3),\  y_4=\zeta_-(-\Lambda_4)$ using the parameterization \eq{eq:zazb}, and also use that by simple geometry the angles $\phi_3,\phi_4$ are related to $\Lambda_3,\Lambda_4$  by
\beq
    e^{\Lambda_4-\Lambda_3}=
    \frac{\sin\frac{\phi-\phi_4+\phi_3}{2}}{\sin\frac{\phi+\phi_4-\phi_3}{2}} \ .
\eeq
Then taking the ratio of \eq{GLL} and \eq{Gcft3} we find after some manipulations
\beqa\nn
    \frac{G}{G_{\rm CFT}}&=&1+
    \hat g^2 \csc \phi  \left[2 \phi  \log \left(\frac{2 \sin ^2\phi }{\cos \delta \phi -\cos
   \phi }\right)+i \text{Li}_2\left(e^{-i \phi } \csc \frac{\delta \phi +\phi
   }{2} \sin \frac{\delta \phi -\phi }{2}\right)
   \right.
   \\ \nn && \left.
   -i \text{Li}_2\left(e^{i
   \phi } \csc \frac{\delta \phi +\phi }{2} \sin \frac{\delta \phi -\phi
   }{2}\right)+i \text{Li}_2\left(e^{-i \phi } \csc \frac{\delta \phi -\phi
   }{2} \sin \frac{\delta \phi +\phi }{2}\right)
   \right. \\ \nn && \left.
   -i \text{Li}_2\left(e^{i
   \phi } \csc \frac{\delta \phi -\phi }{2} \sin \frac{\delta \phi +\phi
   }{2}\right)+2 i \text{Li}_2\left(-e^{-i \phi }\right)-2 i \text{Li}_2\left(-e^{i \phi
   }\right)-4 \phi  \log (\epsilon )\right]\\ 
   &=& 1
    + \hat{g}^2  F_{123}(\phi , \phi_4, \phi_3 )  + \hat g^2 \[ \Delta_0^{(1)} \log \epsilon + \log\(2\cos\frac{\phi}{2}\)+F_{123}\(\phi,\frac{\pi}{2},\frac{\pi}{2}\)\] \label{eq:G15}
\eeqa
where $\Delta_0^{(1)} = 4 \phi \, \csc(\phi)$ is the 1-loop  ground state dimension, $\delta\phi=\phi_4-\phi_3$ and $F_{123}$ is the 1-loop HLL  structure constant  given as a function of the three angles in (\ref{F123}). Remarkably, we see that all spacetime dependence (involving $\Lambda_3,\Lambda_4$) has disappeared in the ratio $G/G_{\rm CFT}$~! What remains in \eq{eq:G15} is a function only of the regulator $\epsilon$ and the angles $\phi,\phi_3$ and $\phi_4$ which characterise the three cusp operators whose correlator we are computing.  Furthermore, the term in square brackets in  \eq{eq:G15} precisely matches the 2pt  normalization factor from \eq{NDelta} at 1 loop. If we divide by this factor in order to get the normalized correlator, what is left is precisely the HLL structure constant for three ground states  $C^{\bullet oo}=1+\hat g^2 F_{123}(\phi,\phi_4,\phi_3)$ matching the 1-loop expansion \eq{F123} of our exact result. 

Thus we have verified at 1 loop that in the limit when two points collide we recover perfectly the 3pt correlator from the 4pt function, including the correct normalization and spacetime dependence. This is a direct 1-loop check of our all-loop result for the HLL correlator.

\end{document}